# Benchmarking Cylindrical Blast Wave Theory Against the OSIRIS-REx Sample Return Capsule Reentry

Elizabeth A. Silber[1,*]

Sandia National Laboratories, Albuquerque, NM, 87123

*Corresponding author: esilbe[at]sandia.gov

**Accepted for publication on 12 May 2026 in Pure and Applied Geophysics**
**Special issue: Shaping the Future of Nuclear Explosion Monitoring and Verification**

## Abstract

Weak shock theory based on cylindrical blast waves has been used to interpret meteor infrasound, but it has not been systematically benchmarked against a non-ablating hypersonic source with independently known parameters. The objective of this study is not to propose a new theoretical framework, but to evaluate the operational validity of the existing suite of blast radius formulations against a high-fidelity ground truth dataset. The OSIRIS-REx Sample Return Capsule reentry on 24 September 2023 provides such a benchmark because the capsule geometry, trajectory, and infrasound emission points are constrained from mission data and ray tracing, reducing source-side uncertainty associated with ablation. Using observations from 39 infrasound stations, this benchmarking study evaluates six published blast radius ($R_0$) formulations and three weak-shock transition coefficients ($C$) within a stratified atmospheric propagation model to predict signal period and peak overpressure. The benchmarking identifies the Sakurai formulation as the best-performing formulation for non-ablating bodies, with the Jones/Plooster formulation performing comparably when a physically appropriate $C$ is adopted. Sakurai and Jones/Plooster yield linear-period median absolute percentage residuals of 9% and 11%, respectively. The period predictions show only weak sensitivity to $C$ at these propagation distances. The Mach-diameter approximation commonly used in meteor studies overestimates $R_0$ by more than a factor of 3 in the absence of ablation. These results establish a performance baseline for applying cylindrical blast wave theory to non-ablating hypersonic bodies and demonstrate that the signal period is a robust observable for constraining $R_0$.

## 1. Introduction

The Origins, Spectral Interpretation, Resource Identification, Security, Regolith Explorer (OSIRIS-REx) Sample Return Capsule (SRC) reentry on 24 September 2023 served as a rare "ground truth" experiment for the geophysical community (Silber et al., 2024). Unlike natural meteoroids, whose dimensions, density, and composition must be inferred indirectly, the SRC had well-constrained physical and kinematic parameters known *a priori* (Francis et al., 2024; Lauretta et al., 2017). This work leverages infrasound signals recorded at 39 stations, each sampling a different emission point along the trajectory (Silber and Bowman, 2025), to systematically evaluate cylindrical blast wave formulations and isolate the underlying physics from the unpredictable effects of mass loss and fragmentation. Complementary infrasound source analyses of the OSIRIS-REx SRC based on Burgers-equation propagation and an independent observational dataset have also been reported (Bishop et al., 2025). The present work focuses on benchmarking cylindrical blast-wave formulations within the weak-shock treatment and addresses a different set of questions. The topic is also timely in the context of renewed deep-space capsule returns. The April 2026 return of the Orion spacecraft during the Artemis II mission of the National Aeronautics and Space Administration (NASA) (Dooren, 2026) illustrates the continuing operational relevance of interpreting high-speed reentries and their atmospheric signatures. Although Orion occupies a different size and thermal-protection regime from the OSIRIS-REx SRC (e.g., Thomas, 2026), both belong to the broader class of geometrically well-constrained spacecraft returns for which physically informed acoustic signature interpretation is increasingly relevant.

Hypersonic objects traversing the Earth's atmosphere generate shock waves that decay through distinct propagation regimes: from the strongly nonlinear near the source, through a weak shock regime where the waveform retains its N-wave character, and ultimately into the linear acoustic domain where classical wave theory applies (Lin, 1954; Plooster, 1970; ReVelle, 1976; Sachdev, 1972; Silber et al., 2018). As these shock waves decay, they produce infrasound, sub-audible acoustic waves with frequencies below 20 Hz that propagate efficiently over long distances with minimal attenuation (Evans et al., 1972). Infrasound is one of the four sensing modalities employed by the Comprehensive Nuclear-Test-Ban Treaty Organization (CTBTO) International Monitoring System (IMS) for detecting atmospheric explosions (Campus and Christie, 2009; Christie and Campus, 2010; Hupe et al., 2022), and it has been widely applied to the detection and characterization of bolides (e.g., Ott et al., 2019; Pilger et al., 2019; Silber et al., 2011; Wilson et al., 2025) and other atmospheric events (e.g., Evers et al., 2018; Green et al., 2011; Nippress and Green, 2017; Vergoz et al., 2022). Physically benchmarked source models for hypersonic atmospheric entries therefore have broader monitoring value, because they help separate source physics from propagation effects and improve interpretation of impulsive infrasound signals in operational atmospheric-monitoring datasets. The characteristic blast radius, $R_0$, defines the spatial scale of the strongly nonlinear region near the source and is determined by the energy deposited per unit path length, $E_0$ (in J/m), and the ambient pressure, $p_0$ (in Pa), through various formulations that differ in their normalization constants (e.g., ReVelle, 1976; Silber and Brown, 2019).

The weak shock approach stems from cylindrical blast wave theory and was developed by ReVelle (1974) for meteoroid entries through a stratified, inhomogeneous atmosphere, under the assumption that the source is a non-fragmenting body without strong ablation. The approach provides closed-form expressions for the evolution of signal period and overpressure amplitude as functions of propagation distance from the source. A critical parameter in this formulation is the distance which determines where the transition from weak shock to linear acoustic propagation occurs. This parameter is important both for accurately predicting the infrasound signal properties (period and amplitude) observed at ground-level receivers and for inverting observed signals to recover source function. The weak-shock transition distance is controlled by a dimensionless coefficient, $C$, which has been assigned different values by different authors: Morse and Ingard (1968) derived $C$ = 5.38, while Towne (1967), as cited in ReVelle (1974), proposed $C$ = 34.3.

Silber et al. (2015) provided the first systematic observational test of the weak shock approach using a dataset of 24 centimeter-sized meteoroids simultaneously detected by video and infrasound at regional distances (< 300 km). Notably, those meteoroids were strongly ablating, a regime where ReVelle's original assumptions are not strictly satisfied. Using optical photometry as ground truth for energy deposition, Silber et al. (2015) found that the weak shock approach correctly predicts blast radii from infrasound periods but systematically underpredicts $R_0$ when derived from pressure amplitude. They also found that the effect of gravity wave perturbations to the wind field on $R_0$ is relatively small (<10%). By adjusting the weak-shock transition distance coefficient, Silber et al. (2015) obtained an approximate empirically derived coefficient that in most cases fits both period and amplitude for a single $R_0$. The weak shock approach has been applied almost exclusively to ablating meteoroids, which present fundamental challenges for validation: the meteoroid's physical dimensions, density, and shape are generally not directly measured but must be inferred from optical photometry and entry modeling (e.g., Brown et al., 2011; Koschny et al., 2017; Silber et al., 2015). Furthermore, ablation introduces time-varying energy deposition, fragmentation, and mass loss that complicate the interpretation of blast radius values (Silber and Brown, 2019). Although ReVelle applied a version of weak shock approach to the Stardust SRC reentry (ReVelle and Edwards, 2006), that analysis was limited in scope.

As part of the largest geophysical observational campaign for a hypersonic reentry to date (Silber et al., 2024), the SRC reentry was detected by 39 infrasound stations deployed across Nevada and Utah, each sampling a different emission point along the trajectory (Silber and Bowman, 2025). This multi-station dataset, combined with the known capsule parameters and raytracing-derived emission points, provides the inputs needed to evaluate the weak shock approach without ambiguity: $R_0$ can be computed from independently known source conditions at each station. This contrasts significantly with meteor studies, where the source altitude and velocity at the emission point must typically be inferred jointly with optical observation and entry modeling (e.g., Ceplecha et al., 1998; Devillepoix et al., 2020; Silber et al., 2015).

Operationally, weak shock analyses of meteors and reentry vehicles hinge on two coupled modeling choices that are rarely benchmarked: the blast radius normalization used to compute $R_0$, and the weak shock transition distance coefficient $C$ that controls the weak shock to linear transition. The OSIRIS-REx SRC is uniquely suited to isolate this ambiguity because it is a rigid, effectively non-ablating body with independently known geometry and trajectory, and because raytracing constrains the emission point associated with each of the 39 infrasound stations. In this benchmarking study, success is defined by period agreement at the network level, specifically a network-median absolute period residual of ≤15%, while peak overpressure is evaluated but treated as a secondary diagnostic due to its greater sensitivity to propagation effects and transition assumptions.

Here, the weak shock formulation is applied to the OSIRIS-REx SRC reentry data to address several interrelated questions. Unlike previous studies (ReVelle, 1974, 1976; Silber et al., 2015), which employed a single $R_0$ formulation, this study systematically benchmarks six formulations of the blast radius: Few (1969), Jones et al. (1968)/Plooster (1970), Sakurai (1965), Tsikulin (1970) (standard and modified), and the Mach-diameter approximation (ReVelle, 1974, 1976), to determine whether any formulation is applicable as-is or whether modifications are required. Second, three weak shock transition (WST) distance coefficients ($C$ = 5.38, 34.3, and 57.2) are tested to determine whether any existing coefficient reproduces the observations. The study also examines whether a unique solution exists for the $R_0$ and $C$ combination, evaluates the relative utility of period versus amplitude as observational constraints, and discusses implications of the findings for the physical interpretation of $R_0$ and the weak shock-to-linear transition for non-ablating hypersonic sources. In summary, this paper provides three deliverables: a formulation-level ranking that identifies which $R_0$ normalization best reproduces the observed infrasound period for a rigid body, a determination that the signal period is the operationally reliable observable while the amplitude is not, and a quantitative characterization of the systematic along-trajectory residual trend, which defines the altitude range over which the best-performing formulations remain within the adopted error limits and identifies where a source-condition-dependent bias might emerge.

## 2. Theoretical Background

Shock waves are broadly classified by their strength, defined as the ratio of the absolute pressure $p$ immediately behind the shock front to the undisturbed ambient pressure $p_0$: $\zeta = p/p_0$ (Krehl, 2009). In the strong shock regime ($\zeta \gg 1$), the wave propagates supersonically and exhibits strongly nonlinear behavior. As the wave decays and $\zeta$ approaches unity, it enters the weak shock regime, where the overpressure is small but finite and the propagation velocity approaches the ambient sound speed (Needham, 2018; Sakurai, 1965).

In the weak shock regime, the waveform takes on the characteristic N-wave pressure signature (Dumond et al., 1946), so named because the pressure-time profile resembles the letter "N". As shown independently by Landau (1945) and Dumond et al. (1946), cylindrical and spherical N-waves decay more rapidly than predicted by geometrical spreading alone, while the wave period increases with propagation distance. These second-order nonlinear effects, cumulative waveform steepening and enhanced decay, persist over many thousands of wavelengths and govern the evolution of the signal from the source to the receiver (Maglieri and Plotkin, 1991; Wright, 1983).

The mathematical treatment of blast wave propagation was advanced through the similarity principle (Sedov, 1946; Taylor, 1950), which reduces the governing partial differential equations to ordinary differential equations while preserving the essential nonlinear behavior (Sachdev, 2004; Sakurai, 1965). A body moving at hypersonic speeds generates a conical bow shock with Mach angle $\eta = \sin^{-1}(1/M)$, where $M$ is Mach number, defined as the ratio of the body speed $v$ to the local sound speed $c_s$, that is, $M = v/c_s$. For meteoroids, which enter the atmosphere at Mach numbers ranging from ~35 to 270 (e.g., Ceplecha et al., 1998), the Mach cone is extremely narrow ($\eta$ < 1.7°). Under such conditions, the shock can be approximated as a cylinder, leading to the treatment of the shock as an instantaneous cylindrical line source (ReVelle, 1976; Tsikulin, 1970).

The complete mathematical formulation of the weak shock model, including all expressions for propagation through a stratified, windy atmosphere with absorption, is given in Silber et al. (2015) and Silber and Brown (2019). Only the expressions essential for interpreting the results of this study are presented here.

## 2.1 Cylindrical Blast Wave Theory and the Blast Radius

A hypersonic body traversing the atmosphere deposits energy along its flight path, generating a cylindrical shock wave that expands radially from the trajectory (e.g., Lin, 1954; Plooster, 1968; Tsikulin, 1970). The blast radius $R_0$ characterizes the radial extent of the strongly nonlinear region near the source and is related to the energy deposited per unit path length $E_0$ and the ambient pressure $p_0$ through various formulations that differ in their normalization constants (see **Table 1**). The energy deposited per unit path length for a body experiencing pure aerodynamic drag is given by:

$$E_0 = \frac{1}{2}\rho_0 v^2 C_D S, \qquad (1)$$

where $\rho_0$ is the ambient air density (kg/m$^3$), $v$ is the velocity (m/s), $C_D$ is the drag coefficient, and $S$ is the reference cross-sectional area (m$^2$) (ReVelle, 1974). For a sphere or hemisphere, $S = \pi\left(\frac{d_m}{2}\right)^2$, where $d_m$ is the effective body diameter. For the OSIRIS-REx SRC with $d_m$ = 0.81, this expression gives $S$~0.515 m$^2$.

Multiple formulations of $R_0$ exist in literature, reflecting different normalizations of the cylindrical blast wave solution. These are summarized in **Table 1**, following the catalog presented in Silber and Brown (2019).

For Mach numbers much greater than unity ($M \gg 1$), ReVelle (1974) introduced a convenient approximation relating $R_0$ directly to the Mach number $M$ and the effective body diameter $d_m$ of the source body (i.e., $R_0 \sim M d_m$); here $d_m$ denotes the characteristic physical diameter of the body, taken for the OSIRIS-REx SRC as the known capsule diameter of 0.81 m. However, the Mach-diameter approximation deserves scrutiny for non-ablating bodies. The $M d_m$ scaling was derived empirically from ablating meteoroid observations, where the total energy budget includes both aerodynamic drag and ablative mass loss (Bronshten, 1983; Silber et al., 2018). For an ablating body, the vaporized and fragmented material expands into the wake, depositing additional energy that enlarges the effective blast radius beyond what aerodynamic drag alone would produce. In this sense, $M d_m$ is an empirical scaling law calibrated to 'dirty' shocks, those with ablation products, rather than a first-principles physical relationship. For a rigid, non-ablating object such as the SRC, the energy source is strictly aerodynamic drag (Eq. (1)) with no ablative contribution. Consequently, $M d_m$ is expected to systematically overestimate $R_0$ for the SRC, a prediction confirmed quantitatively in **Section 4**.

**Table 1.** Blast radius formulations from the literature. $E_0$ is energy deposited per unit path length and $p_0$ is ambient pressure.

| Formulation | Expression | Reference | Equation |
|---|---|---|---|
| Few | $R_0 = \left(\frac{E_0}{\pi p_0}\right)^{1/2}$ | Few (1969) | (2a) |
| Jones/Plooster | $R_0 = \left(\frac{E_0}{\gamma b p_0}\right)^{1/2}$<br>$b$ = 3.94 and $\gamma$ = 1.4 | Jones et al. (1968)<br>Plooster (1970) | (2b) |
| Sakurai | $R_0 = \left(\frac{E_0}{2\pi p_0}\right)^{1/2}$ | Sakurai (1965) | (2c) |
| Tsikulin (std) | $R_0 = \left(\frac{E_0}{p_0}\right)^{1/2}$ | Tsikulin (1970)<br>standard definition | (2d) |
| Tsikulin (mod) | $R_0 = \left(\frac{2E_0}{p_0}\right)^{1/2}$ | Tsikulin (1970)<br>modified definition | (2e) |
| Mach-diameter | $R_0 \sim M d_m$ | ReVelle (1974) | (2f) |

## 2.2 Weak Shock Propagation

In the weak shock regime, the N-wave generated by the cylindrical blast wave decays according to the analytical relationships derived by ReVelle (1974). The signal period $\tau$ and overpressure amplitude $\Delta p$ evolve with the scaled distance $x = r/R_0$ from the source trajectory. In practice, $r$ is approximated by the slant range along the raytracing path from the emission point to the station, projected perpendicular to the trajectory. The fundamental period at the source is:

$$\tau_0 = 2.81 \frac{R_0}{c_s}, \quad (3)$$

where $c_s$ is the sound speed at the source altitude. As the N-wave propagates, the period increases:

$$\tau(x) = 0.562\, \tau_0\, x^{1/4}. \quad (4)$$

The weak shock overpressure ratio decays as:

$$\frac{\Delta p}{p_0} = \frac{2(\gamma+1)}{\gamma} \left(\frac{3}{8}\right)^{-\frac{3}{5}} \left\{ \left[ 1 + \left(\frac{8}{3}\right)^{\frac{8}{5}} x^2 \right]^{\frac{3}{8}} - 1 \right\}^{-1}. \quad (5)$$

In the limits $x \ll 1$and $x \gg 1$, Eq. (5) approaches $\Delta p/p_0 \propto x^{-2}$ and $\Delta p/p_0 \propto x^{-3/4}$, respectively, recovering the expected near-field strong-shock and far-field weak-shock cylindrical decays. Equations (4) and (5) describe the weak-shock evolution in a homogeneous medium as functions of the scaled distance $x = r/R_0$. In the full implementation used here, these relations are applied stepwise through a stratified atmosphere, with local density, sound speed, wind, and absorption updated at each propagation step along the source-to-station path (ReVelle, 1974, 1976; Silber et al., 2015; Sutherland and Bass, 2004). The complete mathematical treatment for propagation in a vertically varying atmosphere is given in Silber et al. (2015) and Silber and Brown (2019). The expressions above are retained here because they define the underlying functional dependence of period and overpressure on scaled distance that governs the physical interpretation of the results presented here. In the forward model, wind effects are incorporated through the effective sound speed computed from the user-selected atmospheric specification. Atmospheric absorption was treated using either the classical formulation of ReVelle (1974) or the frequency-dependent model of Sutherland and Bass (2004).

For the cylindrical blast wave analogy to hypersonic flow to be valid (Pan and Sotomayer, 1972), several conditions beyond $M \gg 1$ must be satisfied: the energy release must be effectively instantaneous, and the Mach angle must remain small many body diameters behind the source (Tsikulin, 1970). These conditions are well satisfied for meteoroids, which encounter the atmosphere at $v > 11$ km/s. For the OSIRIS-REx SRC, which decelerates substantially along its trajectory, the conditions are evaluated locally at each emission point rather than globally: the local Mach number ranges from ~8 at the lowest observed source altitude ($\eta \sim 6°$) to ~35 at the highest ($\eta < 2°$), satisfying $v >> c_s$ at every point along the observed portion of the trajectory. The line source is also assumed to radiate in free field, independent of reflections from finite boundaries such as topographic features (ReVelle, 1974).

It is important to note that the ReVelle (1974) weak shock formulation, as applied here, is strictly valid for direct acoustic arrivals, those propagating along refracted ray paths from the source to the station without intermediate reflections or ducting. At greater distances, atmospheric dynamics including stratospheric and thermospheric refractions, ducting, caustic formation, and diffraction into geometrical-acoustics shadow zones may produce arrivals whose amplitude and waveform character depart from the direct-path weak-shock predictions. Caustic regions are zones of ray convergence, where simple geometrical acoustics predicts strong local amplification and finite-frequency effects can distort the received waveform. These conditions lie outside the assumptions of the ReVelle weak shock treatment, and signals of this type should therefore be excluded from the analysis.

The 39 stations used in this study are all within direct-arrival range (slant distances of ~43 to 76 km), and the raytracing confirms direct propagation paths for all stations. Moreover, the weak shock model is semi-analytical: it combines closed-form cylindrical blast wave similarity solutions for the evolution of the signal period and amplitude (ReVelle, 1974, 1976) with numerical integration through a realistically stratified atmosphere, incorporating altitude-dependent absorption (Sutherland and Bass, 2004), the weak-shock-to-linear transition criterion governed by the WST coefficient $C$. This approach preserves the physical transparency of the analytical blast wave scaling while accommodating the altitude-dependent atmospheric structure that precludes a single closed-form solution.

In addition to forward modeling (propagating a known $R_0$ to predict receiver-side observables such as signal period and peak overpressure), the weak shock formulation can be inverted to derive $R_0$ from the observed signal period and amplitude (**Section 3**). For details, see Silber et al. (2015).

### 2.3 Weak Shock Transition Coefficient and the Linear Regime

The propagation of the blast wave is governed by the weak shock formulation until the signal transitions to the linear acoustic regime. This transition is controlled by a dimensionless coefficient $C$ that enters the weak-shock-to-linear transition criterion through the ratio

$$d = \frac{c_s \tau}{C \frac{\Delta p}{p_0}}, \qquad (6)$$

where $c_s$ is the local sound speed, and $\tau$ is the instantaneous period, and $\Delta p/p_0$ is the local overpressure ratio, all evaluated at the current propagation step.

Eq. (6) has the same basic dependence on sound speed, period, and overpressure ratio as the classical nonlinear distortion or shock-formation scale familiar from nonlinear acoustics, namely $c\tau/(\Delta p/p_0)$ (e.g., Pierce, 2019; Rudenko and Soluyan, 1977). In the present study, however, $d$ is used operationally within the ReVelle weak-shock treatment as the transition length scale governing when the weak shock solution is replaced by linear propagation. Because the classical shock-formation distance is derived for a plane sinusoid in a quasi-homogeneous medium, whereas the ReVelle (1974) treatment applies these scales to a cylindrical N-wave propagating through a stratified atmosphere, a more formal connection between these quantities is left for future work.

The quantity $d$ has units of length and serves here as the operational nonlinear distortion scale used to determine when the weak shock solution is abandoned in favor of linear propagation. At each propagation step, $d$ is compared to the remaining slant range; when $d$ exceeds the remaining distance, the period is frozen at its current value and propagated linearly to the receiver. Concurrently, the amplitude is propagated in the linear regime to the receiver under geometric spreading and absorption. For physical interpretation, propagation distances are sometimes expressed as multiples of $R_0$ throughout this paper (e.g., 10,000 $R_0$).

Three published values of $C$ are tested, representing two distinct families of physical criterion. Morse and Ingard (1968) obtained $C$ = 5.38 from the characteristic nonlinear-distortion length, which is the distance scale over which finite-amplitude effects first produce a shock. Towne (1967), as adopted by ReVelle (1974), proposed $C$ = 34.3 by defining the distortion distance (ReVelle's terminology for the equivalent of $d$ in that specific context) as the distance over which cumulative nonlinear distortion produces a 10% change in the period of the cylindrical weak shock (under the weak-shock evolution $\tau \propto x^{1/4}$). Silber et al. (2015) reported an average period change of approximately 6% in their optical infrasound-meteor dataset; applying the same scaling ($C f_{pc} \approx constant$, where $f_{pc}$ is the fractional period change that sets $d$ in this criterion family) to the 6% observation of Silber et al. (2015) yields $C$ = 34.3 × 10 / 6 ≈ 57.2, which is the third value adopted in the present benchmarking. The Morse and Ingard (1968) value therefore derives from a shock-formation-distance criterion, whereas the Towne (1967) value and the Silber-based value derive from a period-change criterion applied to ReVelle's distortion distance. Because the three values originate from distinct physical definitions, they are treated here as candidate scalars entering Eq. (6) through the same functional form; accordingly, $C$ is referred to hereafter as the weak shock transition (WST) coefficient, and the associated length scale $d$ as the WST distance. The benchmarking reported in **Section 4** is empirical in nature: it identifies which value of $C$, taken as a single operational scalar in Eq. (6), produces predictions closest to the 39-station observations. A larger $C$ produces a smaller WST distance d at every propagation step (since $d \propto 1/C$). Because the weak-shock-to-linear transition is declared only when $d$ exceeds the remaining slant distance to the station, a smaller $d$ delays satisfaction of this criterion and the wave therefore remains in the weak-shock regime over a longer portion of the propagation path. The transition consequently occurs later along the path, at a lower altitude and closer to the station, producing greater period growth and a greater reduction of the overpressure ratio $\Delta p/p_0$ before the signal enters the linear regime.

Once the transition is identified, the period freezes at its value at the transition altitude, while the amplitude continues to evolve under geometric spreading and linear absorption. For a non-ablating body such as the OSIRIS-REx SRC, the transition to the linear regime may occur sooner (at higher altitude, farther from the station) than for an equivalently sized ablating meteoroid. This difference arises from a fundamental contrast in how energy is deposited along the trail. An ablating meteoroid continuously sheds mass, and the vaporized material expands radially into the wake, producing an ablationally amplified recompression shock (Silber et al., 2018) that sustains the cylindrical shock structure by feeding additional energy into the wavefront over an extended length of the trail. The energy deposition from a fragmenting meteoroid is further distributed spatially, creating a diffuse, time-varying source that reinforces the shock over a broad region. A non-ablating body, by contrast, deposits energy primarily through aerodynamic drag and the associated bow-shock flow field that encloses the hypersonic flow field around the vehicle, without the additional distributed energy input from ablation products. The structure and strength of this shock are governed by classical hypersonic gas dynamics and depend closely on the object's shape (geometry), diameter, and velocity, with no additional energy injection from ablation products, vapor expansion, or wake-deposited momentum downstream of the body. Because the energy source is confined to the immediate vicinity of the bow shock rather than distributed along an extended wake, the resulting cylindrical blast wave is expected to decay more rapidly toward the linear acoustic regime. Additionally, the rigid, fixed geometry of the SRC produces a consistent, coherent shock front at every point along the trajectory, in contrast to the evolving cross-section of an ablating body whose effective diameter changes continuously as mass is lost. The shock is therefore expected to decay more rapidly toward the linear regime.

It is worth noting that at the propagation distances relevant to the OSIRIS-REx SRC observations (slant distances of approximately 43 to 76 km from the source trajectory; see Silber and Bowman (2025)), the difference between the weak shock period ($\tau_{ws}$) and the linear period ($\tau_{lin}$) is small. This arises because the period evolves slowly with distance (as $x^{1/4}$), and the transition altitude, while influencing the amplitude, produces only modest differences in the predicted period. This observation has implications for the relative utility of period versus amplitude as constraints on $R_0$.

## 3. Data and Methods

### 3.1 OSIRIS-REx SRC Parameters

The OSIRIS-REx SRC is a blunt-body hemispherical capsule with the following engineering specifications relevant to this analysis: diameter $d_m$ = 0.81 m, mass $m$ = 46 kg (including the sample), and atmospheric entry velocity $v \approx$ 12.36 km/s at the entry interface (Francis et al., 2024). The entry angle was nearly horizontal, ~8.2°, resulting in a long atmospheric flight path and extended interaction with the upper atmosphere. The capsule features a PICA (Phenolic Impregnated Carbon Ablator) heat shield (Covington, 2005; Tran et al., 1996), but the heat shield was not expected to undergo significant mass loss during the hypersonic phase relevant to infrasound generation (above ~40 km altitude) (Francis et al., 2024; Lauretta et al., 2017; Silber et al., 2024). The shape factor for a hemisphere is $A_s$ = 1.92, and the wave drag coefficient for a blunt body at hypersonic speeds is $C_D \approx 1.0$ (ReVelle, 1974, 1976; ReVelle and Edwards, 2006; Silber and Brown, 2019; Silber et al., 2015). Aerodynamic analysis of the geometrically identical Stardust SRC by Mitcheltree et al. (1999) yields axial force coefficients of ~1.48 to 1.56 in the hypersonic continuum regime using thermochemical nonequilibrium computational fluid dynamics (CFD), with little variation across the Mach 7 to 43 range. The value $C_D$ = 1.0 adopted here is therefore smaller than the total aerodynamic drag coefficient and is best interpreted as an effective parameter controlling the fraction of aerodynamic energy coupled into the cylindrical blast wave. Because all energy-based $R_0$ formulations scale identically as $C_D^{1/2}$, this assumption does not affect the relative ranking among formulations.

The shape factor describes the relation between body mass and effective cross-sectional area, while the drag coefficient enters the drag-based energy-deposition estimate by converting dynamic pressure into aerodynamic drag and hence into energy deposition per unit path length (Ceplecha et al., 1998). For the rigid SRC, both the effective cross-sectional area and $C_D$ are treated as constant along the portion of the trajectory analyzed here. Under that assumption, the energy deposition per unit path length is taken to arise entirely from aerodynamic drag, with no contribution from ablative mass loss. **Figure 1** shows the thermoaerodynamic environment along the SRC descent trajectory from the entry, descent, and landing (EDL) entry state (Francis et al., 2024).

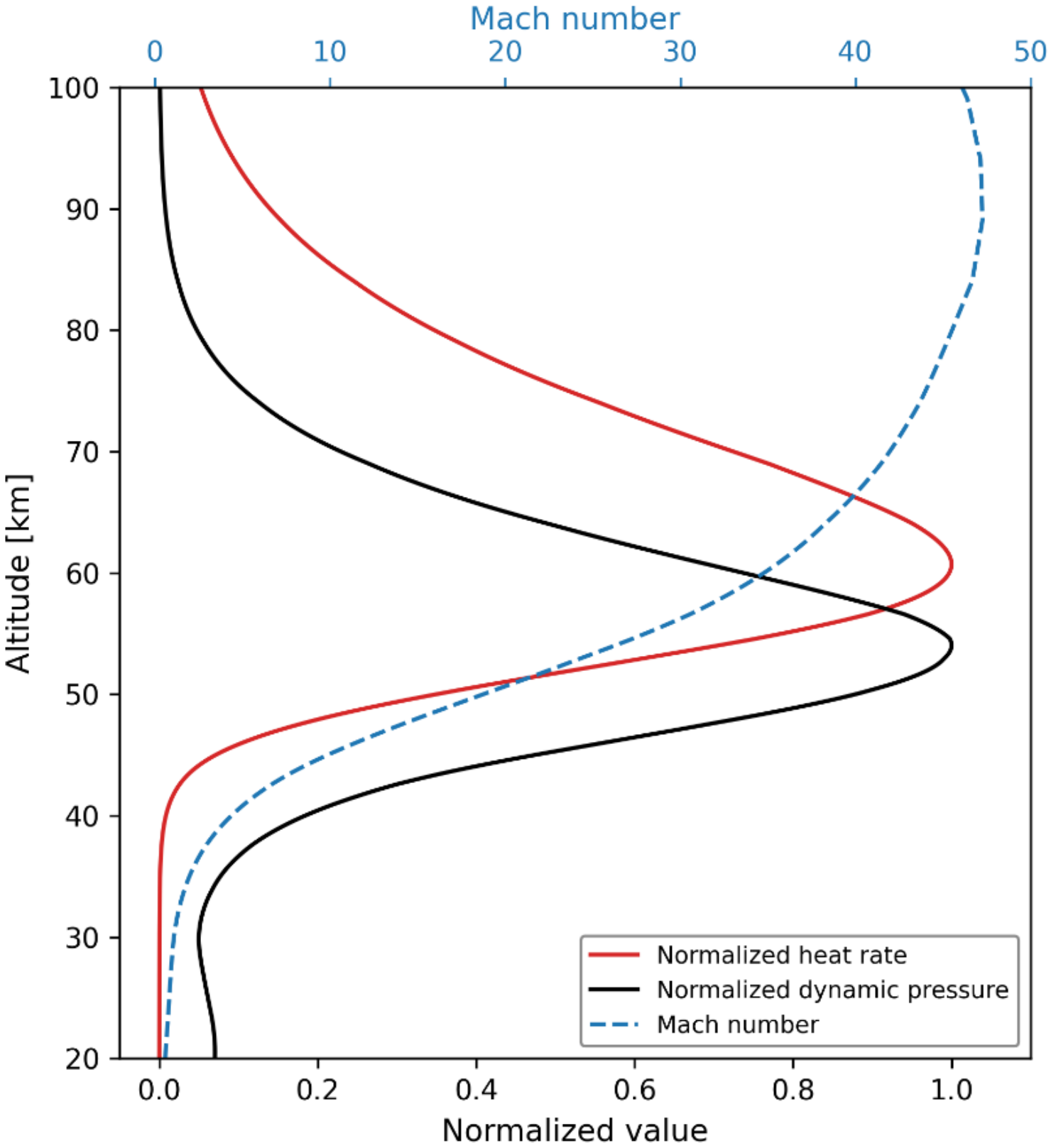

**Figure 1:** Normalized aerothermal environment along the OSIRIS-REx SRC trajectory between 20 and 100 km altitude. Heat rate (red, solid) and dynamic pressure (black, solid) are each normalized to their respective maxima. Mach number (blue, dashed; upper axis) is shown on its native scale. Peak dynamic pressure occurs near 55 km, while peak convective heating occurs at a slightly higher altitude (~60 km), consistent with the stronger velocity dependence of heat rate ($\sim v^3$) compared to dynamic pressure ($\sim v^2$).

## 3.2 Observational Data and Atmospheric Specifications

Infrasound signals from the OSIRIS-REx SRC reentry were recorded at a dense network of 39 single-sensor infrasound stations deployed across Utah and Nevada, as described in detail by Silber and Bowman (2025). That study provides the event-specific observational inputs adopted here: (1) sensor locations (latitude, longitude, elevation, (2) measured signal periods and amplitudes, and (3) station-associated emission points along the trajectory (latitude, longitude, altitude). For completeness, a brief summary of the observational dataset, the period-picking procedure, and the atmospheric specification underlying those source-point assignments is provided below. All subsequent weak-shock modeling, derived quantities, benchmarking comparisons, and statistical analyses presented in this paper were carried out in the present study.

The geographic configuration of the station network and the SRC ground track are shown in **Figure 2**. The stations are organized into three lines, denoted A, T, and C, and span the portion of the trajectory from which the direct infrasound arrivals analyzed in this work were identified. The corresponding station-associated source points sample reentry altitudes of approximately 44–62 km, providing a distributed set of along-trajectory constraints for the benchmarking analysis.

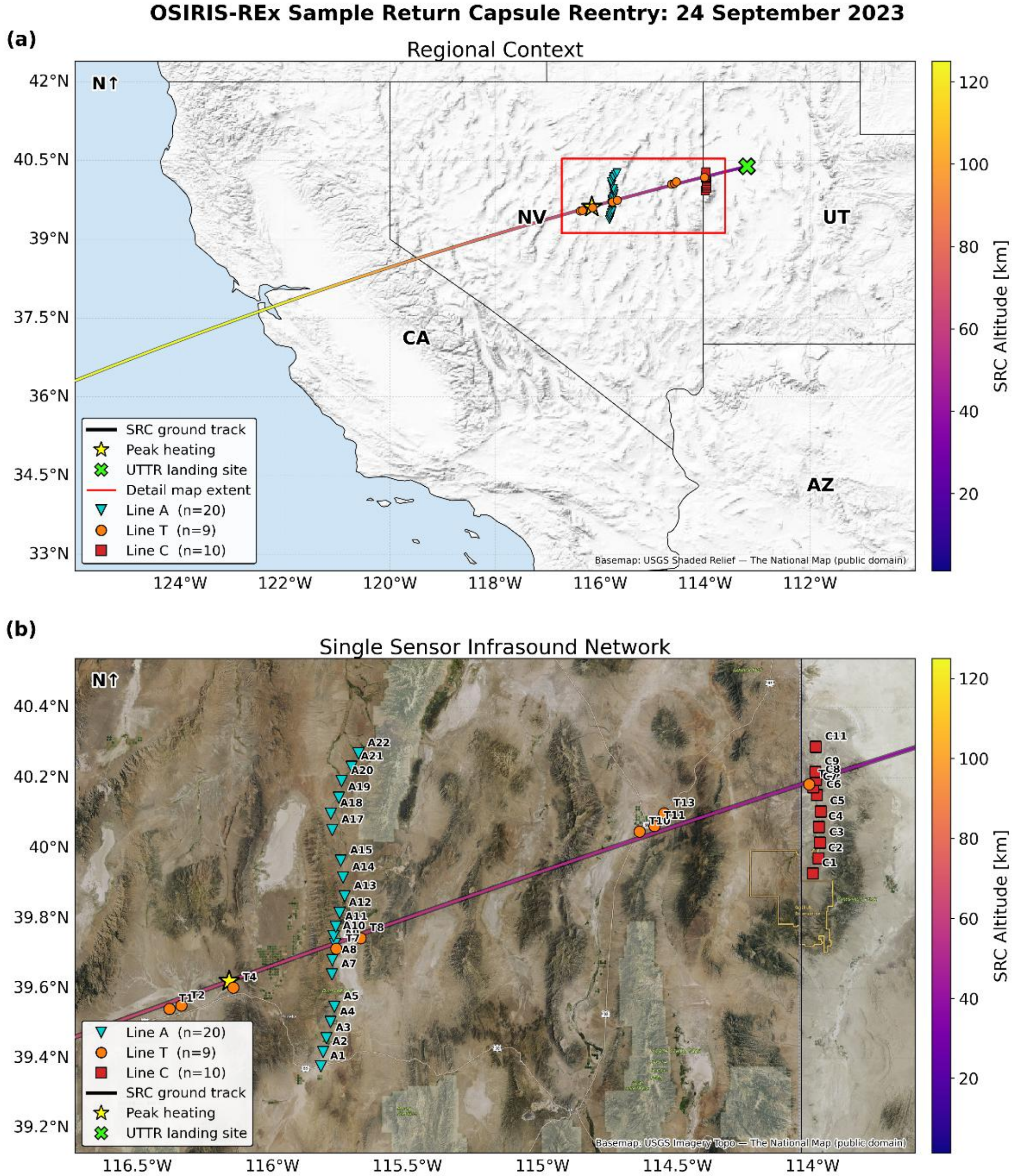

**Figure 2:** Station network and SRC ground track for the OSIRIS-REx reentry used in this study. (a) Regional context map showing the full projected SRC ground track across the western United States. The star denotes the location of peak heating and the green symbol indicates the Utah Test and Training Range (UTTR) landing site. (b) Map of the infrasound network, showing the three station lines (A, T, and C) relative to the sampled segment of the SRC ground track. The track is colored by SRC altitude.

Silber and Bowman (2025) constrained the source altitudes (infrasound emission points along the trajectory) using eigenray modeling with infraGA, an open-source geometric ray-tracing software package (Blom, 2014; Blom and Waxler, 2017). Representative examples of the resulting eigenray families are shown in **Figure 3** for one station from line A and one station from line C. These panels illustrate the direct-arrival geometry underlying the adopted source-point assignments; the full eigenray analysis is reported in Silber and Bowman (2025) and is not repeated here.

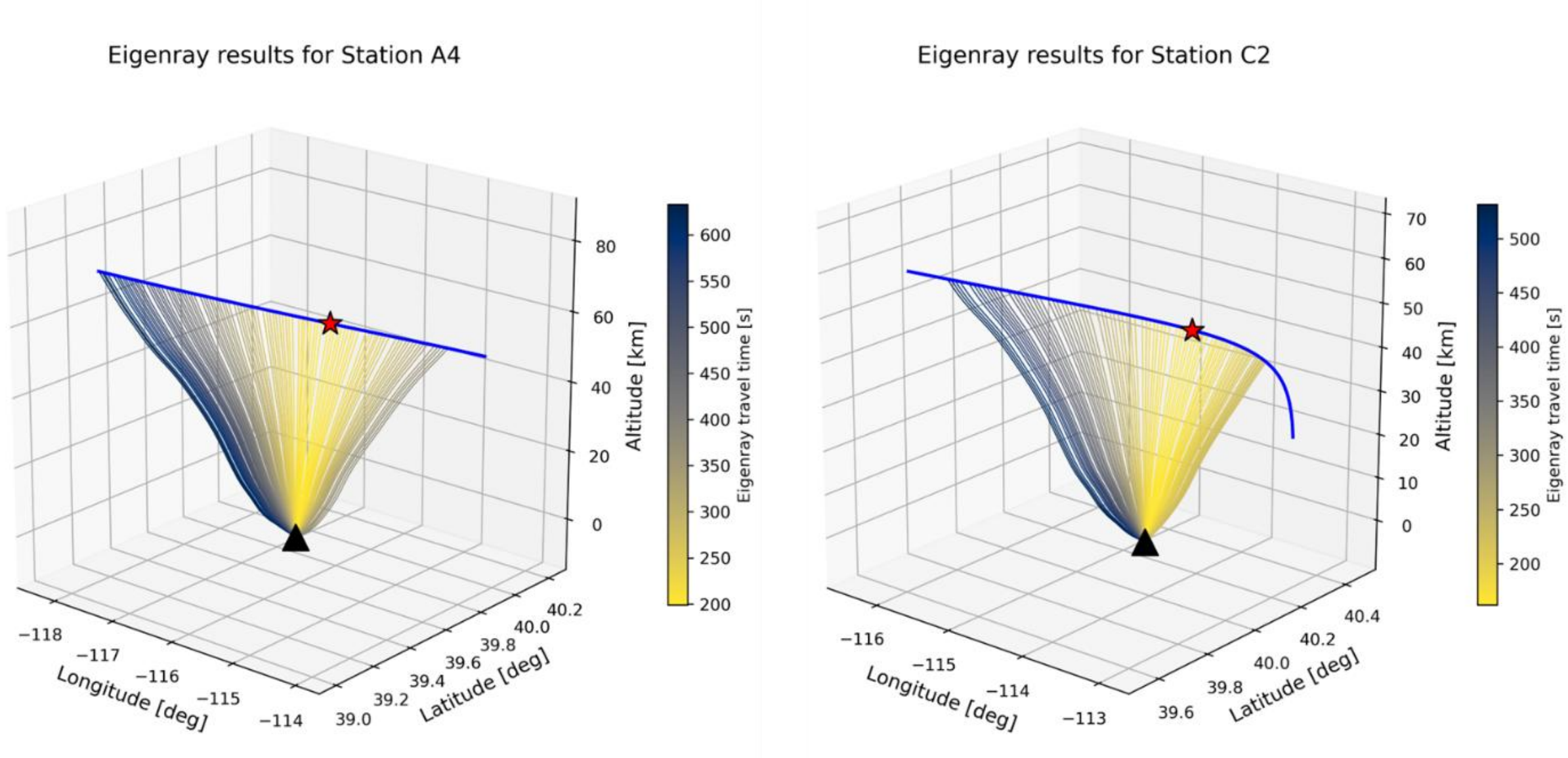

**Figure 3:** Representative eigenray families used in the source-point analysis of Silber and Bowman (2025). Shown are example results for station A04 (left) and station C02 (right). In each panel, the black triangle marks the station location, the blue curve shows the relevant segment of the OSIRIS-REx EDL trajectory, and the red star marks the station-associated emission point adopted for the present study. Ray paths are colored by eigenray travel time. These examples illustrate the direct-arrival geometry underlying the source-point assignments used here.

The ray tracing used Ground-to-Space (G2S) atmospheric specifications (Drob et al., 2003), provided by the National Center for Physical Acoustics (NCPA) at the University of Mississippi (Hetzer, 2024), for the date, time, and location of the reentry. These profiles provide altitude-dependent temperature, pressure, density, and zonal and meridional wind components for infrasound propagation analysis. G2S is a composite atmospheric product that can incorporate lower- and middle-atmosphere meteorological fields from products such as the Global Forecast System (GFS) (National Centers for Environmental Prediction, 2025) and the Modern-Era Retrospective analysis for Research and Applications, Version 2 (MERRA-2) (e.g., Randles et al., 2017), together with upper-atmosphere models such as the Naval Research Laboratory Mass Spectrometer and Incoherent Scatter Radar Extended-2000 (NRLMSISE-00) empirical atmosphere and the Horizontal Wind Model (HWM) (Drob et al., 2013; Drob et al., 2003; Picone, 2002). The same atmospheric specifications used by Silber and Bowman (2025) were employed here for consistency. The wind profile and an example of the sound speed profile are shown in **Figure 4**.

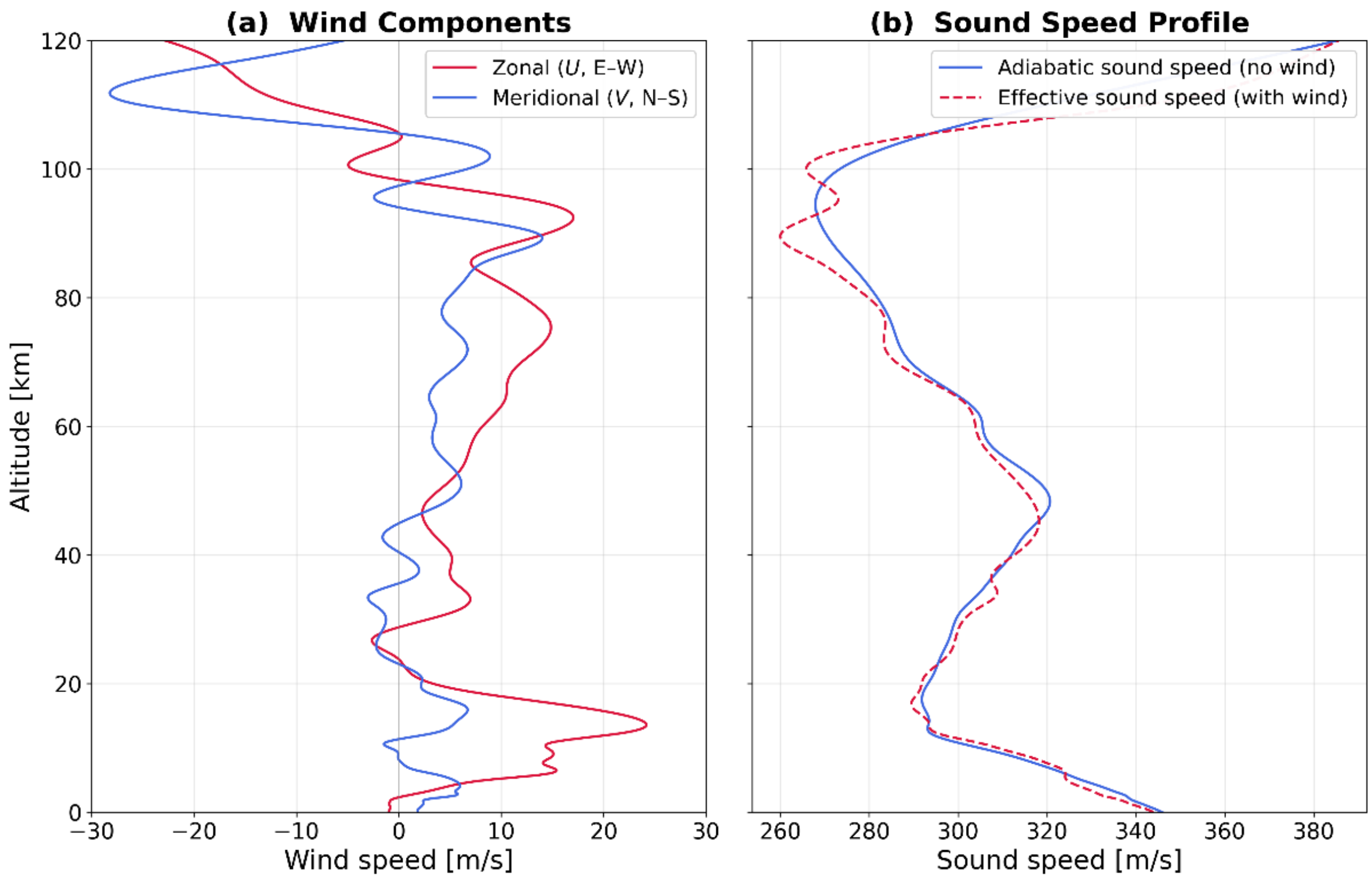

**Figure 4:** (a) Atmospheric wind profile at the Eureka Airport. Zonal ($U$) and meridional ($V$) wind components from the G2S model as functions of altitude from the surface to 120 km for the date and location of the OSIRIS-REx SRC reentry. (b) The adiabatic and effective sound speed at Station A3.

Representative example signals are shown in **Figure 5**, which presents time-domain waveforms and spectrograms for one station from line A and one station from line C. These examples illustrate the impulsive character of the OSIRIS-REx arrivals, the leading compressional phase used by Silber and Bowman (2025) for the published period measurements adopted here, and the trailing coda that motivates the chosen period-picking convention.

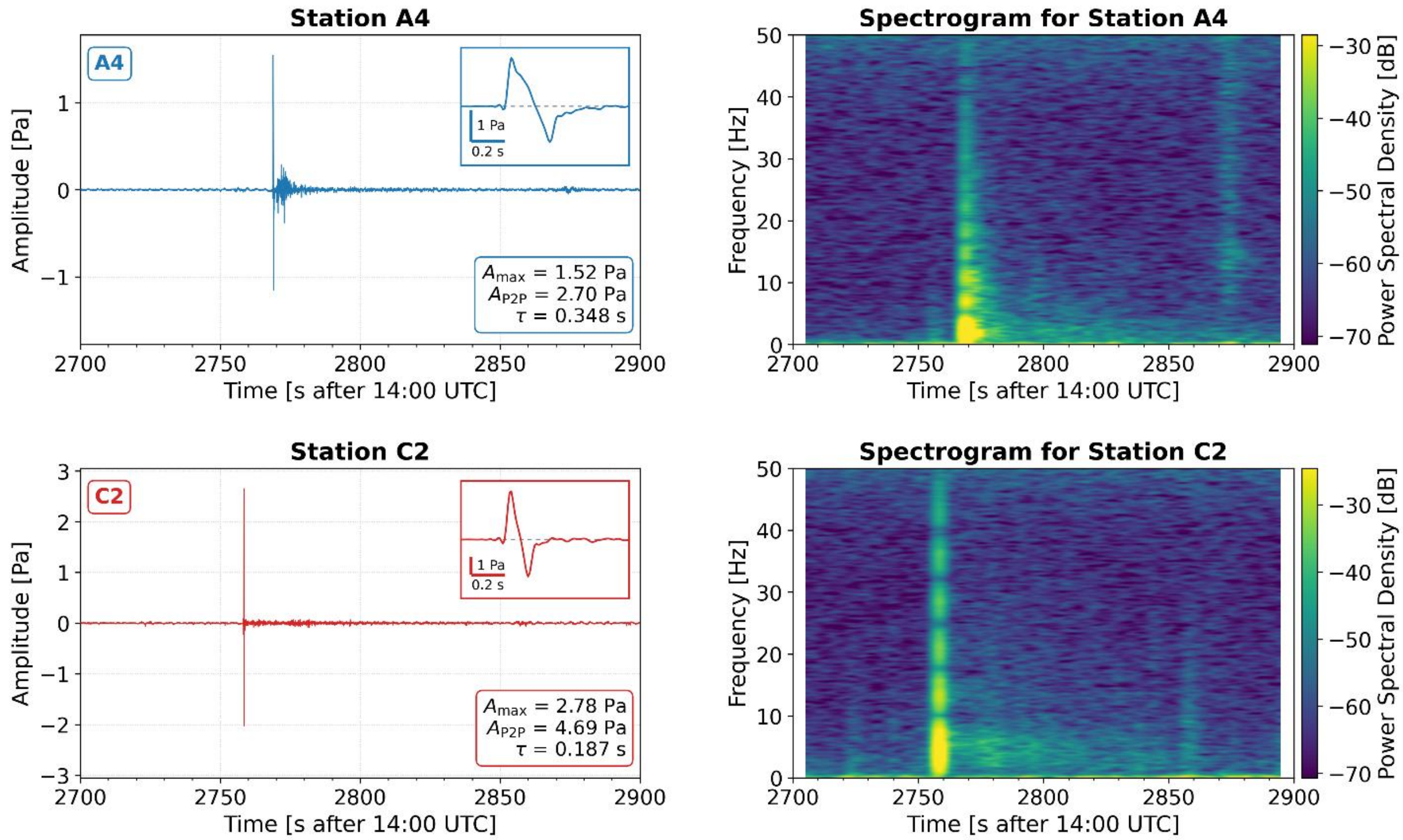

**Figure 5:** Representative OSIRIS-REx infrasound signals from station A4 (top) and station C2 (bottom). Left panels show the time-domain waveform over a broad time window, with inset panels demonstrating the impulsive arrival shape used in the adopted period and amplitude measurements. Right panels show the corresponding spectrograms, illustrating the short-duration, broadband character of the arrivals and the lower-amplitude trailing signal energy. These examples are representative of the signal morphology underlying the published period and peak-overpressure values adopted from Silber and Bowman (2025).

Signal periods and peak overpressures at each of the 39 stations were measured by Silber and Bowman (2025) and are adopted here as event-specific observational inputs. Signal periods were obtained from the time-domain waveform using the zero-crossing procedure originally developed for nuclear explosion infrasound and later adapted to bolide-generated signals by ReVelle (1997). The dominant period may also be estimated in the frequency domain from the dominant spectral peak, and for bolides, these period estimates are typically comparable at about the 10% level (e.g., Ens et al., 2012; Gi and Brown, 2017; Silber and Brown, 2014).

For the OSIRIS-REx dataset, Silber and Bowman (2025) introduced and applied the two-zero-crossing convention for signal period measurement at maximum amplitude. Two successive zero crossings, $t_{ZC1}$ and $t_{ZC2}$, are selected such that they bracket the first positive, or compressional, half-cycle of the infrasound arrival at maximum amplitude, and the dominant signal period is defined as $\tau = 2(t_{ZC2} - t_{ZC1})$. This process is described Supplemental **Section S1** (Supplemental Materials). For many impulsive infrasound arrivals, a full-cycle zero-crossing pick is adequate because the positive and negative phases remain sufficiently coherent to support a stable period estimate. In the OSIRIS-REx signals, however, and in some similar short-duration impulsive cases (Silber et al., 2026), the second half of the cycle can be more readily contaminated by trailing coda and other propagation-related effects. Under those conditions, a full-cycle pick can introduce additional ambiguity into the interpretation of the dominant period. Restricting the measurement to the leading compressional half-cycle thus provides a more repeatable period estimate for the arrivals analyzed here. Scamfer et al. (2026) performed a side-by-side comparison and found that the standard zero-crossing and two-zero-crossing period estimates for meteor-generated infrasound signals are generally comparable to within about 10%, supporting the use of the present convention. Signal-energy- or power-based observables may provide a useful complementary measure in future work.

The stations span source heights ranging from approximately 44 to 62 km altitude, with the nearly horizontal trajectory producing a wide spatial distribution of infrasound emission points. For each station, the raytracing-derived source altitude identifies the specific trajectory point from which the detected infrasound originated. The capsule velocity and Mach number at each emission point are known from the EDL trajectory (Francis et al., 2024), and the local atmospheric conditions (density, pressure, temperature) are obtained from the G2S model. Combined with the known capsule dimensions and drag properties, every quantity required to compute $R_0$ and evaluate the weak shock approach is independently constrained. The observed signal periods and maximum amplitudes at each station then provide the observational constraints against which forward-modeled weak shock predictions can be evaluated.

### 3.3 Trajectory-Specific $R_0$ Calculations

The six $R_0$ formulations (**Table 1**) are evaluated at each trajectory-specific source point, using the local atmospheric conditions and capsule velocity at that altitude. Because $R_0$ depends on the local atmospheric density and pressure as well as the capsule velocity, it varies along the trajectory and is therefore altitude-dependent.

### 3.4 Weak Shock Implementation

A Python implementation of the weak shock formulation described in Silber et al. (2015) and Silber and Brown (2019) was used for all forward and inverse modeling. The implementation includes propagation through a stratified, inhomogeneous atmosphere with altitude-dependent density, temperature, and wind profiles, as well as frequency-dependent atmospheric absorption following Sutherland and Bass (2004). In practice, the weak shock period and overpressure relations are applied stepwise through the atmospheric profile, using the local propagation conditions at each altitude level along the source-to-station path. Wind effects on the effective sound speed are introduced through the adopted atmospheric specification; here, event-specific Ground-to-Space (G2S) specifications were used for consistency with Silber and Bowman (2025). The adiabatic sound speed was calculated from the temperature profile using the relation $c = \sqrt{\gamma RT/M_{mol}}$, where $\gamma$ is the ratio of specific heats for air (1.4), $R$ is the universal gas constant (8.314 J/mol K), and $M_{mol}$ is the mean molar mass of dry air (0.0289644 kg/mol) (e.g., Pierce, 2019). The effective sound speed accounts for the projection of horizontal winds along the propagation direction, and the Doppler shift due to winds along the propagation path is included in the period and amplitude calculations (Silber et al., 2015).

### *3.4.1 Forward Modeling Procedure*

The complete mathematical treatment of the weak shock model, including the governing equations for propagation through a stratified, windy atmosphere with frequency-dependent absorption, is given in Silber et al. (2015) and Silber and Brown (2019); only the procedural steps essential for interpreting the present results are summarized here.

For each station, the forward model computes the theoretical blast radius $R_0$ from the known capsule parameters and the atmospheric conditions at the raytracing-derived source height, using each of the six formulations in **Table 1**. The fundamental period $\tau_0$ and initial overpressure ratio are calculated at the source altitude. The altitude array from the source to the station is constructed, and the weak shock period and overpressure are propagated downward through the stratified atmosphere, accounting for density and sound speed variations, wind effects on the effective sound speed, and atmospheric absorption. The weak shock transition distance is evaluated at each altitude step to determine whether the transition to the linear regime occurs before the wave reaches the station. If a transition is identified, the period freezes at the transition value and the amplitude decays according to geometric spreading and linear absorption below the transition altitude. The output for each $R_0$ formulation and each WST coefficient $C$ consists of four predicted observables: weak shock period ($\tau_{ws}$), linear period ($\tau_{lin}$), weak shock amplitude ($\Delta p_{ws}$), and linear amplitude ($\Delta p_{lin}$). Each forward model run is performed three times, using $C$ = 5.38 (Morse and Ingard, 1968), $C$ = 34.3 (Towne, 1967), and $C$ = 57.2 (Silber et al., 2015). This produces a total of 18 combinations (6 $R_0$ formulations × 3 WST coefficients) for each station. The predicted observables are compared against the measured period and amplitude to identify the combination that minimizes the misfit. **Figure 6** illustrates the forward-model output for a representative station (A4), together with the alternative propagation branches considered by the calculation: the weak-shock branch, the linear-propagation branch selected once the weak-shock-to-linear transition criterion is satisfied, and the corresponding hypothetical weak-shock continuation if no transition were imposed. The transition point is governed by the weak shock transition coefficient $C$, as described in **Section 2.3**.

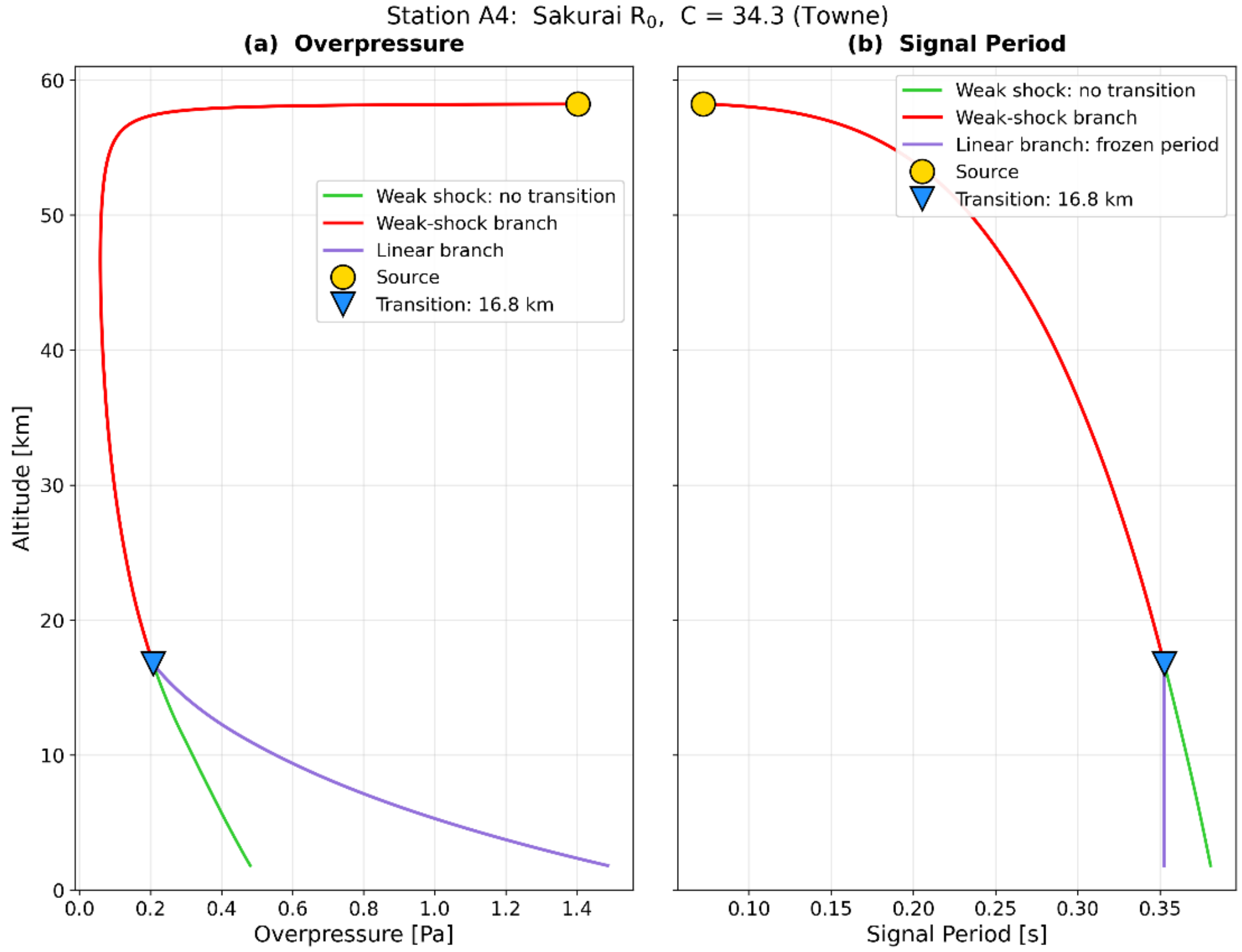

**Figure 6:** Forward model output for station A4 using the Sakurai $R_0$ formulation and Towne $C$ = 34.3. Altitude versus modeled signal **(a)** overpressure and **(b)** period showing the evolution of the weak-shock branch (red) from the source downward to the transition point, the linear-propagation continuation below the transition altitude (purple), and the corresponding hypothetical weak shock continuation if no transition were imposed (green). The blue triangle marks the weak-shock-to-linear transition altitude, which is determined by the criterion governed by the selected value of $C$. Because the wave propagates through a stratified atmosphere, the modeled overpressure along these branches does not necessarily vary monotonically with decreasing altitude.

### *3.4.2 Inversion*

For the inversion, the observed signal period $\tau_{obs}$ and amplitude $\Delta p_{obs}$ at each station are used to derive $R_0$ under each of the three WST coefficient assumptions. The period inversion is performed by numerically solving for the $R_0$ that, when propagated through the forward model to the station altitude, reproduces the observed period. This requires an iterative root-finding procedure (bisection) because the period depends nonlinearly on $R_0$ through both the source period $\tau_0$ and the altitude-dependent propagation path. The amplitude inversion follows an analogous procedure, solving for the $R_0$ that reproduces the observed overpressure. For each $C$ value and each station, four independent $R_0$ estimates are obtained: from the weak shock period ($R_{0,wsP}$), from the linear period ($R_{0,linP}$), from the weak shock amplitude ($R_{0,wsA}$), and from the linear amplitude ($R_{0,linA}$). The consistency among these four estimates and their agreement with the theoretical $R_0$ values from **Table 1** provide diagnostic information about the validity of the underlying assumptions and the appropriateness of the WST coefficient. More broadly, the period inversion establishes a pathway from a single observable quantity (i.e., the infrasound period at a remote station) to an estimate of the blast radius at the source, which in turn constrains the energy deposition per unit path length through the chosen $R_0$ formulation.

### 3.5 Statistical Analysis

The statistical framework is organized into primary and supporting analyses. The primary tools are bootstrap confidence intervals on medians (for bias assessment and uncertainty quantification) and paired Wilcoxon signed-rank tests (for formulation ranking by error magnitude) (Wilcoxon, 1945). These two methods address the central benchmarking questions: whether a formulation is biased and which formulation produces smaller errors. Supporting analyses include the Friedman repeated-measures test (for sensitivity of the ranking to the WST coefficient) and Spearman rank correlation (Spearman, 1904) with permutation testing (for along-trajectory trends in the residuals). Hodges-Lehmann (Hodges Jr and Lehmann, 2011) median differences with confidence intervals are reported alongside the Wilcoxon tests to provide effect size estimates for formulation comparisons.

Model performance is summarized using two complementary metrics computed across the 39 stations. The signed percent residual, defined as (theoretical - observed) / observed × 100%, characterizes bias: positive values indicate overestimation, negative values indicate underestimation. Here, “predicted” refers to the quantity of interest produced by the model (e.g., the forward-predicted period or the theoretically expected $R_0$), and “observed” refers to the corresponding measurement-derived quantity (e.g., the measured period or the inversely determined $R_0$). The median absolute percent residual (MAPR), defined as the median of the absolute values of the station-wise signed residuals, characterizes accuracy regardless of direction. Both metrics are reported throughout **Sections 4** and **5** and the Supplemental Materials.

Bootstrap confidence intervals: Uncertainty on median statistics is quantified using nonparametric bootstrap resampling (Efron and Tibshirani, 1993). For each statistic of interest (MAPR or median signed residual), the 39-station dataset is resampled with replacement 10,000 times, the statistic is recomputed for each resample, and the 2.5th and 97.5th percentiles of the resulting distribution define the 95% confidence interval. This approach makes no distributional assumptions and provides reliable interval estimates for medians even with small or skewed samples. The bootstrap confidence interval on the median signed residual replaces hypothesis testing for bias assessment: if the interval includes zero, the formulation shows no detectable systematic bias at the given sample size.

Paired Wilcoxon signed-rank tests: Differences in error magnitude between $R_0$ formulations are assessed using the Wilcoxon signed-rank test (Wilcoxon, 1945), a nonparametric test for paired samples. Because each station provides a matched pair of absolute residuals (one per formulation evaluated at the same source point), the paired design controls for station-to-station variability in source conditions (altitude, velocity, atmospheric density). The null hypothesis is that the median difference in absolute residuals between two formulations is zero. Tests are applied to pairwise comparisons among the three energy-normalized formulations (Few, Jones/Plooster, Sakurai), yielding three comparisons, with $p$-values corrected for multiple testing using the Holm-Bonferroni sequential procedure (Holm, 1979). For each comparison, the Hodges-Lehmann estimator of the median paired difference is reported as an effect size, together with a bootstrap 95 percent confidence interval.

Friedman repeated-measures test: The sensitivity of the MAPR to the WST coefficient $C$ is assessed using the Friedman test (Friedman, 1937), a nonparametric analog of repeated-measures ANOVA. Each of the 39 stations provides a matched triplet of absolute residuals (one per $C$ value: 5.38, 34.3, and 57.2), and the test evaluates whether the three $C$ values produce systematically different error magnitudes for a given formulation. When the omnibus Friedman test is significant ($p < 0.05$), post-hoc pairwise comparisons between $C$ values are conducted using Wilcoxon signed-rank tests with Holm-Bonferroni correction.

Spearman rank correlation and permutation test: Associations between the signed residual and source parameters (altitude, velocity) are quantified using the Spearman rank correlation coefficient (Spearman, 1904), which measures monotonic association without assuming linearity or bivariate normality. Significance is assessed using a permutation test: the source parameter labels are randomly shuffled 10,000 times, the Spearman correlation is recomputed for each permutation, and the fraction of permuted correlations exceeding the observed value in absolute magnitude gives the distribution-free $p$-value. This approach avoids parametric assumptions about the joint distribution of residuals and source parameters.

All statistical computations use the full set of 39 stations and are performed independently for each WST coefficient where applicable. Statistical analyses were carried out using SciPy v1.11 (Virtanen et al., 2020) with custom bootstrap and permutation implementations in Python.

# 4. Results

## 4.1 Theoretical Blast Radii Along the Trajectory

The trajectory-specific $R_0$ calculations show how the blast radius varies along the SRC's flight path as a function of altitude. Because the capsule decelerates as it descends and the atmospheric density increases, $R_0$ is not monotonic with altitude. **Figure 7** shows the theoretical $R_0$ as continuous functions of altitude for all six formulations, and **Table 2** summarizes the $R_0$ ranges across the observed source heights (~44 to 62 km). The Sakurai formulation produces the smallest values (5.83±2.52 m), while the Tsikulin modified formulation produces the largest (20.65±8.93 m).

The six formulations separate into two distinct families. The energy-normalized formulations (Few (1969), Jones et al. (1968)/Plooster (1970), and Sakurai (1965)) yield mean $R_0$ values of ~5.8 to 8.2 m. These formulations differ primarily in their normalization conventions but share the same underlying energy-per-unit-length scaling. The remaining three (Tsikulin (1970) standard, Tsikulin (1970) modified, and Mach-diameter (ReVelle, 1974)) are scaling-type formulations that yield substantially larger values of approximately 14.6 to 20.7 m, a factor of 2.5–3.5 greater than the energy-normalized group. This separation has a direct consequence for forward modeling: because the initial period $\tau_0$ scales with $R_0$, the larger-$R_0$ formulations will predict proportionally longer periods at the source, leading to systematic overprediction at all stations, a pattern that is confirmed in the forward model results below.

**Table 2.** Summary of theoretical $R_0$ values across observed source altitudes (~44 to 62 km). The second column represents standard deviation (σ).

| Formulation | Mean [m] | $\sigma$ [m] | Min [m] | Max [m] |
|---|---|---|---|---|
| Few (1969) | 8.24 | 3.56 | 2.75 | 12.48 |
| Jones et al. (1968) / Plooster (1970) | 6.22 | 2.69 | 2.07 | 9.41 |
| Sakurai (1965) | 5.83 | 2.52 | 1.94 | 8.82 |
| Tsikulin (1970), standard | 14.61 | 6.32 | 4.87 | 22.11 |
| Tsikulin (1970), modified | 20.65 | 8.93 | 6.89 | 31.27 |
| Mach-diameter (ReVelle, 1974) | 19.70 | 8.52 | 6.57 | 29.82 |

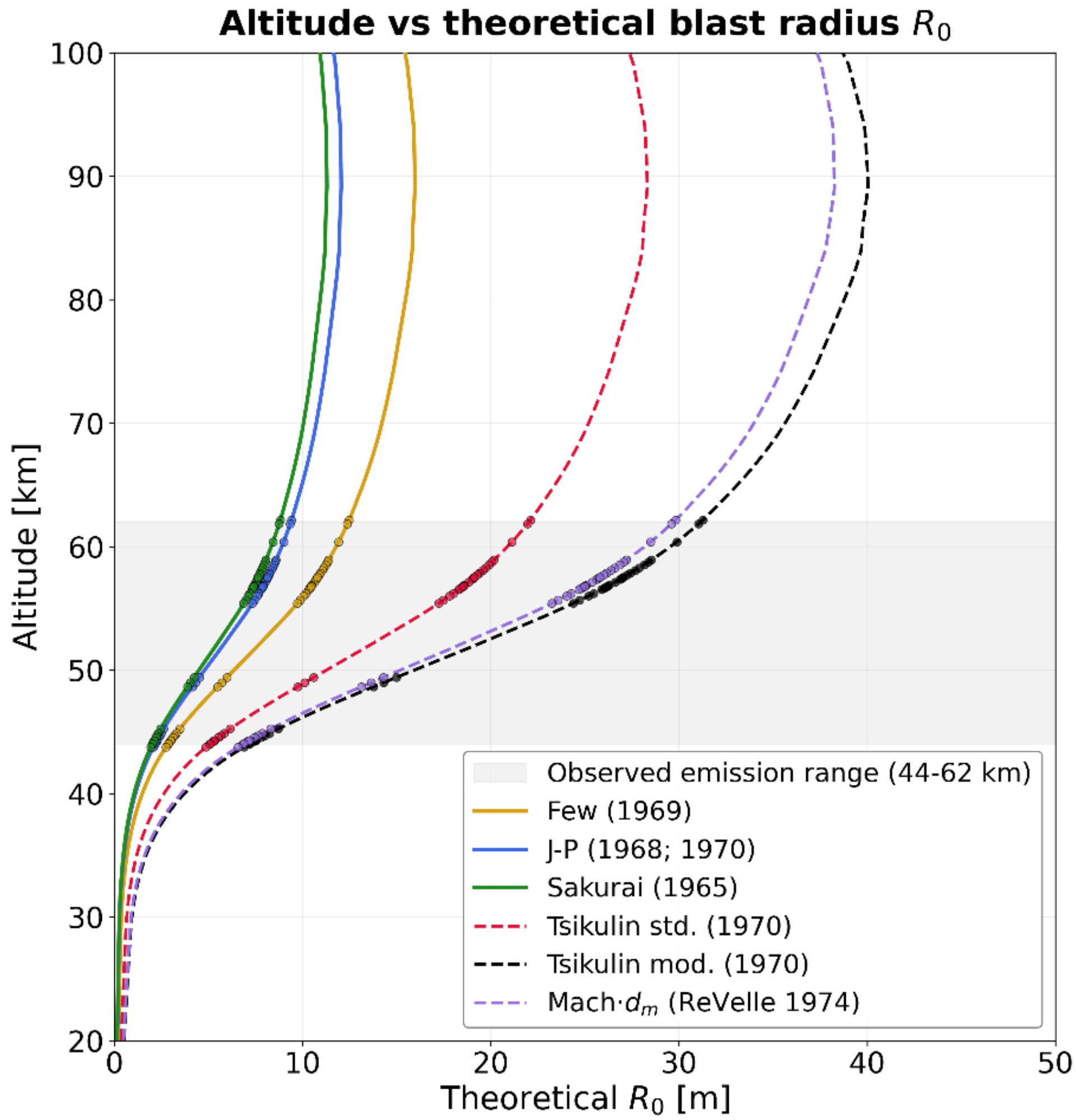

**Figure 7:** Theoretical $R_0$ for all six formulations.

## 4.2 Forward Model Benchmark

The forward model is evaluated for all 18 combinations (6 $R_0$ formulations × 3 WST coefficients) across all 39 stations. For each combination, the theoretical $R_0$ at the station-specific source altitude sets the source-side period and peak overpressure via the cylindrical blast-wave formulas; these acoustic observables are then propagated through the stratified atmosphere to yield the predicted receiver-side $\tau$ and $\Delta p$ at each station. Model performance is quantified using the signed residual and median absolute percentage residual (MAPR) defined in Section 3.5.

The per-station period-residual distributions (Supplementary **Figure S2**, Section S2.1) divide the six $R_0$ formulations into two distinct groups. The three energy-normalized formulations (Sakurai, Jones/Plooster, Few) reproduce the observed period well. The Sakurai–Towne combination yields the lowest station-wise MAPR (9% at $C$ = 34.3) with a near-zero median signed residual (+1%), whereas the three scaling-type formulations (Tsikulin standard, Tsikulin modified, Mach-diameter) systematically overpredict the period and exceed 80% MAPR at every $C$, confirming their unsuitability for the non-ablating SRC source. The amplitude-residual distributions (Supplementary **Figure S3**, Section S2.2) exhibit a qualitatively different pattern. Every $R_0$ formulation underpredicts the weak-shock amplitude $\Delta p_{ws}$: the three energy-normalized formulations by 66 to 74% (signed residuals strictly negative), and the three scaling-type formulations by 31 to 47%. The linear-regime amplitude $\Delta p_{lin}$ is strongly $C$-dependent: the minimum MAPR (28%, Sakurai and Jones/Plooster at $C$ = 34.3) is nonetheless three times the minimum period MAPR, anticipating the period-versus-amplitude asymmetry quantified below.

**Figure 8a** presents the period scorecard, organizing the station-median MAPRs by formulation (columns) and benchmark (rows: the $C$-independent $\tau_{ws}$ row followed by $\tau_{lin}$ at each of the three tested $C$ values). Sakurai (1965) attains the lowest MAPR in every period row; 13% for $\tau_{ws}$ and 9–10% for $\tau_{lin}$ across all three $C$ values, with the single-cell minimum of 9% at the Towne coefficient ($C$ = 34.3) accompanied by a near-zero station-median signed residual of +1 %. Jones et al. (1968) / Plooster (1970) ranks second in every row (MAPRs 11–18%) and Few third (20–42%). The three scaling-type formulations (Tsikulin (1970) standard and modified, Mach-diameter (ReVelle, 1974)) systematically overpredict the period by factors of 1.8–2.8 (MAPRs 82–183%), with the global maximum at Tsikulin (1970) modified $\tau_{ws}$ (183%).

**Figure 8b** presents the amplitude scorecard, which displays a qualitatively different structure in which no single formulation minimizes the residual across all rows. For the weak-shock amplitude $\Delta p_{ws}$ ($C$-independent), the three scaling-type formulations achieve the smallest MAPRs (Tsikulin modified 31%, Mach-diameter 34%, Tsikulin standard 47%) because their larger theoretical $R_0$ values partially offset the systematic amplitude underprediction inherent in the cylindrical blast-wave formulation; the three energy-normalized formulations all underpredict $\Delta p_{ws}$ by 66–74 % (signed residuals strictly negative). For the linear-regime amplitude $\Delta p_{lin}$, the ranking re-sorts with $C$: Sakurai and Jones et al. (1968) / Plooster (1970) jointly achieve the minimum at $C$ = 34.3 (both 28%), Sakurai remains best at $C$ = 5.38 (109%, still a factor-of-two overprediction), and Few is best at $C$ = 57.2 (29%). The minimum amplitude MAPR across all 18 combinations (28%, Sakurai and Jones/Plooster at $C$ = 34.3) is three times the minimum period MAPR (9%, Sakurai at $C$ = 34.3).

Two conclusions follow from the forward benchmark. First, no tested formulation reconciles the period and amplitude constraints simultaneously: the formulations that best reproduce the period (Sakurai and Jones et al. (1968) / Plooster (1970)) underpredict $\Delta p_{ws}$ by 73–74%, while the formulations with the smallest $\Delta p_{ws}$ residuals (Tsikulin modified 31% and Mach-diameter 34%) overpredict $\tau_{ws}$ by 173% and 183%. The tension is systematic rather than accidental, and its physical origin is examined further in **Section 5**. Second, the period prediction is only weakly sensitive to the transition coefficient at $C \geq 34.3$; Sakurai $\tau_{lin}$ varies by one percentage point between $C$ = 34.3 (9%) and $C$ = 57.2 (10%), whereas Sakurai $\Delta p_{lin}$ varies by nearly a factor of four across the tested $C$ range, from 28% ($C$ = 34.3) to 109% ($C$ = 5.38). At the propagation distances sampled here (~7,000–10,000 $R_0$), the forward benchmark therefore identifies the period as the more reliable observable for rigid, non-ablating hypersonic sources, with Sakurai paired with the Towne coefficient as the best-performing combination; the Mach-diameter approximation is not supported for this source class.

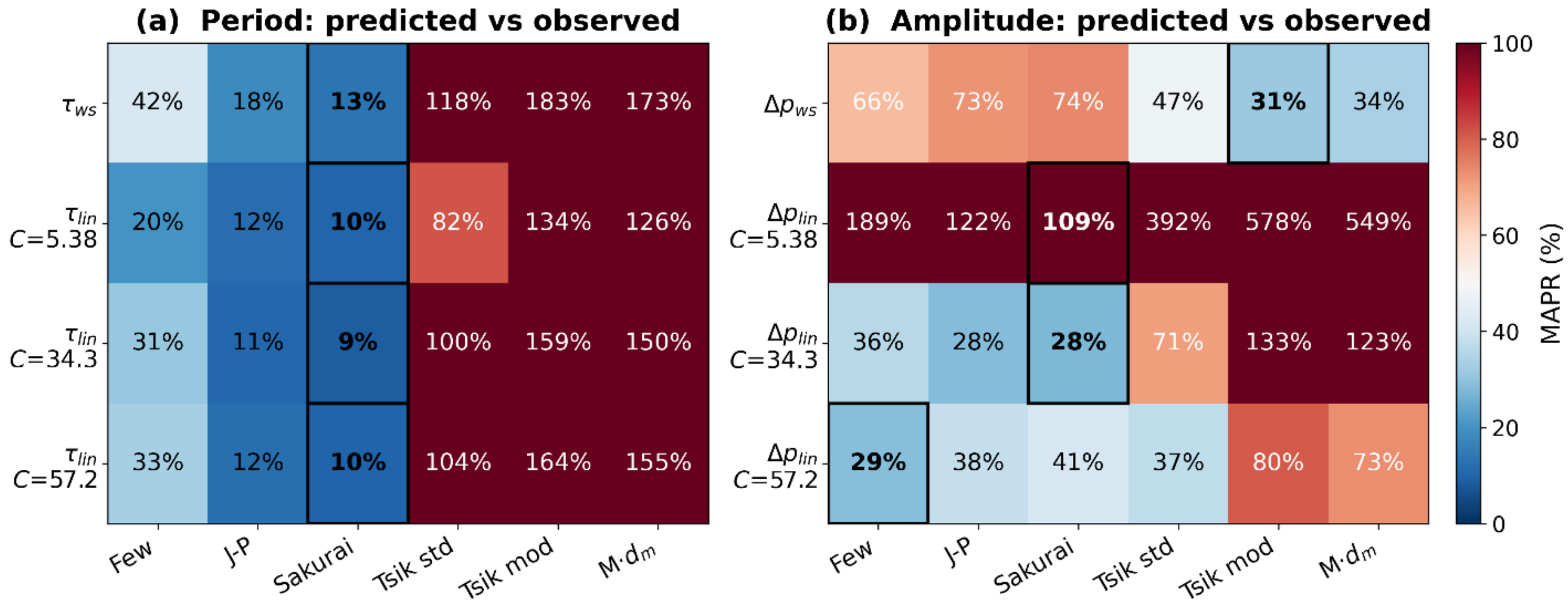

**Figure 8:** Forward model scorecard showing the median absolute percentage residual (MAPR) between predicted and observed values for (a) period and (b) amplitude, across six $R_0$ formulations (columns) and four propagation scenarios (rows). In panel (a), the first row shows the weak shock period ($\tau_{ws}$), which is independent of $C$, followed by the linear period ($\tau_{lin}$) for $C$ = 5.38, 34.3, and 57.2. In panel (b), the first row shows the weak shock amplitude ($\Delta p_{ws}$), which is independent of $C$, followed by the linear amplitude ($\Delta p_{lin}$) for the same three $C$ values. Colors follow a diverging blue-red scale from 0% (perfect agreement) to 100% (factor-of-two error); values exceeding 100% are displayed numerically and saturated at the darkest red. Black borders mark the best-performing formulation (lowest MAPR) in each row. Bold text within bordered cells indicates the row minimum. All residuals are computed over the 39 stations with matched source-to-station geometries from raytracing.

### 4.3 Inversion $R_0$ Benchmark

The inversion procedure yields four $R_0$ estimates per station: the weak-shock and linear-regime period-derived estimates ($R_{0,wsP}$ and $R_{0,linP}$) and the corresponding amplitude-derived estimates ($R_{0,wsA}$ and $R_{0,linA}$), with $R_{0,linP}$ and $R_{0,linA}$ evaluated at each of the three WST coefficients (**Table 3**). Station-mean and station-median values for each of the eight inverted benchmarks are summarized in **Table 3**. The per-station distributions of these inverted values, together with their dependence on $C$, are documented in Supplementary Information Section S3 and **Figure S4**. The main-text discussion below focuses on the inversion benchmark heatmap in **Figure 9**, which reports the station-median MAPR between each theoretical $R_0$ (columns) and each inverted benchmark (rows) on the period side (**Figure 9a**) and the amplitude side (**Figure 9b**).

**Figure 9a** presents the period-derived benchmark. Sakurai achieves the lowest MAPR in three of the four rows: 20% against $R_{0,wsP}$, 11% against $R_{0,linP}$ at $C$ = 34.3, and 13% against $R_{0,linP}$ at $C$ = 57.2. At $C$ = 5.38, Jones/Plooster is marginally better (13%) than Sakurai (14%), with Few following at 27%. Jones/Plooster ranks second in the three remaining rows (26% against $R_{0,wsP}$, 15% at $C$ = 34.3, 17% at $C$ = 57.2), and Few ranks third (58%, 42%, 45%). The three scaling-type formulations exceed 120% MAPR in every period row; the global period-side maximum occurs at Tsikulin modified against $R_{0,wsP}$ (297%).

**Figure 9b** presents the amplitude-derived benchmark, where the formulation ranking differs qualitatively from the period side. Against $R_{0,wsA}$ ($C$-independent), the scaling-type formulations achieve the smallest MAPRs (Tsikulin modified 40%, Mach-diameter 42%, Tsikulin standard 57%), whereas the three energy-normalized formulations all exceed 76%. Against $R_{0,linA}$, the best-in-row entry is strongly $C$-dependent. Jones/Plooster reaches 30% at $C$ = 34.3 (the global minimum on the amplitude side), with Sakurai at 33% and Few at 39% at the same $C$. Few achieves its best amplitude-benchmark MAPR (33%) at $C$ = 57.2. At $C$ = 5.38, every amplitude-benchmark MAPR exceeds 109% (Sakurai minimum), with the scaling-type rows above 400%. The minimum amplitude-benchmark MAPR (30%) is 2.7 times the minimum period-benchmark MAPR (11%).

The station-median inverted values in **Table 3** quantify this asymmetry directly: the period-derived $R_0$ medians fall in the narrow range 6.2–7.8 m across all $C$, whereas the amplitude-derived $R_0$ medians span 3.1–31.3 m, more than a factor of ten. Two features of **Figure 9** stand out. First, the period-derived benchmark is both lower in absolute MAPR and less variable across the heatmap than the amplitude-derived benchmark, identifying $R_{0,linP}$ at $C$ ≥ 34.3 as the stable pathway to inferring the along-trajectory source $R_0$ from infrasound observables. Second, the apparent advantage of the scaling-type formulations against $R_{0,wsA}$ is a compensation effect rather than a physical improvement: the scaling-type theoretical $R_0$ values are systematically larger than those of the energy-normalized formulations, and $R_{0,wsA}$ is itself inflated relative to $R_{0,wsP}$ (by approximately a factor of five at the station median; see Section S3 and **Figure S4**) because the cylindrical blast-wave model underpredicts the weak-shock overpressure. The two biases partially cancel, producing the apparent alignment; neither is physically informative, and the scaling-type formulations remain unsuitable for rigid, non-ablating hypersonic sources such as the SRC.

**Table 3.** Mean inverse-modeled $R_0$ values (m) across 39 stations for three WST coefficients. $R_{0,wsP}$ = weak shock period; $R_{0,linP}$ = linear period; $R_{0,wsA}$ = weak shock amplitude; $R_{0,linA}$ = linear amplitude.

| C value | $R_{0,wsP}$ [m] | $R_{0,linP}$ [m] | $R_{0,wsA}$ [m] | $R_{0,linA}$ [m] |
|---|---|---|---|---|
| 5.38 (Morse and Ingard, 1968) | 5.41±1.52 | 6.69±2.02 | 31.75±12.78 | 3.00±0.62 |
| 34.3 (Towne, 1967) | 5.41±1.52 | 5.94±1.78 | 31.75±12.78 | 8.66±1.92 |
| 57.2 (Silber et al., 2015) | 5.41±1.52 | 5.85±1.76 | 31.75±12.78 | 11.21±2.57 |

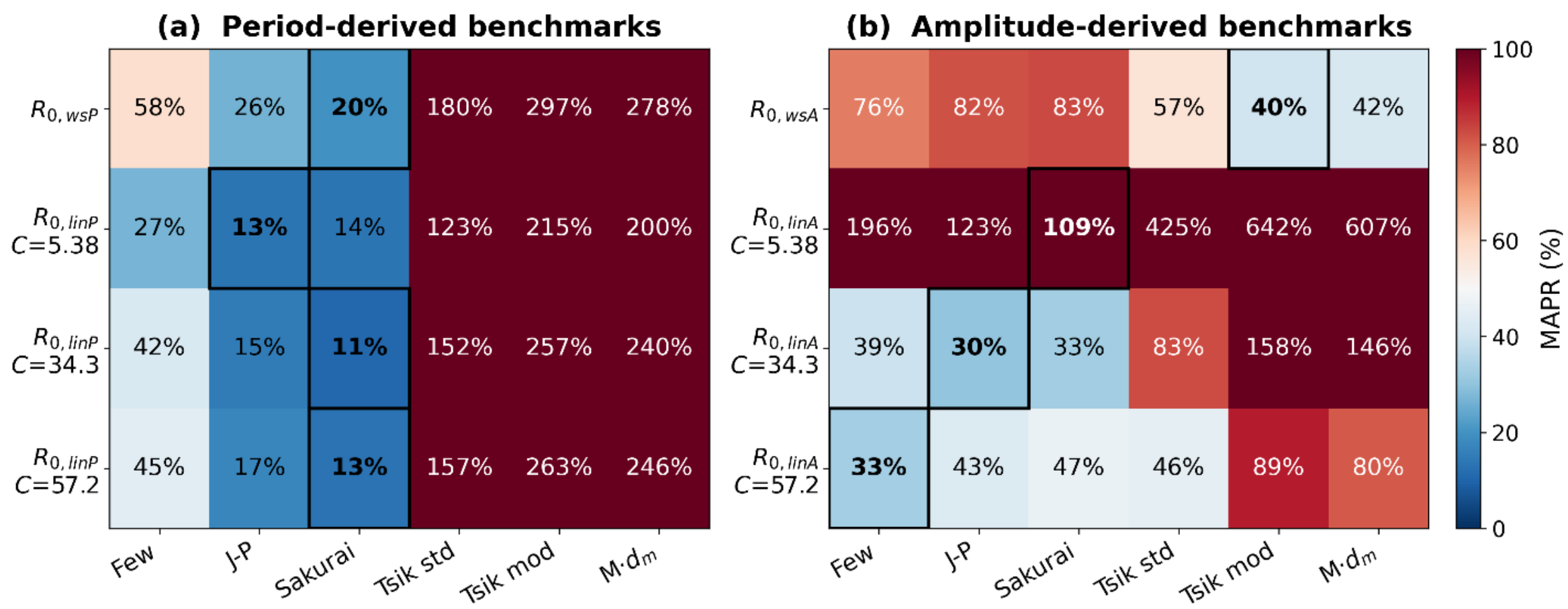

**Figure 9:** Median absolute percentage residual (MAPR) quantifying the agreement between inverted and theoretical $R_0$ (columns) and observation-derived $R_0$ (rows) for six blast radius formulations across 39 stations. Panel (a) shows period-derived benchmarks: $R_{0,wsP}$ (weak shock period, independent of $C$) and $R_{0,linP}$ (linear period) for $C$ = 5.38, 34.3, and 57.2. Panel (b) shows amplitude-derived benchmarks: $R_{0,wsA}$ (weak shock amplitude, independent of $C$) and $R_{0,linA}$ (linear amplitude) for the same three $C$ values. Colors follow a diverging blue-red scale from 0% (perfect agreement) to 100%; values exceeding 100% are displayed numerically and saturated at the darkest red. Black borders mark the best-performing formulation in each row.

### 4.4 Statistical Ranking

**Figure 10** reports bootstrap 95% confidence intervals (10,000 resamples, percentile method) on the station-median MAPR for each formulation against four period benchmarks: the weak-shock period $R_{0,wsP}$ ($C$-independent) and the linear-regime period $R_{0,linP}$ at $C$ = 5.38, 34.3, and 57.2. Against $R_{0,wsP}$, Sakurai attains the lowest MAPR (20%, 95% CI 16–26 %), followed by Jones/Plooster (26%, 23–29%) and Few (58%, 28–67%). Pairwise Wilcoxon signed-rank tests on the station-wise absolute residuals, Holm-corrected across the three formulation comparisons, reject the null hypothesis of zero median difference at $p$ = 0.0001 for Sakurai versus Jones/Plooster and at $p$ < 0.001 for Sakurai versus Few and for Jones/Plooster versus Few. Hodges-Lehmann effect sizes corroborate the ranking, with station-median pairwise differences of −4.6 pp for Sakurai versus Jones/Plooster (95% CI −9.9 to +3.3 pp), −27.2 pp for Sakurai versus Few (CI −42.3 to −11.5 pp), and −26.1 pp for Jones/Plooster versus Few (CI −38.4 to −7.7 pp). The Sakurai-versus-Jones/Plooster ranking is itself $C$-dependent when evaluated against $R_{0,linP}$: the two formulations are practically indistinguishable at $C$ = 5.38 (MAPRs of 13.6% and 13.4%, within 0.2 percentage points), whereas Sakurai outperforms Jones/Plooster at $C$ ≥ 34.3 (MAPR 11–13% versus 15–17%, Wilcoxon Holm-adjusted $p$ ≤ 0.003 across the three $C$ comparisons). The full pairwise Wilcoxon $p$-values, Hodges-Lehmann confidence intervals, bootstrap signed-bias intervals, and the consolidated formulation-ranking summary are provided in Supplementary Information Sections S4 and S5 and **Tables S1** and **S2**. The physical interpretation of the $C$-dependent ranking, namely that smaller $C$ values are more appropriate for the rigid, non-ablating SRC source because ablation-driven energy deposition (the rationale for the Silber et al. (2015) value $C$ = 57.2) is absent, is developed in **Section 5**.

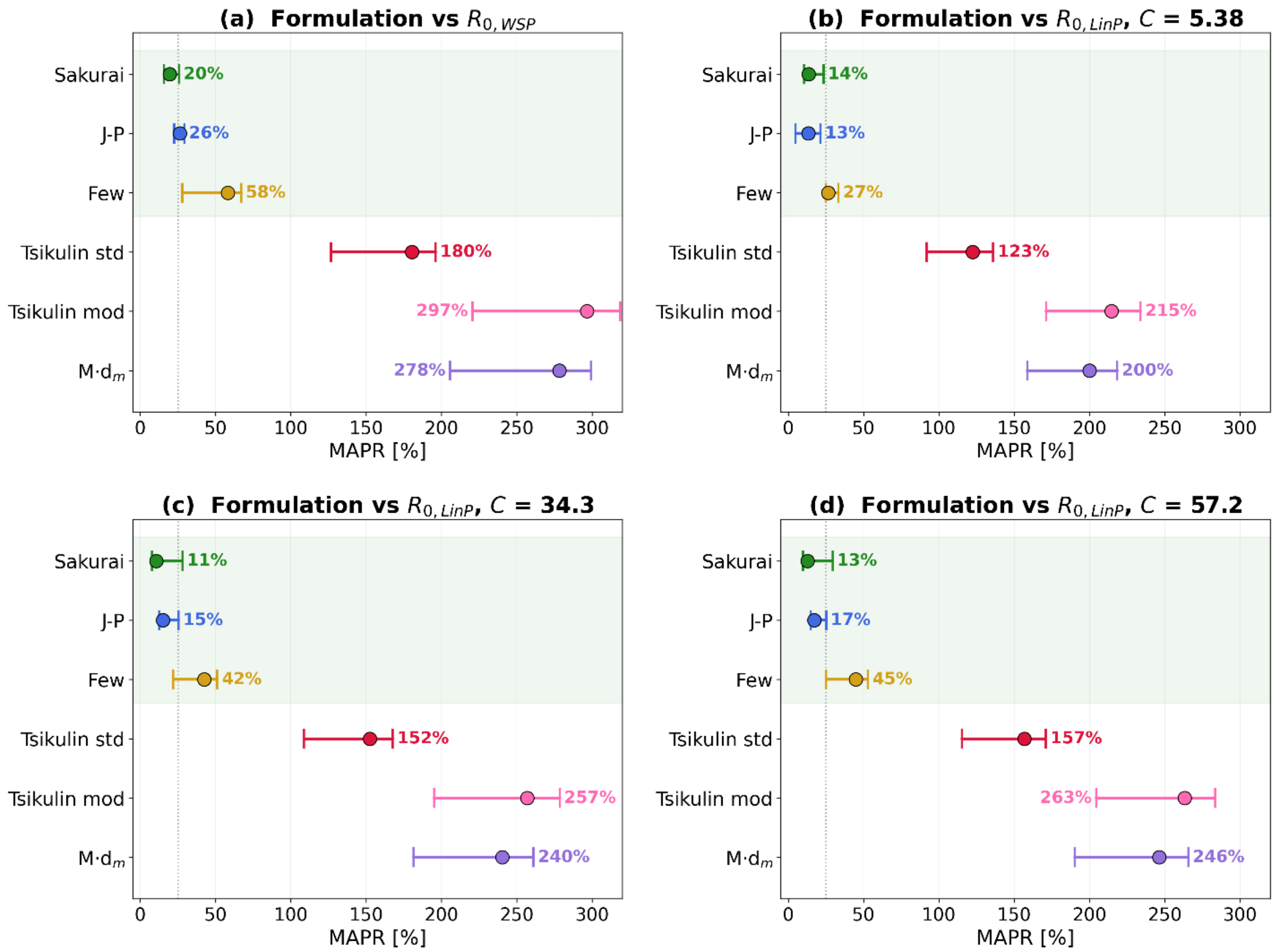

**Figure 10:** Bootstrap confidence intervals on the median absolute percentage residual (MAPR) for six $R_0$ formulations, evaluated against four period-derived benchmarks: (a) $R_{0,wsP}$ (weak shock period, independent of $C$), (b) $R_{0,linP}$ at $C$ = 5.38, (c) $R_{0,linP}$ at $C$ = 34.3, and (d) $R_{0,linP}$ at $C$ = 57.2. Markers show the observed MAPR and horizontal bars span the 95% confidence interval from 10,000 bootstrap resamples (percentile method) over the 39 stations. The vertical dotted line at 25% marks a reference threshold. In all four panels, the Sakurai (1965) and Jones et al. (1968)/ Plooster (1970) formulations cluster below 50% with narrow intervals relative to the scaling-type formulations (Tsikulin standard, Tsikulin modified, Mach-diameter approximation), which exceed 120% in every panel. The Few (1969) formulation occupies an intermediate position. The lowest MAPR occurs for Sakurai against $R_{0,linP}$ at $C$ = 34.3 (panel c), where the MAPR is 11% (95% CI: 8-28%); the wide upper bound reflects the right-skewed residual distribution produced by along-trajectory variability in source conditions.

### 4.5 Sensitivity to the WST Coefficient

The WST coefficient $C$ enters only the linear-regime inversion; the weak-shock branch, and therefore $R_{0,wsP}$ and $R_{0,wsA}$, are $C$-independent by construction. **Figure 11** summarizes the $C$-sensitivity of the linear-regime inverted $R_0$ estimates, and **Figure 12** quantifies the station-wise agreement between the two period-branch $R_0$ estimates. The Friedman repeated-measures tests formalizing the $C$-sensitivity of the station-wise MAPR against $R_{0,linP}$ are reported in Supplementary Information Section S6 and **Table S3**.

The period-derived estimates are weakly $C$-dependent (**Figure 11a**). The station median of $R_{0,wsP}$ is 6.2 m, independent of $C$ by construction. $R_{0,linP}$ has station medians of 7.8 m at $C$ = 5.38, 6.8 m at $C$ = 34.3, and 6.6 m at $C$ = 57.2. As $C$ increases, the transition distance $C \cdot R_0$ moves farther from the source, the linear-regime propagation path between transition and station shortens, and $R_{0,linP}$ converges toward $R_{0,wsP}$. All period-derived station medians fall within the 6 to 8 m window, bracketed by the theoretical Sakurai (5.8 m) and Few (8.2 m) values. For the Sakurai (1965) formulation, the predicted linear-regime period at the station shifts by ~11% across the full tested $C$ range (5.38 ≤ $C$ ≤ 57.2), with most of this shift occurring between $C$ = 5.38 and $C$ = 34.3; the further change from $C$ = 34.3 to $C$ = 57.2 is ~3%.

The amplitude-derived estimates display the opposite pattern (**Figure 11b**). $R_{0,wsA}$ is $C$-independent, with a station median of 31.3 m, approximately five times $R_{0,wsP}$, and an interquartile range of ~20–45 m that far exceeds any theoretical $R_0$ value. $R_{0,linA}$ is strongly $C$-dependent: the station median rises from 3.1 m at $C$ = 5.38 to 8.6 m at $C$ = 34.3 to 10.9 m at $C$ = 57.2, a factor of 3.5 across the tested $C$ range. Across the same range, $R_{0,linP}$ varies by only a factor of 1.2. The amplitude-derived $R_0$ therefore does not provide a constraint robust to the assumed WST coefficient.

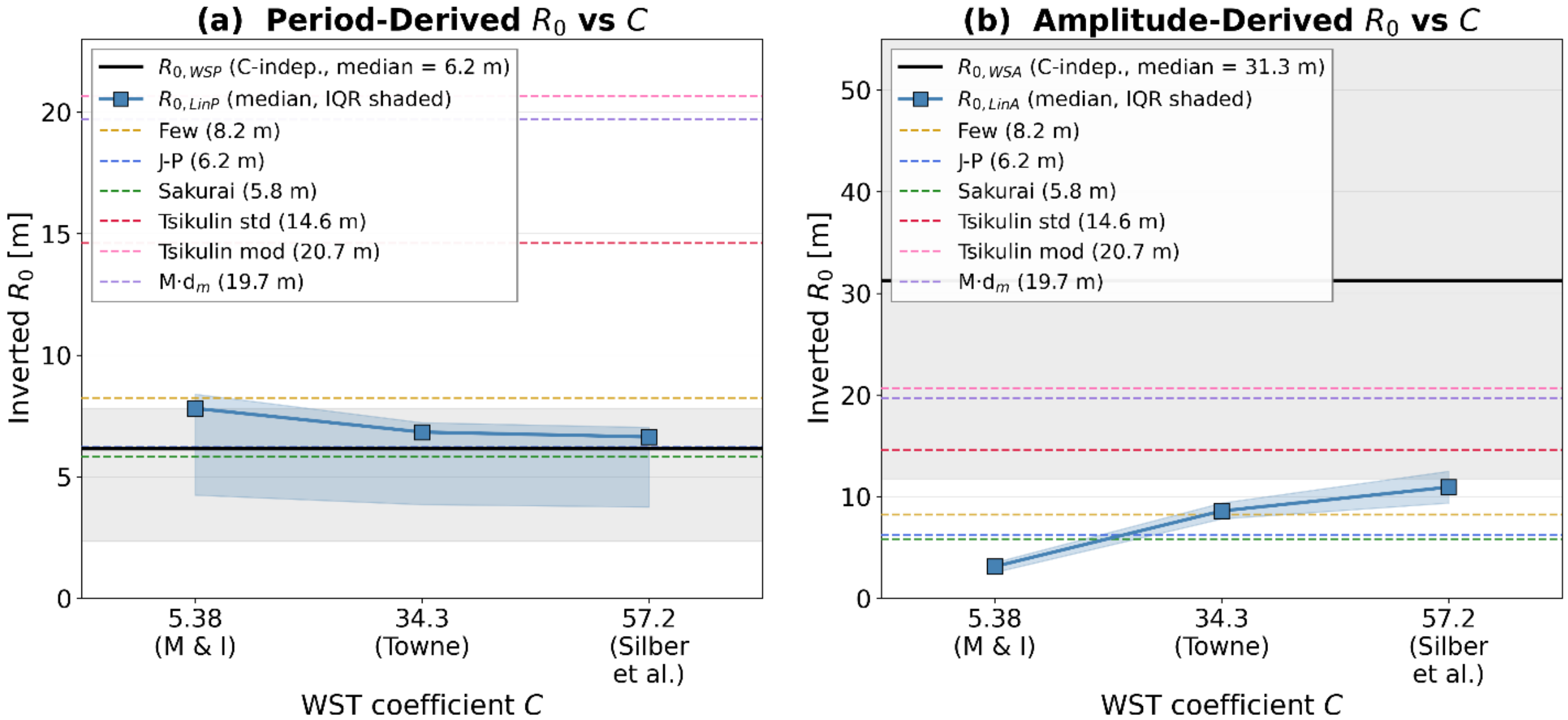

**Figure 11:** Sensitivity of inverted $R_0$ to the WST coefficient. (a) Period-derived: $R_{0,wsP}$ (black line, $C$-independent) and $R_{0,linP}$ (blue squares) with interquartile range shaded (blue) and full station range shown (gray). (b) Amplitude-derived: $R_{0,wsA}$ (black line, $C$-independent) and $R_{0,linA}$ (blue squares) with the same shading convention. Horizontal dashed lines indicate mean theoretical $R_0$ from each formulation. Note the different vertical scales.

**Figure 12** quantifies the station-wise agreement between the two period-branch $R_0$ estimates introduced above. The histograms show the distribution of $(R_{0,linP} - R_{0,wsP})$ / $R_{0,wsP}$ × 100% across the 39 stations at each $C$. At $C$ = 34.3 the median divergence is 10% (bootstrap 95% CI 8 to 11%), and all 39 stations lie within ±15%. At $C$ = 57.2 the median is 7.9 % (CI 6–10%), again with 100% of stations within ±15%. At $C$ = 5.38 the median rises to 25%, with all stations within 30% but only 15% of stations within ±15%. Pooling the two larger coefficients ($n$ = 78 station evaluations), the median absolute divergence is 9% and the 95th percentile is 13%. The modest divergence at $C \geq 34.3$ reflects the intermediate propagation geometry of the SRC event: the stations are neither so close to the source that the signal remains entirely within the weak-shock regime (where both branches would agree by construction), nor so far that the linear-regime evolution has accumulated enough path length to diverge substantially from the weak-shock extrapolation. In this intermediate regime the two period branches yield closely matched inverted $R_0$ estimates, and the period-derived $R_0$ is robust to the assumed WST coefficient.

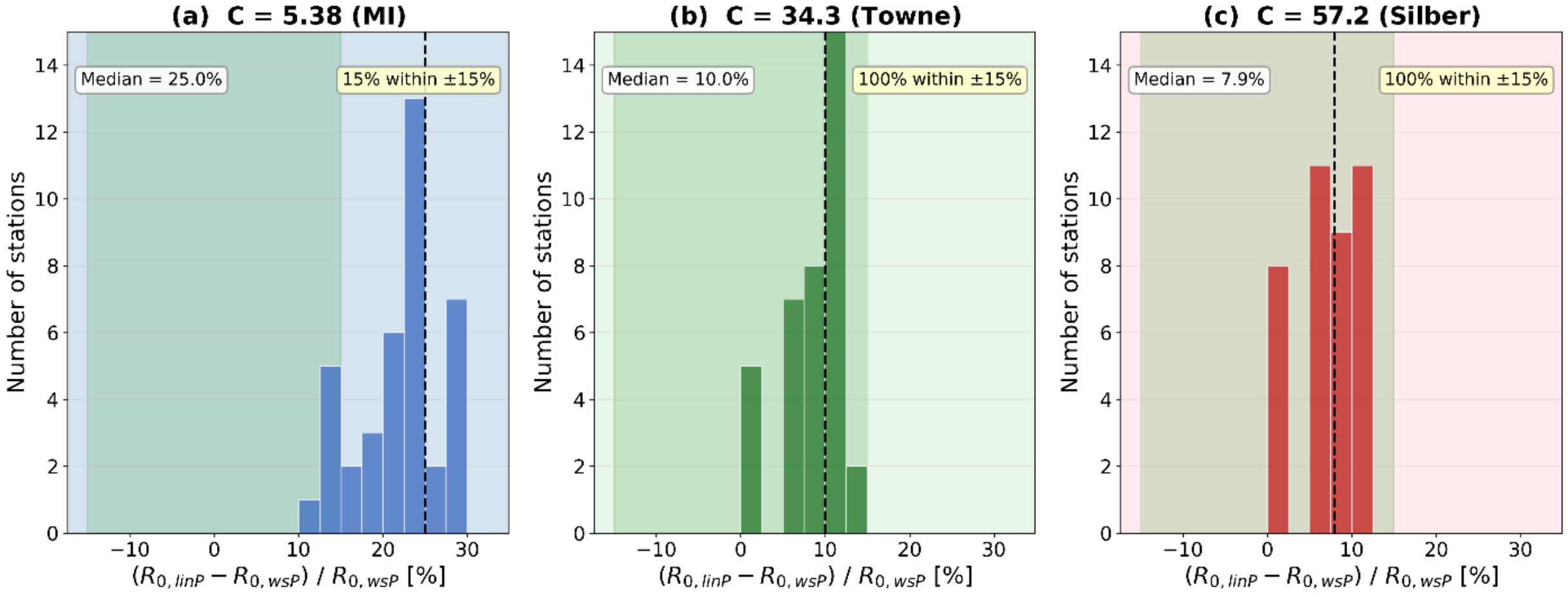

**Figure 12:** Histograms of the percentage difference between the linear and weak shock period-derived $R_0$ estimates, $(R_{0,linP} - R_{0,wsP})$ / $R_{0,wsP}$ × 100%, for all 39 stations at each tested WST coefficient: (a) $C$ = 5.38 (Morse and Ingard), (b) $C$ = 34.3 (Towne), and (c) $C$ = 57.2 (Silber et al.). Dashed vertical lines indicate the median. The green shaded region marks the ±15% convergence zone. For $C \geq 34.3$, 100% of stations fall within ±15%, demonstrating that the weak shock and linear period predictions have effectively converged at the propagation distances of this study. For $C$ = 5.38, the earlier weak shock-to-linear transition produces larger differences (median 25%), but all stations remain within 30%.

Friedman repeated-measures tests applied to the station-wise MAPR against $R_{0,linP}$ corroborate the asymmetric sensitivity at the formulation-resolved level. For the Sakurai formulation the omnibus test is significant ($\chi^2$ = 7.2, $p$ = 0.027), although no individual Holm-corrected pairwise Wilcoxon comparison reaches the α = 0.05 threshold (Holm-adjusted $p$-values of 0.180, 0.255, and 0.255 for the three $C$-pairs), consistent with a MAPR shift of only 1−3 percentage points across the tested $C$ values. For the Jones/Plooster formulation the omnibus test is not significant ($\chi^2$ = 1.2, $p$ = 0.56), indicating that the Jones/Plooster station-wise residuals are effectively insensitive to $C$ within the tested range. The full post-hoc pairwise tables for the Friedman tests above are provided in Supplementary Information Section S6 and **Table S3**.

### 4.6 Along-Trajectory Source-Condition Dependence

Because the SRC follows a single decelerating reentry trajectory, the source altitude, velocity, Mach number, and ambient density are all rank-correlated at Spearman $r_s$ = 1.00 across the 39 stations. The source altitude is therefore adopted as a proxy for this co-varying bundle of source conditions. **Figure 13** examines the dependence of the period residuals on source altitude across the 43 to 62 km range sampled by the stations, on both the inverse side (panel a) and the forward side (panel b). Two altitude-resolved complementary views of the inverted $R_0$ itself and of the period-branch offset are provided in Supplementary Information Section S7 and **Figure S5**.

All three energy-normalized formulations exhibit a strong monotonic trend in the signed per-station residual of the inverted $R_{0,wsP}$ against source altitude (**Figure 13a**), with Spearman rank correlations $r_s$ = 0.85 and slopes of 4.0%/km (Sakurai), 4.3%/km (Jones/Plooster), and 5.7%/km (Few). The residuals cross zero near 52 km source altitude. This altitude-dependent structure explains why the station-median signed residual can be centered near zero (Sakurai +12%, bootstrap 95% CI −10 to +18%, spanning zero) while the per-station residuals are far from randomly distributed: the theoretical $R_0$ is systematically too small at low-Mach/high-density conditions (low altitude) and too large at high-Mach/low-density conditions (high altitude).

The same altitude trend is reproduced independently on the forward side (**Figure 13b**), with slopes of 3.0 %/km (Sakurai), 3.1 %/km (Jones/Plooster), and 3.8 %/km (Few) and Spearman $r_s$ = 0.84 for all three energy-normalized formulations. The consistency between the inverse-side and forward-side trends demonstrates that the altitude dependence is embedded in the $R_0$ formulations themselves rather than introduced by the inversion procedure; the trend is a structural property of the cylindrical blast wave model applied to a decelerating hypersonic source, not an artifact of the benchmarking.

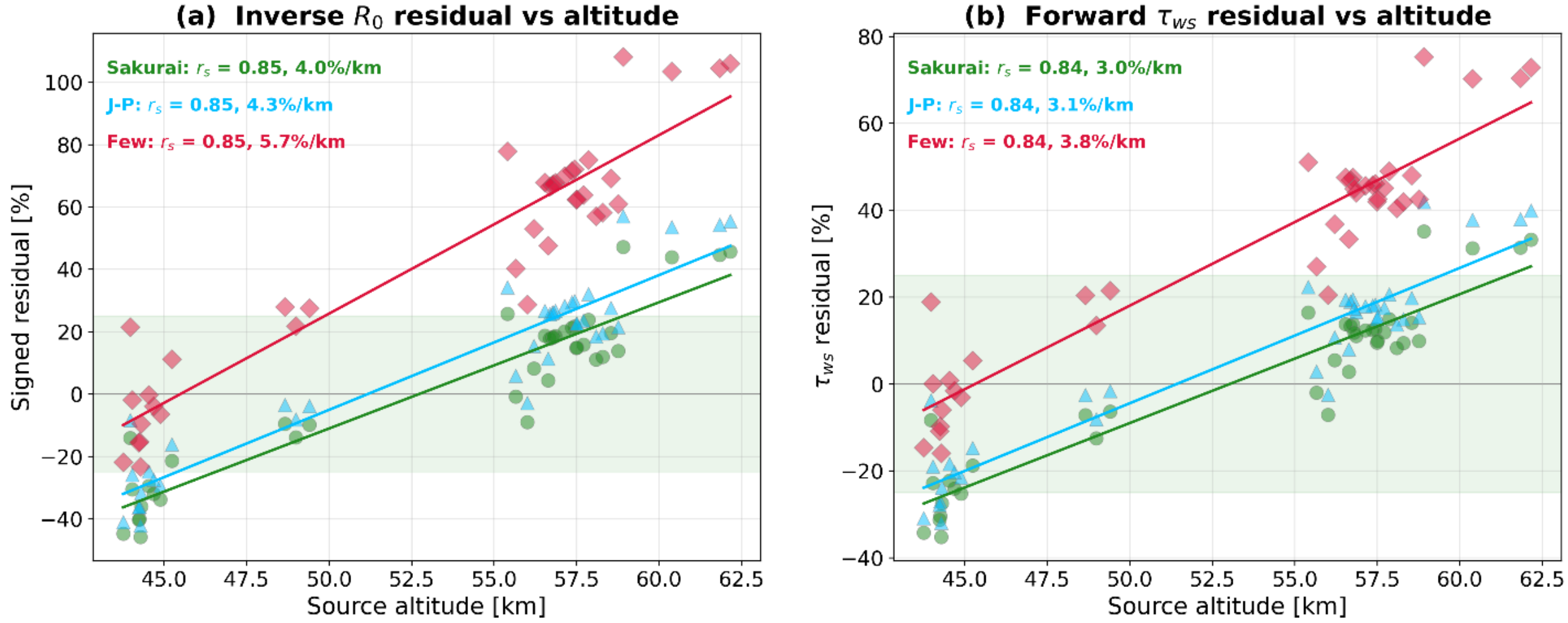

**Figure 13:** Dependence of model residuals on along-trajectory source conditions. Source altitude serves as a proxy for the co-varying bundle of velocity, Mach number, and ambient density along the OSIRIS-REx SRC trajectory (Spearman rank correlation between altitude and velocity: $r_s$ = 1.00). (a) Signed percentage residual (theoretical minus inverted, normalized by inverted) for each formulation, with linear regression lines and Spearman correlations annotated. The green band marks ±25%. All three formulations share $r_s$ = 0.85 and cross zero near 52 km, producing negative residuals (underprediction) at low altitude and positive residuals (overprediction) at high altitude. (b) Forward model confirmation: signed percentage residual of the predicted weak shock period ($\tau_{ws}$) relative to the observed period, showing the same monotonic trend ($r_s$ = 0.84 for all formulations).

# 5. Discussion

## 5.1 Context and Motivation

The weak shock formulation has been applied to meteor infrasound for several decades (ReVelle, 1974, 1976), and the observational validation by Silber et al. (2015) demonstrated that the model reasonably predicts blast radii from infrasound periods for centimeter-sized ablating meteoroids. However, in all meteor applications, the source parameters (mass, velocity, fragmentation state, ablation rate) must be inferred indirectly from optical observations, introducing substantial uncertainty into any comparison between theoretical and observed $R_0$. As a result, no previous study has been able to evaluate the performance of the $R_0$ formulations themselves from the uncertainties in the source characterization. The OSIRIS-REx SRC, with its well-known dimensions, mass, trajectory, and velocity at each point along the flight path, removes this ambiguity and enables the first controlled benchmarking of blast radius formulations against a non-ablating hypersonic source. ReVelle and Edwards (2006) applied a weak shock theory to the Stardust SRC reentry using infrasound recorded at a single four-element array ~33 km from the trajectory, with the analysis based on the Mach-diameter approximation for $R_0$ and a single emission point near 43 km altitude. The present work extends that approach to a 39-station network sampling emission points across an 18 km altitude range, enabling both a formulation-level comparison and the detection of source-condition-dependent trends that a single observation point cannot resolve. Several recent studies have examined the OSIRIS-REx SRC reentry from complementary perspectives (e.g., Bishop et al., 2025; KC et al., 2025; Nishikawa et al., 2026; Silber and Bowman, 2025). Building on the infrasound observations reported by Silber and Bowman (2025), the present work focuses on benchmarking cylindrical blast wave formulations within the weak shock framework by applying a systematic forward and inverse modeling approach. Whereas previous applications of weak shock theory to both meteors and SRC reentries were limited to single-formulation, single-coefficient analyses (e.g., Edwards, 2009; ReVelle and Edwards, 2006; Silber et al., 2015), the present work evaluates six formulations (Jones et al., 1968; Plooster, 1970; ReVelle, 1974; Sakurai, 1965; Tsikulin, 1970) across three WST coefficients (Morse and Ingard, 1968; Silber et al., 2015; Towne, 1967) using both forward and inverse approaches, with rigorous nonparametric statistical testing. This systematic comparison, yielding a formulation-level ranking, a quantitative assessment of period versus amplitude reliability, and an along-trajectory residual characterization, has not been attempted previously for any atmospheric entry source, ablating or non-ablating.

### 5.2 Benchmarking $R_0$ Formulations for Non-Ablating Bodies

The forward and inverse scorecards (**Figures 8 and 9**; summarized in **Table S1**) provide compelling evidence that the Sakurai (1965) formulation is the most appropriate for non-ablating bodies. In the forward analysis, Sakurai achieves the lowest period MAPR in every row of the scorecard, with the best single-cell value of 9% ($\tau_{lin}$ at $C$ = 34.3). In the inverse analysis, Sakurai achieves the lowest MAPR in three of four period rows, with the best value of 11% ($R_{0,linP}$ at $C$ = 34.3). Against $R_{0,wsP}$, pairwise Wilcoxon signed-rank tests confirm that Sakurai yields significantly smaller residuals than Jones-Plooster ($p$ = 0.0001, Holm corrected). However, this ranking is $C$-dependent in the linear period regime: at $C$ = 5.38, the two formulations are effectively indistinguishable (MAPR 13.6% versus 13.4%), whereas at $C$ ≥ 34.3 Sakurai outperforms Jones/Plooster. Because a rigid, non-ablating body is better described by a smaller WST coefficient, the practical difference between the two formulations is modest under the most physically appropriate $C$ choice. The bootstrap confidence interval on the Sakurai signed residual spans zero (+12%, CI: −10% to +18%), indicating no detectable aggregate bias.

The scaling-type formulations (Tsikulin standard, Tsikulin modified, Mach-diameter) overestimate $R_0$ by 120–297% across all period benchmarks. Because the characteristic source period is proportional to $R_0$, the inflated blast radius values produced by these formulations yield source-side periods that already exceed the observed periods, and the discrepancy persists under propagation to the receiver. The Mach-diameter approximation ($R_0 \approx Md_m$) is particularly inappropriate for non-ablating bodies because the absence of ablation products reduces the energy coupled into the wake relative to what would be expected for an ablating meteoroid of comparable size and velocity.

However, this overall assessment masks a systematic, monotonic trend in the residuals with along-trajectory source conditions. All six formulations show a strong positive correlation between residual and source altitude ($r_s$ = 0.85, **Figure 13a**), driven by the co-varying capsule velocity and Mach number. The Sakurai formulation achieves residuals within ±25% for stations with source altitudes between approximately 46 and 58 km, defining a validity envelope for this formulation applied to non-ablating sources. At the trajectory extremes, the constant-normalization $R_0$ systematically underpredicts at low Mach/high density and overpredicts at high Mach/low density. The aggregate near-zero bias arises because the residuals cross zero near 52 km altitude (**Figure 13a**), not because the formulation is uniformly accurate. The Sakurai formulation underpredicts $R_0$ by 30-45% at the lowest source altitudes and overpredicts by 20-50% at the highest, and it is the fortuitous cancellation of these opposing biases across the station ensemble that produces the aggregate result of no detectable systematic offset.

### 5.3 Period-Branch Agreement and the Case Against Amplitude

A central result of this study is that the weak shock period ($\tau_{ws}$) and the linear period ($\tau_{lin}$) branches yield closely matched $R_0$ estimates at the propagation distances characteristic of the OSIRIS-REx observations. For $C$ = 34.3 and $C$ = 57.2, 100% of stations show $R_{0,linP}$ within 15% of $R_{0,wsP}$, with a median difference of 9% (95th percentile: 13%) (**Figure 12**). This branch agreement is altitude-independent (**Figure S5b**) and implies that, at these ranges (~7,000–10,000 blast radii), the period-based $R_0$ estimate is effectively independent of both the WST coefficient and the choice of propagation regime.

The amplitude, by contrast, does not provide a self-consistent constraint on $R_0$. The amplitude-derived $R_{0,wsA}$ exceeds $R_{0,wsP}$ by a station-median factor of 5.9 (range 2.8 to 10.0; Wilcoxon signed-rank $p < 10^{-7}$). The linear amplitude estimate $R_{0,linA}$ varies by a factor of 3.5 across the tested $C$ range, compared to only 16% variation for $R_{0,linP}$. The forward scorecard confirms this asymmetry: the minimum amplitude MAPR (28%, Sakurai $\Delta p_{lin}$ at $C$ = 34.3) is three times larger than the minimum period MAPR (9%). Furthermore, the formulations that best predict the period perform poorly for the weak shock amplitude, and vice versa, indicating that the amplitude agreement of the scaling-type formulations is partially fortuitous, arising because their larger $R_0$ values compensate for the systematic amplitude underprediction of the present cylindrical blast-wave treatment.

The amplitude misfit likely arises from multiple sources. First, frequency-dependent atmospheric absorption along the propagation path (Sutherland and Bass, 2004), wind-dependent refraction and multipathing, scattering from atmospheric fine structure, and local site amplification or de-amplification effects all modify the amplitude in ways not captured by the cylindrical blast wave formulation (Chunchuzov et al., 2011; Chunchuzov et al., 2025; Chunchuzov et al., 2026; de Groot-Hedlin et al., 2010; Drob et al., 2010; Ostashev et al., 2005). Second, the weak shock model assumes an ideal N-wave pressure signature, which may not hold for a blunt body such as the SRC (e.g., Cheng, 1959, 1963; Hayes et al., 1968). The detached bow shock geometry of the capsule produces a near-field pressure signature that might differ from the idealized cylindrical blast wave of an infinitesimally thin line source, and these near-field departures may persist into the far field as amplitude anomalies even after the period has converged to the cylindrical solution. Third, infrasound amplitudes are inherently more susceptible to station-specific effects than the period: local topography, wind noise, sensor coupling, and site response all introduce amplitude variability that might not affect the general temporal structure of the waveform (e.g., Averiyanov et al., 2011; Bass et al., 1991; Bird et al., 2022; Ens et al., 2012; Lacanna et al., 2014; Tahira and Donn, 1983). The period, as a temporal feature of the N-wave, is largely independent of these station-specific gain factors, making it a fundamentally more stable observable for constraining $R_0$. A rigorous treatment of the amplitude would require full-waveform propagation modeling, which is beyond the scope of this study. While this study demonstrates that overpressure amplitude remains poorly constrained by all tested formulations at these propagation distances, providing a theoretical resolution to this multi-factor misfit requires future numerical investigation and is outside the present scope.

### 5.4 Dependence on Source Conditions

The strong monotonic dependence of the residuals on along-trajectory source conditions (**Figure 13a,b**) reveals a systematic limitation of the constant-normalization cylindrical blast wave scaling. At the low-altitude end of the trajectory (43–46 km), where the capsule is decelerating rapidly through dense atmosphere at Mach 8–10, all three energy-normalized formulations produce $R_0$ values that are too small relative to the period-inverted benchmark. At the high-altitude end (57–62 km), where the capsule travels at Mach 30–35 in thin atmosphere, the formulations produce $R_0$ values that are too large. This pattern is consistent across both inverse and forward analyses and is shared by all energy-normalized formulations (identical $r_s$ = 0.85), indicating that it reflects a structural feature of the cylindrical blast wave approach rather than a deficiency of any single normalization convention.

This along-trajectory bias points to a missing dependence in the effective $R_0$ mapping. The theoretical $R_0$ assumes that the fraction of kinetic energy coupled into the cylindrical shock is constant along the trajectory, but the coupling efficiency may in fact depend on Mach number, on the rate of deceleration (which affects the coherence of the line source approximation), or on the ambient density profile. The results suggest that $R_0$ may require a source-condition-dependent normalization factor $\alpha$ to bridge the low and high altitude regimes, such that the effective blast radius becomes $R_{0,eff} = \alpha R_{0,Sakurai}$, where $\alpha$ varies monotonically from values greater than unity at low Mach numbers to values less than unity at high Mach numbers. Whether $\alpha$ is primarily a function of Mach number, ambient density, or deceleration rate is difficult to isolate from the present dataset because these parameters co-vary tightly along a single decelerating trajectory. Although the different functional dependencies of density (exponential with altitude), velocity (governed by the drag equation), and Mach number (velocity divided by the altitude-dependent sound speed) are in principle distinguishable, the 39-station sample provides limited statistical power to discriminate among competing predictors. Definitive attribution would benefit from observations of multiple entry events with different trajectory geometries, where the co-variation structure differs. The OSIRIS-REx dataset nonetheless establishes the existence and magnitude of the bias (~3-6% per km of source altitude) and defines a validity envelope for the Sakurai formulation.

### 5.5 Weak-Shock Transition Distance: Source-Type Dependence

The WST coefficient $C$ governs the distance over which nonlinear waveform steepening is effective before the wave transitions to the linear regime. Although the three candidate values of $C$ originate from different physical criteria (a shock-formation-distance criterion for Morse and Ingard (1968), and a period-change criterion applied to ReVelle's distortion distance for Towne (1967) and for the Silber et al. (2015)-based value adopted here), they are compared empirically because Eq. (6) applies them identically as scalar thresholds on the WST distance; the benchmarking reported here addresses the question of which threshold is best suited to a rigid, non-ablating source, rather than which underlying physical criterion is most fundamental.

For the OSIRIS-REx SRC, the period-based $R_0$ is only weakly sensitive to $C$: $R_{0,wsP}$ is identical across all three tested values, and $R_{0,linP}$ varies by only 16% across the full range from $C$ = 5.38 to $C$ = 57.2. This insensitivity arises because the SRC stations occupy an intermediate propagation regime in which the two period branches give closely matched predictions regardless of the precise location of the weak-shock-to-linear transition: the accumulated divergence between the two branches, integrated from the transition point to each station, is relatively small at all three tested $C$ values. The implication is that the period alone cannot discriminate among WST coefficient values at these propagation distances. Because the amplitude, which is the observable most sensitive to $C$, is not reliably predicted by the cylindrical blast wave formulation, constraining $C$ for non-ablating bodies will require additional investigation, whether through higher-fidelity propagation models, further analysis of the present dataset, or observations from future controlled reentry events.

### 5.6 Implications for Non-Ablating Hypersonic Sources

The results of this study have direct implications for the interpretation of infrasound from non-ablating hypersonic sources, a category that encompasses sample return capsules, space debris, reentry vehicles, and, at the upper size limit, small near-Earth asteroids that penetrate the atmosphere without significant ablation. Each of these source types generates a cylindrical shock governed by the same blast wave physics examined here, but the physical parameters (body size, velocity, entry angle, ablation state) span a wide range.

For sample return capsule-like objects, the benchmarked Sakurai formulation and the period-based inversion framework provide a means to estimate the source function and energy deposited per unit path length from infrasound observations alone, without requiring a complementary sensing modality as a prerequisite for first-order source characterization. This is relevant for post-event characterization of reentries where only infrasound data are available (e.g., Silber et al., 2011), as may occur for capsules returning over oceanic or remote continental areas where dedicated detection infrastructure is absent. The Stardust SRC reentry (ReVelle and Edwards, 2006) and Hayabusa reentries (Yamamoto et al., 2011) were both observed by infrasound, but the blast radius formulations used in those analyses were not benchmarked against the known capsule parameters in the systematic manner presented here. The present calibration, which establishes that the Sakurai formulation achieves 9–20% MAPR for a capsule of diameter 0.81 m entering at 3–11 km/s over 43–62 km altitude, provides a quantitative baseline for reanalyzing those events and for planning infrasound observation campaigns for future space missions.

The present problem also has broader relevance to current and forthcoming exploration-class reentries. The April 2026 Artemis II mission returned the Orion spacecraft to Earth following a crewed lunar flyby, providing a recent example of continuing operational interest in high-speed capsule return and recovery (e.g., Dooren, 2026; Woffinden et al., 2023). In this broader context, the OSIRIS-REx benchmark is relevant to blunt-body reentry capsules whose geometry and trajectory are independently constrained and for which geophysical observations may support physically grounded interpretation of source behavior during entry. The Orion crew module occupies a different aerodynamic and thermal-protection regime from OSIRIS-REx, including a substantially larger characteristic size and an ablative Avcoat heat shield (Thomas, 2026), but both missions illustrate the importance of benchmarked atmospheric-entry models for interpreting the signatures of modern spacecraft returns.

Because the blast radius is directly linked to the energy deposited per unit path length, the period-based inversion provides a practical route to estimating the source function from infrasound periods alone. For a chosen normalization (for example, the Sakurai formulation), a period-inverted blast radius at a given emission altitude can be mapped to an observational estimate of $E_0$ through the local ambient pressure, yielding a line-source strength associated with that trajectory segment. When a network of stations samples different emission points along the path, the resulting set of period-derived $E_0$ values defines a discretized source function $E_0(z)$ that reveals where along the trajectory the atmospheric coupling is strongest. If an independent velocity profile is available, $E_0(z)$ can also be converted to a time-domain source function via the deposited power per unit time, enabling direct comparison with drag-only expectations and providing a basis for identifying deviations consistent with changing coupling efficiency, ablation, or fragmentation. This period-only pathway is particularly useful in operational settings because the period-based $R_0$ estimate is robust to transition assumptions at the ranges considered here, whereas amplitude-based constraints are not.

More broadly, the cylindrical blast wave formulation as calibrated here is applicable to any non-ablating or weakly ablating body for which the line source approximation remains valid. The critical requirement is that the body diameter be small relative to the blast radius ($d_m \ll R_0$), so that the shock geometry is effectively cylindrical rather than dominated by finite-source effects. For the OSIRIS-REx SRC, $d_m/R_0 \approx 0.1$–$0.4$ depending on altitude, indicating that the line source assumption is marginally satisfied at low altitudes and well satisfied at high altitudes. For larger bodies or lower velocities where $d_m/R_0$ approaches unity, the cylindrical line-source approximation becomes less appropriate and a treatment that accounts explicitly for finite source geometry would be required. For fragmenting sources, the problem may instead require a distributed-source or multi-source treatment, depending on how the energy deposition is partitioned in space and time (e.g., Silber et al., 2025; Trigo-Rodríguez et al., 2021).

The along-trajectory bias carries a practical caution: applying a single constant-normalization $R_0$ formulation across a wide range of source conditions introduces systematic errors of 3–6% per km of source altitude (inverse-side slopes 4.0 to 5.7 %/km; forward-side slopes 3.0 to 3.8 %/km), even for the best-performing formulation. For applications that require accuracy better than ±25% (e.g., energy estimation for planetary defense), altitude- or Mach-dependent corrections may be necessary. The monotonic, well-characterized nature of the bias suggests that an empirical correction could be derived from the present dataset, although its generalizability to other entry events would need to be verified.

Finally, the finding that the period is robust while the amplitude is unreliable at propagation distances exceeding several thousand blast radii has a practical consequence for infrasound network design. For atmospheric entry monitoring, the period can be measured reliably from a single station without requiring amplitude calibration, station- specific site response corrections, or knowledge of the propagation path attenuation. This makes period-based inversion a more operationally practical approach for post-event characterization of atmospheric entry events.

### 5.7 Limitations and Future Work

The weak shock model treats the hypersonic shock as originating from a cylindrical line source, an idealization that neglects the detailed near-field structure of the detached bow shock and wake. For a non-ablating body such as the SRC, the shock coupling is governed by aerodynamic drag, whereas for ablating meteors additional energy is deposited through mass loss and vapor-phase interactions. The extent to which these differences affect the far-field weak shock evolution at thousands of blast radii has not been quantified through numerical simulation for this specific case. The atmospheric profile is based on the G2S specifications, which do not resolve small-scale variability such as gravity waves; however, Silber et al. (2015) found that gravity wave perturbations affect $R_0$ by only ~10%, well below the formulation-level biases of 20–50% documented here, and therefore do not warrant additional treatment in the present analysis. Because source parameters (altitude, velocity, Mach number, density) co-vary along a single decelerating trajectory, the along-trajectory trend cannot be confidently attributed to any single physical parameter from this dataset alone, although the differing functional forms of the co-varying parameters leave open the possibility that future analysis may partially disentangle them.

The present analysis is restricted to direct acoustic arrivals, for which the weak shock formulation is applicable. At distances beyond the direct-arrival range, atmospheric dynamics play an increasingly important role: stratospheric and thermospheric refracted or ducted arrivals, caustic effects, and diffraction into geometrical-acoustics shadow zones can produce secondary arrivals whose amplitude, waveform character, and propagation losses differ substantially from the direct-path predictions (e.g., Chunchuzov et al., 2011; Chunchuzov et al., 2025; Chunchuzov et al., 2026; de Groot-Hedlin et al., 2010; Drob et al., 2010; Ostashev et al., 2005). The amplitude misfit may be partially attributable to atmospheric effects that modify the waveform even along nominally direct paths, including frequency-dependent absorption, refraction through mesoscale wind and temperature structures, and scattering from gravity wave-induced perturbations (e.g., Averbuch et al., 2022a; Averbuch et al., 2022b; Averiyanov et al., 2011; Silber and Bowman, 2023). Extending the weak shock framework to non-direct arrivals would require coupling to a full-waveform propagation model that accounts for these atmospheric effects, which represents a natural next step for this line of research.

Several avenues for future work follow from these results. First, strengthening the attribution of the along-trajectory bias to specific physical parameters would benefit from observations of multiple entry events with different trajectory geometries-, either from future sample return missions or from well-characterized fireball events with independent trajectory solutions. Second, the systematic along-trajectory bias identified here serves as a quantitative benchmark for the limitations of the constant-normalization approach; a physical resolution to this trend requires future Mach-dependent or density-dependent theoretical development, which could be tested against numerical blast wave simulations. Third, a full-waveform propagation treatment incorporating frequency-dependent absorption, refraction, and scattering is needed to explain the amplitude misfit and to determine whether the amplitude can be made a useful observable at these propagation distances.

A more formal reconciliation between the ReVelle (1974) weak shock transition criterion and the classical nonlinear-acoustics language of shock-formation and distortion lengths would also be worthwhile but lies beyond the scope of the present benchmarking study.

## 6. Conclusions

This study provides the first systematic benchmarking of the ReVelle (1974) semi-analytical weak-shock treatment of cylindrical blast waves against a non-ablating hypersonic source with independently known geometry, trajectory, and velocity at each emission point. Six blast radius formulations and three WST coefficients were evaluated through forward and inverse modeling of infrasound observations at 39 stations during the OSIRIS-REx Sample Return Capsule reentry. The principal findings are as follows.

i. The Sakurai (1965) formulation is recommended for non-ablating hypersonic bodies, achieving a forward-model period MAPR of 9% at $C$ = 34.3 with a near-zero median signed residual. The Jones et al. (1968)/Plooster (1970) formulation performs comparably (MAPR 11%) and is practically indistinguishable from Sakurai when the physically appropriate Morse and Ingard (1968) WST coefficient is adopted. The three scaling-type formulations systematically overestimate $R_0$ and are not supported for this source class. The Mach-diameter approximation, calibrated to ablating meteoroids, overestimates $R_0$ by a factor of ~3.4 for the rigid, non-ablating SRC-like body.

ii. The weak-shock period ($\tau_{ws}$) and linear period ($\tau_{lin}$) branches yield closely matched $R_0$ estimates at the propagation distances characteristic of this study (approximately 7,000 to 10,000 $R_0$). For $C \geq 34.3$, all 39 stations show period-derived $R_0$ differences within 15%. The period-based $R_0$ estimate is therefore only weakly dependent on $C$ at the assumed propagation regime at these ranges, making the observed infrasound period a robust single-observable pathway from receiver-side measurement to source-level energy deposition per unit path length for the OSIRIS-REx SRC geometry. For non-ablating sources of comparable size and velocity observed at similar scaled distances, this weak sensitivity may simplify source-function recovery from infrasound observations by reducing the dependence on prior knowledge of the WST coefficient.

iii. From an operational standpoint, the weak $C$-sensitivity suggests that, for non-ablating sources of comparable size and velocity observed at similar scaled distances, the period alone may provide a practical constraint on $R_0$ with reduced dependence on prior knowledge of $C$, simplifying source-function recovery from single-station infrasound observations. Whether this weak $C$-sensitivity extends to sources with substantially different $R_0$ or ablation state has not been tested and should not be assumed, because the degree of branch separation at a given scaled distance is controlled by the interplay of $C$, the source-side blast radius, and the atmospheric structure along the propagation path, all of which may differ for other source types and geometries.

iv. The peak overpressure amplitude does not provide a self-consistent constraint on $R_0$ at these propagation distances. The amplitude-derived $R_0$ exceeds the period-derived $R_0$ by a median factor of ~6, is strongly sensitive to $C$, and exhibits substantial station-to-station variability. No tested formulation reconciles the period and amplitude constraints simultaneously. The amplitude misfit is attributed to a combination of unmodeled propagation effects, near-field departures from the idealized cylindrical line-source geometry, and station-specific site response, none of which comparably affect the period.

v. Residuals for all energy-normalized formulations exhibit a strong monotonic dependence on source altitude (Spearman $r_s$ = 0.85), transitioning from underprediction at low altitudes (Mach 8 to 10) to overprediction at high altitudes (Mach 30 to 35). This trend is confirmed independently by both the inverse and forward analyses and represents a structural property of the constant-normalization $R_0$ formulations applied to a decelerating source. The Sakurai (1965) formulation achieves residuals within ±25% over the approximate source-altitude range of 46 to 58 km, defining a quantitative validity envelope for non-ablating bodies. Because the co-varying source parameters cannot be disentangled along a single trajectory, definitive attribution would benefit from observations of multiple entry events with different trajectory geometries.

vi. The three tested WST coefficients originate from two distinct families of physical criterion: a shock-formation-distance criterion (Morse and Ingard, 1968) and a period-change-threshold criterion applied to the distortion distance (Towne (1967) at 10%; the Silber et al. (2015)-based value at ~6%). Both families enter the model identically through Eq. (6), and the comparison is empirical. A smaller $C$ is expected to be physically more appropriate for the non-ablating SRC, because the absence of continuous ablation-driven energy deposition allows the shock to weaken more rapidly toward the linear regime, but quantitative determination of the optimal $C$ for this source class will require either additional observational constraints or higher-fidelity propagation modeling.

vii. The OSIRIS-REx SRC reentry provides a controlled test case that eliminates the source-parameter ambiguities inherent in meteor studies. A physical resolution of the along-trajectory trend, whether through source-condition-dependent normalization corrections or numerical blast wave simulation, would represent the most direct improvement to the weak-shock treatment for decelerating hypersonic sources. A rigorous treatment of the amplitude misfit will require full-waveform propagation modeling that accounts for frequency-dependent absorption, refraction, and scattering not captured by the present semi-analytical framework.

This benchmarking effort establishes a quantitative performance baseline for applying weak-shock theory to non-ablating hypersonic reentries and demonstrates that the signal period is the operationally reliable observable for constraining source-level blast radii at propagation distances of several thousand $R_0$ and beyond. The Sakurai (1965) formulation provides the best overall agreement with period observations, with the Jones et al. (1968)/Plooster (1970) formulation performing comparably under the physically appropriate WST coefficient. A comprehensive theoretical resolution for the observed amplitude misfit and the altitude-dependent residual structure remains an open challenge that will require future numerical and multi-event observational investigation.

## Data availability

The infrasound signal periods, amplitudes, and station locations analyzed in this study are available in Silber and Bowman (2025). The weak-shock modeling framework follows the formulations described in Silber and Brown (2019) and Silber et al. (2015). Statistical analyses were conducted using standard Python scientific computing libraries. Derived quantities, including the blast radius ($R_0$) values, are presented in the article. Supplementary material providing additional statistical detail and supporting figures accompanies this article and is available in the electronic supplement.

## Acknowledgments

The author thanks Sven Peter Näsholm and an anonymous reviewer for their constructive comments that helped improve this paper. This work was supported by the Nuclear Arms Control Technology (NACT) program at the Defense Threat Reduction Agency (DTRA). Cleared for release.

## Conflict of Interest

The author declares no conflict of interest.

# Supporting Materials

# for

# Benchmarking Cylindrical Blast Wave Theory Against the OSIRIS-REx Sample Return Capsule Reentry

Elizabeth A. Silber

Sandia National Laboratories, Albuquerque, NM, 87123

## Contents

## Description

This Supplementary Information documents the two-zero-crossing period-picking convention applied to the OSIRIS-REx arrivals (S1, Figures S1 and S2), the forward-model residual distributions for period and amplitude across the 39 stations (S2, Figures S3 and S4), the distributions of inverted $R_0$values (S3, Figure S5), the full pairwise Wilcoxon and Hodges-Lehmann formulation comparisons (S4, Table S1), the bootstrap signed-bias confidence intervals (S5, Table S2), the Friedman tests of weak-shock-transition-coefficient sensitivity (S6, Table S3), and the altitude-resolved supplementary along-trajectory plots (S7, Figure S6).

## S1. Time-Domain Period Picking for the OSIRIS-REx Arrivals

The period and peak-overpressure values used in this study were taken from the station-by-station measurements reported by Silber and Bowman (2025). The period measure adopted for the Origins, Spectral Interpretation, Resource Identification, and Security–Regolith Explorer (OSIRIS-REx) Sample Return Capsule (SRC) arrivals is a modified version of time-domain zero-crossing estimate. The main paper summarizes the measurement philosophy only briefly; the purpose of this section is to document the period-picking convention more explicitly.

A typical zero-crossing picking technique, described in Ens et al. (2012) and Silber (2024), and schematically depicted in **Figure S1**, uses four successive zero (ZC) crossings at maximum amplitude to derive the dominant period ($\tau_{mean} \pm \Delta\tau$). Here, $\tau_{mean} = \frac{\tau_1 + \tau_2}{2}$, where $\tau_1 = |t_{ZC3} - t_{ZC1}|$, and $\tau_2 = |t_{ZC4} - t_{ZC2}|$. The associated uncertainty is estimated as $\Delta\tau = \frac{s_\tau}{\sqrt{2}}$, where is $s_\tau$ the standard deviation of the two period estimates.

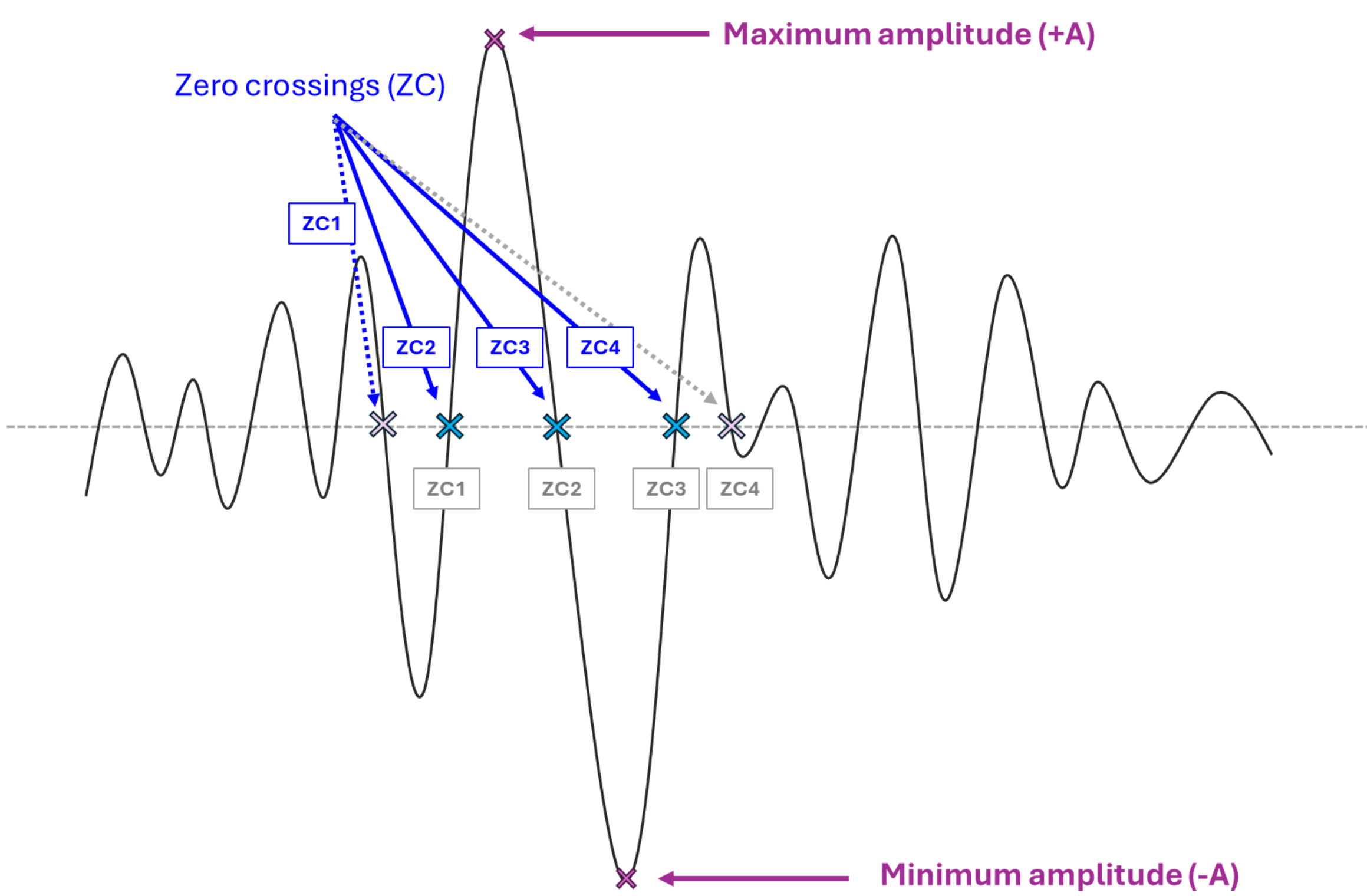

**Figure S1:** Schematic representation of zero crossing time picking, adapted from Figure B1 in Silber et al. (2025).

The dominant period is often cross-validated against the peak frequency ($f_{peak}$) from spectral analysis. For regional bright fireballs producing short-duration impulsive arrivals, Silber and Brown (2014) found that time-domain period estimates and reciprocal dominant-frequency estimates were generally consistent to within about 10%. At larger energies and greater ranges, dominant-frequency-based estimates expressed through $\tau \approx 1/f_{peak}$, have likewise been shown to be of comparable scale to time-domain period estimates for many bolide signals, commonly at the 10% level (Ens et al., 2012; Gi and Brown, 2017). This correspondence may become less straightforward for broader-band arrivals, where the dominant spectral peak can be less sharply defined.

For OSIRIS-REx-like arrivals, the principal difficulty in period picking lies in the signal morphology. These waveforms commonly consist of a short, high-contrast leading pulse followed by a trailing segment that is more distorted, lower in amplitude, and more structurally complex. The trailing portion of the waveform may contain coda and other secondary structure and can remain below the zero-pressure baseline for part of the signal before returning across it. In other cases, it may re-cross the baseline in short, low-amplitude excursions. This morphology introduces ambiguity in identifying the later zero crossings, making conventional four-zero-crossing period estimates more susceptible to subjective analyst interpretation and less reproducible across independent analyses. The difficulty is compounded by propagation effects and site-dependent waveform distortion, which further degrade the coherence of the trailing phase relative to the leading pulse.

To mitigate this, Silber and Bowman (2025) defined the period from the leading compressional pulse by identifying the two zero crossings that bound that half-cycle. If those crossings are denoted $t_{ZC1}$ and $t_{ZC2}$, the corresponding period estimate is $\tau = 2(t_{ZC2} - t_{ZC1})$. The construction is illustrated in **Figure S2**. Because the leading positive excursion is generally the most stable and repeatable portion of the arrival, the two-zero-crossing method reduces sensitivity to analyst judgment and yields a more robust period estimate than a full-cycle pick under these conditions. A full-cycle pick can still be appropriate when both positive and negative phases remain coherent (**Figure S1**), but for signals of this morphological character the two-zero-crossing definition provides a more objective and reproducible event-specific measure while remaining broadly comparable to conventional period estimates. The definition adopted here is therefore intended as a practical event-specific measure for signals of this general character, rather than as a universal prescription for all infrasound waveforms. It emphasizes the most repeatable part of the arrival while retaining a period estimate that remains broadly comparable to more conventional measures. A separate side-by-side study by Scamfer et al. (2026), using the Alaska fireball, found that standard zero-crossing and two-zero-crossing estimates were typically comparable to within about 10%. The same leading-half-cycle logic has also been used outside the bolide context. In their regional ground-explosion study, Silber et al. (2026) applied the two-zero-crossing method to measure infrasound signal periods. These studies support the present convention as a practical and repeatable period measure for some short-duration impulsive infrasound arrivals. More integrative observables, including signal-energy or power-based measures, may provide useful complementary constraints in future work.

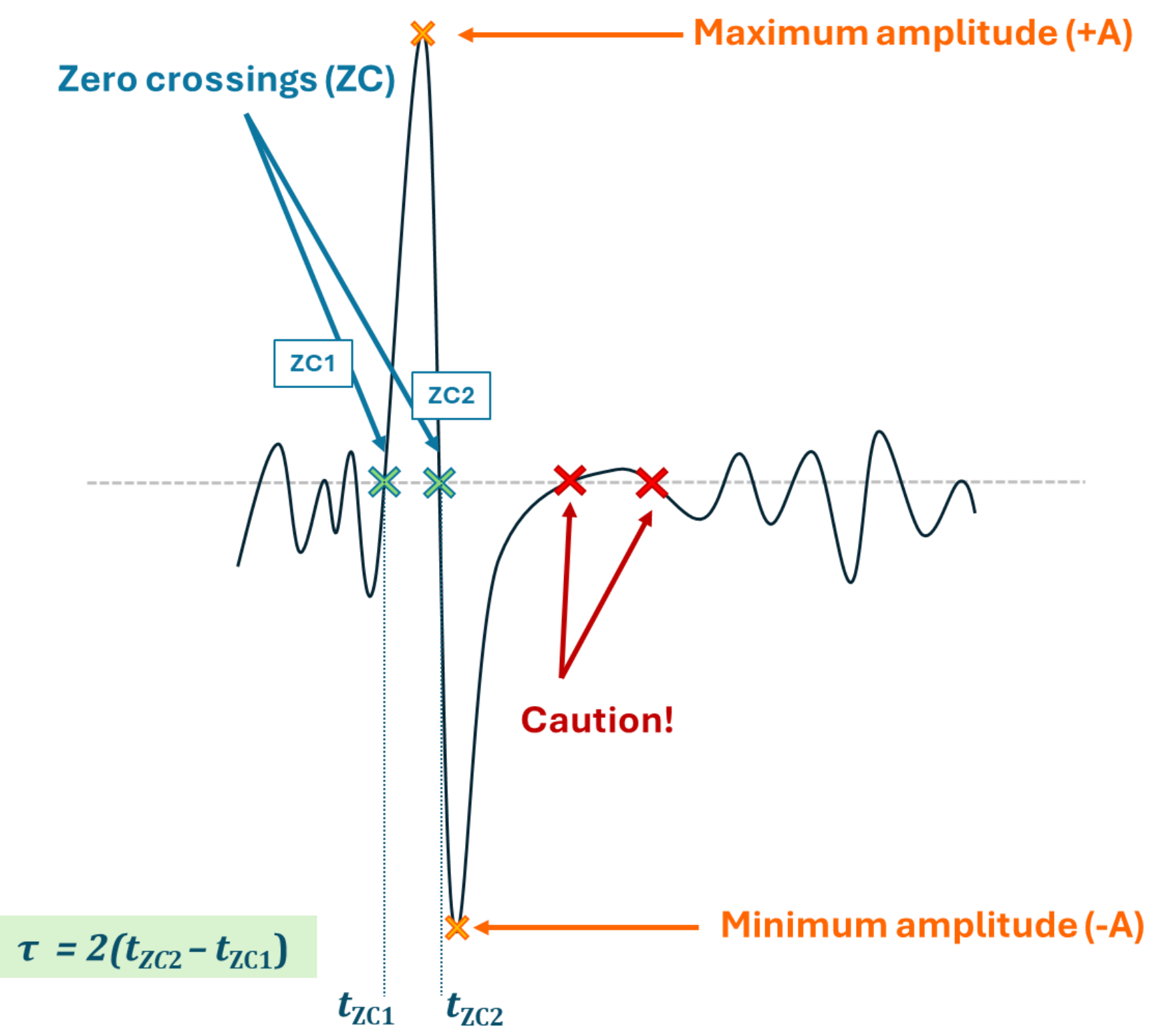

**Figure S2:** Schematic of the two-zero-crossing period measurement approach used for the OSIRIS-REx signals. The times $t_{ZC1}$ and $t_{ZC2}$ mark the zero crossings that bound the leading compressional half-cycle of the waveform, and the corresponding dominant period is computed as $\tau = 2(t_{ZC2} - t_{ZC1})$. This definition emphasizes the earliest and typically most stable part of the arrival. In some short-duration impulsive signals, the later waveform can contain more complex secondary structure and trailing coda, making a full-cycle period estimate less straightforward to interpret. The definition adopted here is intended for signals of this general type.

## S2. Forward Model Residual Distributions

This section describes the distributions of the forward-model period and amplitude residuals across the 39 stations for all six $R_0$ formulations and three weak-shock-transition coefficients. The weak-shock period branch is independent of $C$. These station-wise distributions provide the distributional context for the scorecard summary presented in the main text.

### S2.1 Period-Residual Distributions

**Figure S3** presents the period residuals in a four-panel layout: panel (a) shows the weak shock period ($\tau_{ws}$), which is independent of the weak shock transition (WST) coefficient, while panels (b), (c), and (d) show the linear period ($\tau_{lin}$) for $C$ = 5.38 (Morse and Ingard, 1968), $C$ = 34.3 (Towne, 1967), and $C$ = 57.2 (Silber et al., 2015), respectively. The results divide the six $R_0$ formulations into two distinct groups. The energy-normalized formulations produce the smallest errors: for $\tau_{ws}$, the Sakurai (1965) MAPR is 13% and the Jones et al. (1968)/Plooster (1970) MAPR is 18% (**Figure S3a**). The Few (1969) formulation shows a larger positive bias (median signed residual +42%, MAPR 42%). The scaling-type formulations systematically overpredict the period, with MAPRs of 118% (Tsikulin (1970) standard), 183% (Tsikulin (1970) modified), and 173% (Mach-diameter, ReVelle (1974)).

The linear period ($\tau_{lin}$) shows a modest dependence on the WST coefficient. For the Sakurai formulation, the Towne (1967) coefficient ($C$ = 34.3) yields the lowest MAPR of 9% with a near-zero median signed residual of +1% (**Figure S3c**), consistent with the unbiased performance confirmed by the bootstrap confidence interval on the median signed residual. The Jones/Plooster formulation achieves a MAPR of 11% at the same $C$. Across both formulations, the median signed residuals are near zero and the interquartile ranges straddle zero, confirming that the cylindrical blast wave model captures the major features of period evolution for the energy-normalized $R_0$ group. The scaling-type formulations remain above 80% MAPR regardless of $C$. Across all 18 $R_0/C$ combinations, the period MAPR ranges from 9% (Sakurai (1965), $C$ = 34.3) to 183% (Tsikulin (1970) modified, $C$ = 57.2).

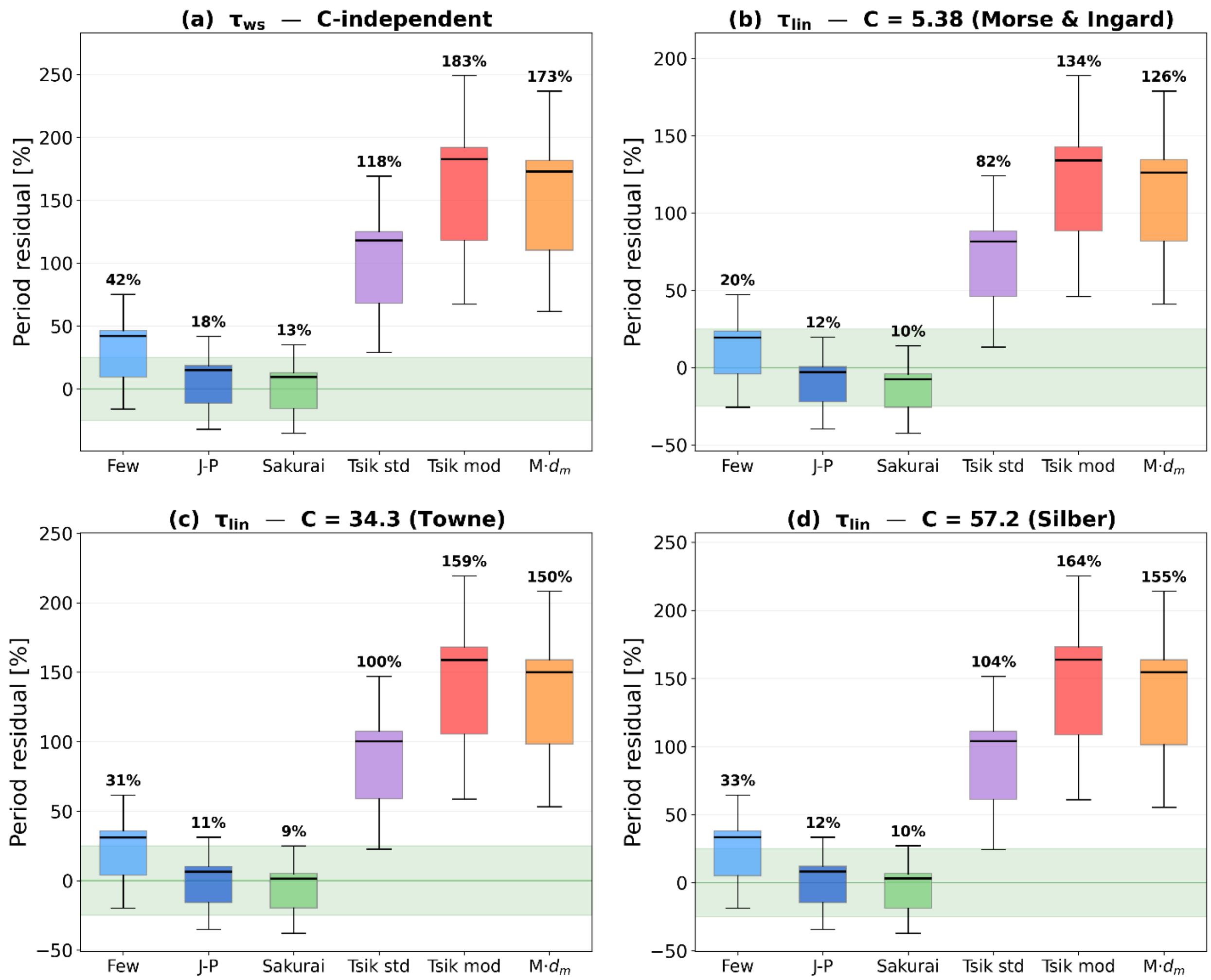

**Figure S3:** $\tau_{\text{lin}}$ and $\tau_{\text{ws}}$ residuals across six $R_0$ formulations and three $C$ values. **(a)** Weak-shock period, $\tau_{ws}$, which is independent of $C$. **(b–d)** Linear-regime period, $\tau_{lin}$, for (b) $C = 5.38$ (Morse and Ingard, 1968), **(c)** $C = 34.3$ (Towne, 1967, as adopted by ReVelle, 1974), and **(d)** $C = 57.2$ (Silber et al., 2015). Residuals are computed as (predicted-observed)/observed×100% over the 39 stations. Positive values indicate overprediction and negative values indicate underprediction.

## S2.2. Amplitude-Residual Distributions

**Figure S4** presents the amplitude residuals, which reveal a pattern that differs fundamentally from the period results. The weak shock amplitude (**Figure S4a**) is systematically underpredicted for all $R_0$ formulations: the model produces amplitudes that are too small relative to observations. The energy-normalized formulations show the largest underprediction, with MAPRs of 66% (Few, 1969), 73% (Jones et al. (1968)/Plooster (1970)), and 74% (Sakurai, 1965). The scaling-type formulations show smaller underprediction (MAPRs of 31–47% for the three scaling-type formulations), but these are the same formulations that overpredict the period. The linear amplitude ($\Delta p_{lin}$) is strongly dependent on both $R_0$ and $C$. For $C$ = 5.38 (**Figure S4b**), all formulations overpredict the amplitude, with MAPRs ranging from 109% (Sakurai) to 578% (Tsikulin modified). The Towne coefficient ($C$ = 34.3; **Figure S4c**) reduces the errors substantially, with the Jones/Plooster and Sakurai formulations achieving MAPRs of 28%, the lowest amplitude values observed across all combinations. For $C$ = 57.2 (**Figure S4d**), the Few formulation achieves its best amplitude MAPR (29%) while the Tsikulin standard achieves 37%.

These results reveal a fundamental tension. The $R_0$ formulations that best reproduce the period (Sakurai and Jones/Plooster, with $\tau_{lin}$ MAPRs of 9–12%) underpredict the weak shock amplitude by 73–74% ($C$-independent) and show $C$-dependent $\Delta p_{lin}$ MAPRs of 28–122%. Conversely, the formulations with the smallest $\Delta p_{ws}$ MAPRs, Tsikulin modified and Mach-diameter, at 31–34%, overpredict the period by 173–183%. Across all 18 combinations, the period MAPR spans 9–183%, whereas the amplitude MAPR spans 28–578%. No single $R_0$ formulation simultaneously satisfies both the period and amplitude constraints. The Towne coefficient ($C$ = 34.3) combined with the Jones/Plooster or Sakurai formulation offers the best joint performance for $\tau_{lin}$ and $\Delta p_{lin}$ (both near 28% MAPR), but the amplitude errors remain two to three times larger than the period errors for the same combination.

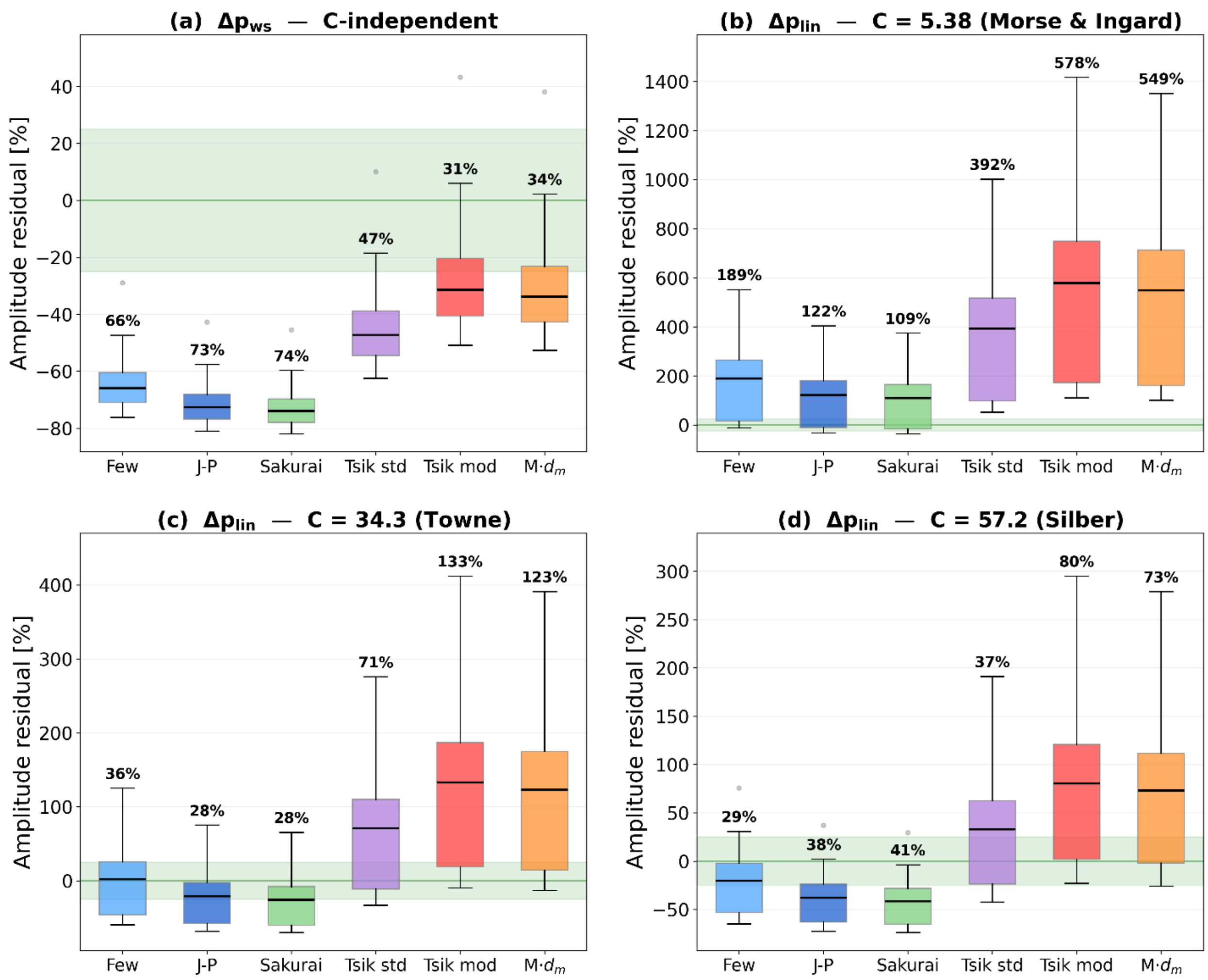

**Figure S4:** $\Delta p_{lin}$ and $\Delta p_{ws}$ residuals across six $R_0$ formulations and three $C$ values. **(a)** Weak-shock overpressure amplitude, $\Delta p_{ws}$, which is independent of $C$. **(b–d)** Linear-regime overpressure amplitude, $\Delta p_{lin}$, for (b) $C = 5.38$ (Morse and Ingard, 1968), **(c)** $C = 34.3$ (Towne, 1967, as adopted by ReVelle, 1974), and **(d)** $C = 57.2$ (Silber et al., 2015). Residuals are computed as (predicted-observed)/observed×100% over the 39 stations. Positive values indicate overprediction and negative values indicate underprediction.

## S3. Inversion $R_0$ Comparison

**Figure S5** presents the distributions of inverted $R_0$ values across all 39 stations, with panel (a) showing period-derived estimates and panel (b) showing amplitude-derived estimates. Horizontal dashed lines indicate the mean theoretical $R_0$ from each formulation for reference. The period-derived $R_0$ values (**Figure S5a**) are tightly clustered relative to the amplitude-derived values. The weak shock period estimate ($R_{0,wsP}$), which is independent of $C$, has a median of 6.2 m, falling between the theoretical Jones et al. (1968)/Plooster (1970) (6.2 m) and Sakurai (5.8 m) values. The linear period estimates show a modest $C$-dependence: $R_{0,linP}$ has a median of 7.8 m for $C$ = 5.38 (Morse and Ingard, 1968), 6.8 m for $C$ = 34.3 (Towne, 1967), and 6.6 m for $C$ = 57.2 (Silber et al., 2015). As $C$ increases, $R_{0,linP}$ converges toward $R_{0,wsP}$, consistent with the expectation that larger $C$ values place the weak-shock-to-linear transition closer to the station and therefore reduce the divergence between the linear-period and weak-shock-period branches. All period-derived medians fall in the range of 6–8 m, bracketed by the Few (1969) (8.2 m) and Sakurai (1965) (5.8 m) theoretical values.

The amplitude-derived $R_0$ values (**Figure S5b**) reveal a strikingly different picture. The weak shock amplitude estimate ($R_{0,wsA}$) has a median of 31.3 m with a large interquartile range extending from approximately 20 to 45 m, far exceeding any theoretical $R_0$ value. This $R_{0,wsA}$ value is approximately five times larger than the period-derived $R_{0,wsP}$ value. The linear amplitude estimates are strongly $C$-dependent: $R_{0,linA}$ ranges from a median of 3.1 m ($C$ = 5.38) to 8.6 m ($C$ = 34.3) to 10.9 m ($C$ = 57.2). For $C$ = 5.38, the early transition causes extensive linear decay, producing $R_{0,linA}$ values well below any theoretical formulation. For $C$ = 34.3, $R_{0,linA}$ falls near the Few (1969) and Jones et al. (1968)/Plooster (1970) theoretical values. For $C$ = 57.2, $R_{0,linA}$ exceeds the Few (1969) theoretical value and approaches the Tsikulin (1970) standard range.

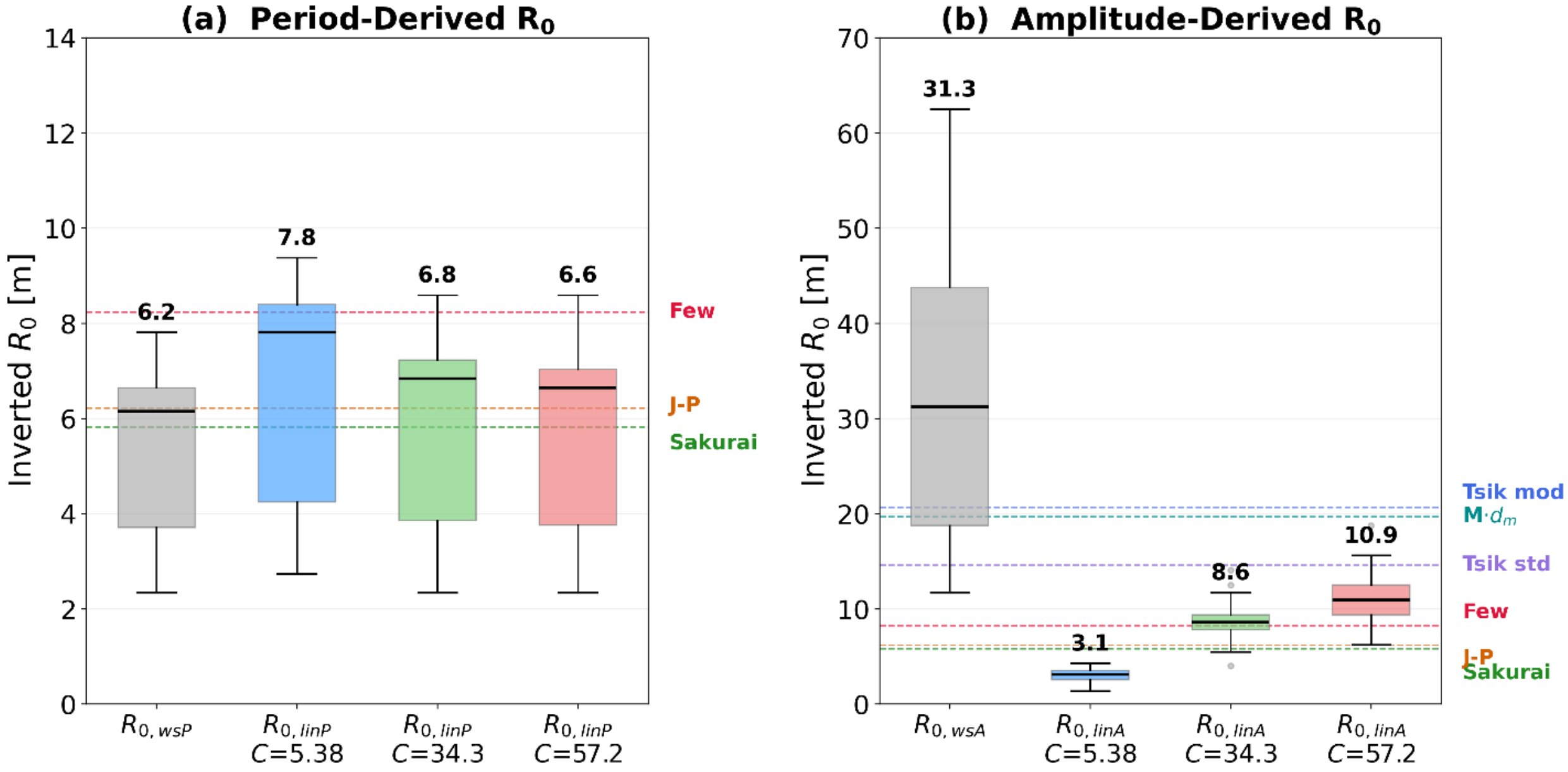

**Figure S5:** Distributions of inverted $R_0$ values across 39 stations. (a) Period-derived $R_0$: weak shock period ($R_{0,wsP}$, independent of $C$) and linear period ($R_{0,linP}$) for three WST coefficients. (b) Amplitude-derived $R_0$: weak shock amplitude ($R_{0,wsA}$, independent of $C$) and linear amplitude ($R_{0,linA}$) for three WST coefficients. Boxplots show the median (annotated), interquartile range, and whiskers extending to 1.5 times the interquartile range. Horizontal dashed lines indicate the mean theoretical $R_0$ from each formulation computed along the trajectory. Note the different vertical scales between panels.

## S4. Pairwise Formulation Comparisons

This section reports the full pairwise statistical comparisons among the three energy-normalized formulations (Sakurai (1965), Jones et al. (1968)/Plooster (1970), Few (1969)). The main paper presents the scorecard-style summary of the forward benchmark, whereas this section provides the supporting pairwise Wilcoxon signed-rank tests and Hodges-Lehmann effect sizes.

Wilcoxon signed-rank tests are applied to the station-wise absolute residuals, with Holm correction across the three pairwise comparisons (Sakurai (1965) versus Jones et al. (1968)/Plooster (1970), Sakurai (1965) versus Few (1969), Jones et al. (1968)/Plooster (1970) versus Few (1969)). The Hodges-Lehmann estimator is used to provide an effect size for each comparison in the form of a median pairwise difference in absolute residual, reported with a 95% confidence interval. Results are reported against the weak-shock period benchmark ($R_{0,wsP}$) and against the linear period benchmark ($R_{0,linP}$) at each of the three tested weak shock transition (WST) coefficients. Wilcoxon signed-rank $p$-values are computed via the asymptotic normal approximation (*scipy.stats.wilcoxon*, zero-method "wilcox"), as tied within-pair differences at one or more stations preclude use of the exact permutation distribution. The same convention applies to the post-hoc pairwise tests in ***Section S6***.

### S4.1 Ranking Against $R_{0,wsP}$

Against $R_{0,wsP}$, the three formulations are ranked as follows: Sakurai (1965) yields the smallest absolute residuals, Jones/Plooster is second, and Few (1969) is third. Sakurai residuals are significantly smaller than Jones et al. (1968)/Plooster (1970) ($p$ = 0.0001, Holm-corrected) and smaller than Few ($p$ < 0.001, Holm-corrected); Jones/Plooster residuals are significantly smaller than Few ($p$ < 0.001, Holm-corrected). The Hodges-Lehmann median pairwise difference is −4.6 percentage points (pp) for Sakurai versus Jones/Plooster (95% CI: −9.9 to +3.3 pp), −27.2 pp for Sakurai versus Few (CI: −42.3 to −11.5 pp), and −26.1 pp for Jones/Plooster versus Few (CI: −38.4 to −7.7 pp). These effect sizes indicate that the Sakurai advantage is substantial on the median absolute percentage residual (MAPR) metric used throughout the study, particularly in comparison with Few (1969).

### S4.2 $C$-dependent ranking against $R_{0,linP}$

The Sakurai-versus-Jones/Plooster ranking is itself dependent on the WST coefficient when tested against $R_{0,linP}$. At $C$ = 5.38 (Morse and Ingard, 1968), the two formulations are practically indistinguishable: the period-derived MAPR is 13.6% for Sakurai versus 13.4% for Jones et al. (1968)/Plooster (1970), a difference of 0.2 pp (Wilcoxon $p$ = 0.003, Holm-corrected). At $C$ ≥ 34.3, Sakurai (1965) outperforms Jones et al. (1968)/Plooster (1970) (MAPR 11–13% versus 15–17%; Wilcoxon $p$ ≤ 0.003, Holm-corrected across the three comparisons).

The physical interpretation is geometric. A smaller $C$ places the weak-shock-to-linear transition closer to the source, and therefore the signal propagates in the linear regime over a longer path before reaching the stations. Meanwhile, if no transition took place, a hypothetical weak shock branch propagates over that same distance. The separation between the two period branches is therefore largest at $C$ = 5.38, and the inverted $R_{0,linP}$ sits furthest above $R_{0,wsP}$ (station medians 7.8 m versus 6.2 m). As $C$ increases, the transition moves farther from the source, the linear-regime path shortens, and $R_{0,linP}$ collapses toward $R_{0,wsP}$ (6.8 m at $C$ = 34.3, 6.6 m at $C$ = 57.2). For a rigid, non-ablating body such as the SRC, values of $C$ smaller than the Silber et al. (2015) coefficient ($C$ = 57.2) are physically more appropriate: that coefficient was derived for ablating meteoroids whose continuous energy deposition reinforces the shock along the trajectory, pushing the transition to lower altitude than would be expected for a body without ablative mass loss. The SRC transition should therefore occur at higher altitude, corresponding to a $C$ < 57.2.

The formulation ranking, the Wilcoxon significance of each pairwise comparison, the Hodges-Lehmann effect sizes with their 95% confidence intervals, and the amplitude-side ranking are consolidated in **Table S1**.

**Table S1:** Formulation ranking summary for the three energy-normalized $R_0$ formulations. Scaling-type formulations all exceed 80% period MAPR. The Hodges-Lehmann $\Delta$ reports the median pairwise difference in absolute residual relative to the next-best formulation against $R_{0,wsP}$, with 95% CI in parentheses.

| Formulation | Model $\tau$ vs. observed $\tau$ (best $C$) MAPR | R, theory vs. R0 inverted $\tau$ (best $C$) MAPR | Wilcoxon vs. next-best | Hodges–Lehmann Δ (95% CI) | Amplitude MAPR | Rank |
|---|---|---|---|---|---|---|
| Sakurai | 9% $\tau_{lin}$, $C$=34.3 | 11% $R_{0,linP}$ $C$=34.3 | vs. JP: $p$ = 0.0001 | −4.6 pp (−9.9, +3.3) | 28% $\Delta p_{lin}$ $C$=34.3 | 1 |
| Jones / Plooster | 11% $\tau_{lin}$ $C$=34.3 | 13% $R_{0,linP}$ $C$=5.38 | vs. Few: $p$ < 0.001 | −26.1 pp (−38.4, −7.7) | 28% $\Delta p_{lin}$ $C$=34.3 | 2 |
| Few | 20% $\tau_{lin}$ $C$=5.38 | 27% $R_{0,linP}$ $C$=5.38 | worst of three | — | 29% $\Delta p_{lin}$ $C$=57.2 | 3 |
| Tsikulin (std, mod), $Md_m$ | >80% | >80% | — | — | >120% | N/R |

## S5. Bootstrap Signed-Bias Confidence Intervals

This section reports the bootstrap 95% confidence intervals (CI) on the station-median signed residual, defined as (predicted − observed) / observed × 100%, for each of the three energy-normalized formulations. Intervals are based on 10,000 bootstrap resamples using the percentile method. A bootstrap CI that spans zero indicates no detectable systematic bias at the sample size of this study ($n$ = 39 stations); an interval strictly on one side of zero indicates systematic over- or under-prediction.

### S5.1 Signed Bias Against $R_{0,wsP}$

Against $R_{0,wsP}$, the Sakurai median signed residual is +12% with 95% CI from −10% to +18%, which spans zero and therefore shows no detectable systematic bias at the aggregate scale. Jones et al. (1968)/Plooster (1970) is +19% (CI: −3% to +26%), also spanning zero but with the interval shifted positive. Few (1969) is +58% (CI: +28% to +67%), strictly positive and therefore a systematic overprediction.

### S5.2 Signed Bias Against $R_{0,linP}$

The signed-bias behavior against $R_{0,linP}$ varies with $C$. At $C$ = 34.3 (Towne, 1967), both Sakurai (1965) (+1%, CI: −17% to +7%) and Jones/Plooster (+7%, CI: −11% to +14%) are centred near zero. At $C$ = 57.2 (Silber et al.), the pattern is similar: Sakurai (1965) +2% (CI: −14% to +8%) and Jones/Plooster +9% (CI: −8% to +15%). At $C$ = 5.38 (Morse and Ingard), the bias shifts negative: Sakurai is −11% (CI: −23% to −6%), strictly negative and therefore a systematic underprediction; Jones/Plooster is −5% (CI: −18% to +1%), with the interval spanning zero (no detectable bias at the aggregate scale).

The $C$-dependence of the signed bias mirrors the $C$-dependence of the ranking. At $C$ = 5.38 the weak-shock-to-linear transition is placed closest to the source, so the signal propagates over the longest linear-regime path before reaching the stations. The inverted $R_{0,linP}$ tends to shift upward relative to the theoretical blast radius, making the per-station (predicted−observed)/observed residual negative. Aggregated across the 39 stations, this produces the $C$ = 5.38 negative bias seen above. At $C \geq 34.3$ the transition occurs farther from the source, the linear-regime path length is shorter, and $R_{0,linP}$ stays close to the theoretical $R_0$ at each station. Therefore, the station-median signed bias is centered near zero. The complete bootstrap signed-bias results for all three energy-normalized formulations and all relevant benchmarks are listed in **Table S2**.

**Table S2:** Bootstrap 95% CI on the station-median signed residual for the three energy-normalized formulations, across all combinations of benchmark and $C$ value. The “Bias?” column applies a single rule: if the interval spans zero, the entry reads “No”; if the interval is strictly on one side of zero, the entry reads “Yes” (with the direction noted where informative).

| Formulation | Benchmark | $C$ | Median Signed Residual | 95% CI | Bias? |
|---|---|---|---|---|---|
| Sakurai | $R_{0,wsP}$ | — | +12% | −10% to +18% | No |
| Jones/Plooster | $R_{0,wsP}$ | — | +19% | −3% to +26% | No |
| Few | $R_{0,wsP}$ | — | +58% | +28% to +67% | Yes (positive) |
| Sakurai | $R_{0,linP}$ | 5.38 | −11% | −23% to −6% | Yes (negative) |
| Jones/Plooster | $R_{0,linP}$ | 5.38 | −5% | −18% to +1% | No |
| Sakurai | $R_{0,linP}$ | 34.3 | +1% | −17% to +7% | No |
| Jones/Plooster | $R_{0,linP}$ | 34.3 | +7% | −11% to +14% | No |
| Sakurai | $R_{0,linP}$ | 57.2 | +2% | −14% to +8% | No |
| Jones/Plooster | $R_{0,linP}$ | 57.2 | +9% | −8% to +15% | No |

## S6. Friedman Tests of $C$-Sensitivity

This section reports Friedman repeated-measures tests of the effect of the WST coefficient $C$ on the station-wise MAPR against $R_{0,linP}$, for each of the two leading energy-normalized formulations (Sakurai (1965), Jones et al. (1968)/Plooster (1970)). The Friedman test is chosen because each station contributes three measurements (one per $C$ value) that are not independent, which is the natural matched-sample structure for this comparison. Post-hoc pairwise comparisons between $C$ values use Wilcoxon signed-rank tests with Holm correction across the three pairwise comparisons.

### S6.1 Sakurai Formulation

For the Sakurai formulation, the Friedman test is significant ($\chi^2$ = 7.2, $p$ = 0.027), indicating that the three $C$ values do not produce identical error distributions at the station-wise level.

The Holm-corrected post-hoc Wilcoxon $p$-values are 0.180 for $C = 5.38$ versus $C = 34.3$, 0.255 for $C = 5.38$ versus $C = 57.2$, and 0.255 for $C = 34.3$ versus $C = 57.2$. Thus, although the omnibus test is significant, no individual pairwise comparison reaches the corrected threshold. The omnibus significance combined with the non-significant post-hoc results is consistent with the modest absolute magnitude of the differences: the Sakurai MAPR varies by approximately 1–3 percentage points across the three tested $C$ values, and the omnibus test is sensitive to the direction of the trend across stations rather than to a large pairwise shift between any two $C$ values.

### S6.2 Jones/Plooster Formulation

For the Jones/Plooster formulation, the Friedman test is not significant ($\chi^2$ = 1.2, $p$ = 0.557). The Holm-corrected post-hoc Wilcoxon $p$-values are 0.160 for $C = 5.38$ versus $C = 34.3$, 0.124 for $C = 5.38$ versus $C = 57.2$, and 0.160 for $C = 34.3$ versus $C = 57.2$. These comparisons confirm that the Jones et al. (1968)/Plooster (1970) station-wise residuals are effectively insensitive to $C$ within the tested range.

These results are summarized in **Table S3**. The small-magnitude $C$-sensitivity of the Sakurai (1965) formulation, and the absence of $C$-sensitivity for Jones et al. (1968)/Plooster (1970), are both consistent with the branch-proximity picture described in the main text: at the OSIRIS-REx source–station geometry, the $\tau_{ws}$ and $\tau_{lin}$ modeled branches are close in value at the transition altitudes spanned by the three tested $C$ values, so moving the transition location within this range produces only a modest shift in the inverted $R_{0,linP}$ and therefore only a modest shift in the station-wise residuals.

**Table S3:** Friedman repeated-measures test of the effect of $C$ on the station-wise MAPR against $R_{0,linP}$ for each energy-normalized formulation. Post-hoc pairwise comparisons use Wilcoxon signed-rank tests with Holm correction across the three pairwise comparisons. Post-hoc tests were conducted for both formulations regardless of omnibus significance.

| **Formulation** | **Friedman $\chi^2$** | **Friedman $p$** | **$C$ = 5.38 vs $C$ = 34.3 ($p$)** | **$C$ = 5.38 vs $C$ = 57.2 ($p$)** | **$C$ = 34.3 vs $C$ = 57.2 ($p$)** |
|---|---|---|---|---|---|
| Sakurai | 7.2 | 0.027 | 0.180 | 0.255 | 0.255 |
| Jones/Plooster | 1.2 | 0.557 | 0.160 | 0.124 | 0.160 |

## S7. Supplementary Along-Trajectory Analysis

This section provides two altitude-resolved views that complement the inverse-side and forward-side residual panels shown in main-text Figure 13. **Figure S6a** plots the inverted weak-shock period-derived blast radius $R_{0,wsP}$ (black) alongside the theoretical $R_0$ from each of the three energy-normalized formulations (Sakurai, Jones/Plooster, Few) as a function of source altitude for all 39 stations. Each theoretical curve rises monotonically with altitude, because the upper-trajectory segments of the reentry have higher source velocity and lower ambient density and therefore produce larger theoretical $R_0$. The inverted $R_{0,wsP}$ rises more slowly with altitude than any of the theoretical curves: at the lowest altitudes sampled (around 43 to 45 km) the inverted values lie near or slightly above the Sakurai and Jones/Plooster curves, whereas at the highest altitudes (around 60 to 62 km) the theoretical curves have pulled well above the inverted values, particularly for the Few formulation. This altitude-dependent mismatch is the direct driver of the signed-residual trends quantified in main-text Figure 13.

**Figure S6b** resolves by source altitude the per-station offset $(R_{0,linP} - R_{0,wsP})$ / $R_{0,wsP}$ × 100% at each of the three tested WST coefficients ($C$ = 5.38, 34.3, 57.2). The aggregate histograms of the same quantity are shown in main-text Figure 12, and the summary statistics are reported in Section 4.5. At $C$ = 34.3 and $C$ = 57.2, all 39 stations lie within the shaded ±15% band across the entire 43 to 62 km altitude range, indicating that the period-branch agreement is essentially uniform with respect to source altitude. At $C$ = 5.38, a substantial fraction of stations lies above the band at every altitude, consistent with the 25% aggregate median and the 15%-within-band fraction (6 of 39 stations) reported in ***Section 4.5***.

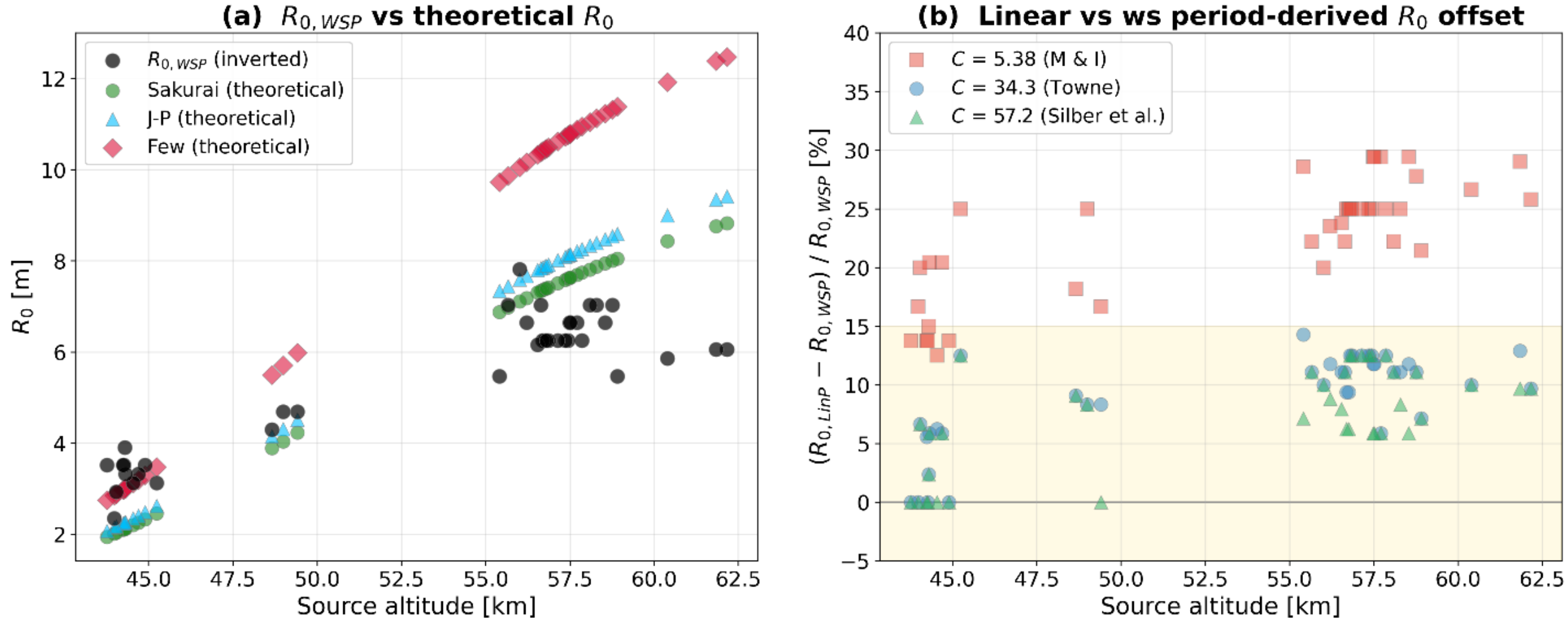

**Figure S6:** Altitude-resolved supplementary plots for the along-trajectory analysis in main-text ***Section 4.6***. (a) Inverted weak-shock period-derived blast radius $R_{0,wsP}$ (black circles) overlaid with the theoretical $R_0$ from the three energy-normalized formulations (Sakurai, Jones/Plooster, Few), plotted against source altitude for all 39 stations. (b) Station-wise offset between the linear-regime and weak-shock period-derived $R_0$, $(R_{0,linP} - R_{0,wsP}) / R_{0,wsP} \times 100\%$, plotted against source altitude at each of the three tested WST coefficients ($C$ = 5.38 Morse and Ingard, $C$ = 34.3 Towne, $C$ = 57.2 Silber et al.). The gold band marks ±15%. Complementary aggregate histograms of the panel (b) quantity appear in main-text Figure 12.

## References

Averbuch, G., Ronac-Giannone, M., Arrowsmith, S., Anderson, J.F., 2022a. Evidence for Short Temporal Atmospheric Variations Observed by Infrasonic Signals: 1. The Troposphere. Earth and Space Science 9, e2021EA002036,10.1029/2021EA002036.

Averbuch, G., Sabatini, R., Arrowsmith, S., 2022b. Evidence for Short Temporal Atmospheric Variations Observed by Infrasonic Signals: 2. The Stratosphere. Earth and Space Science 9, e2022EA002454,10.1029/2022EA002454.

Averiyanov, M., Blanc-Benon, P., Cleveland, R.O., Khokhlova, V., 2011. Nonlinear and diffraction effects in propagation of N-waves in randomly inhomogeneous moving media. The Journal of the Acoustical Society of America 129, 1760-1772,10.1121/1.3557034.

Bass, H.E., Bolen, L.N., Raspet, R., McBride, W., Noble, J., 1991. Acoustic propagation through a turbulent atmosphere: Experimental characterization. The Journal of the Acoustical Society of America 90, 3307-3313,10.1121/1.401441.

Bird, E.J., Lees, J.M., Kero, J., Bowman, D.C., 2022. Topographically Scattered Infrasound Waves Observed on Microbarometer Arrays in the Lower Stratosphere. Earth and Space Science 9, e2022EA002226,10.1029/2022EA002226.

Bishop, J.W., Blom, P., Carr, C., Webster, J., 2025. An infrasound source analysis of the OSIRIS-REx sample return capsule hypersonic re-entry. The Journal of the Acoustical Society of America 158, 4637-4650,10.1121/10.0041857.

Blom, P., 2014. GeoAc: Numerical tools to model acoustic propagation in the geometric limit, Software. Los Alamos National Laboratory. Los Alamos National Laboratory, p. Seismoacoustic software

Blom, P., Waxler, R., 2017. Modeling and observations of an elevated, moving infrasonic source: Eigenray methods. The Journal of the Acoustical Society of America 141, 2681-2692,10.1121/1.4980096.

Bronshten, V.A., 1983. Physics of meteoric phenomena. Fizika Meteornykh Iavlenii, , Moscow, Izdatel'stvo Nauka, 1981 Dordrecht, D. Reidel Publishing Co, Dordrecht, Holland,10.1007/978-94-009-7222-3.

Brown, P., McCausland, P.J.A., Fries, M., Silber, E., Edwards, W.N., Wong, D.K., Weryk, R.J., Fries, J., Krzeminski, Z., 2011. The fall of the Grimsby meteorite—I: Fireball dynamics and orbit from radar, video, and infrasound records. Meteoritics & Planetary Science 46, 339-363,10.1111/j.1945-5100.2010.01167.x.

Campus, P., Christie, D.R., 2009. Worldwide Observations of Infrasonic Waves, in: Le Pichon, A., Blanc, E., Hauchecorne, A. (Eds.), Infrasound Monitoring for Atmospheric Studies. Springer Netherlands, Dordrecht, pp. 185-234,10.1007/978-1-4020-9508-5_6.

Ceplecha, Z., Borovička, J., Elford, W.G., ReVelle, D.O., Hawkes, R.L., Porubčan, V., Šimek, M., 1998. Meteor Phenomena and Bodies. Space Science Reviews 84, 327-471,10.1023/A:1005069928850.

Cheng, H.K., 1959. Similitute of Hypersonic Real-Gas Flows Over Slender Bodies with Blunted Noses. Journal of the Aerospace Sciences 26, 575-585,10.2514/8.8207.

Cheng, H.K., 1963. The blunt-body problem in hypersonic flow at low Reynolds number. Cornell Aeronautical Laboratory, Inc., Buffalo, NY, p. 135

Christie, D.R., Campus, P., 2010. The IMS infrasound network: design and establishment of infrasound stations, Infrasound monitoring for atmospheric studies. Springer, pp. 29-75

Chunchuzov, I.P., Kulichkov, S.N., Popov, O.E., Waxler, R., Assink, J., 2011. Infrasound scattering from atmospheric anisotropic inhomogeneities. Izv. Atmos. Ocean. Phys. 47, 540

Chunchuzov, I.P., Popov, O.E., Silber, E.A., Kulichkov, S.N., 2025. Effect of a Fine-Scale Layered Structure of the Atmosphere on Infrasound Signals from Fragmenting Meteoroids. Pure and Applied Geophysics,10.1007/s00024-025-03835-7.

Chunchuzov, I.P., Popov, O.E., Silber, E.A., Kulichkov, S.N., 2026. Multi-arrival infrasound from meteoroids: Fragmentation signatures versus propagation effects in a fine-scale layered atmosphere. Icarus, 117007,10.1016/j.icarus.2026.117007.

Covington, M., 2005. Performance of a light-weight ablative thermal protection material for the stardust mission sample return capsule, 2nd International Planetary Probe Workshop, Moffett Field, CA, pp. 257-268

de Groot-Hedlin, C.D., Hedlin, M.A.H., Drob, D., 2010. Atmospheric variability and infrasound monitoring, in: A. Le Pichon, E.B., and A. Hauchecorne (Ed.), Infrasound Monitoring for Atmospheric Studies

Devillepoix, H.A.R., Cupák, M., Bland, P.A., Sansom, E.K., Towner, M.C., Howie, R.M., Hartig, B.A.D., Jansen-Sturgeon, T., Shober, P.M., Anderson, S.L., Benedix, G.K., Busan, D., Sayers, R., Jenniskens, P., Albers, J., Herd, C.D.K., Hill, P.J.A., Brown, P.G., Krzeminski, Z., Osinski, G.R., Aoudjehane, H.C., Benkhaldoun, Z., Jabiri, A., Guennoun, M., Barka, A., Darhmaoui, H., Daly, L., Collins, G.S., McMullan, S., Suttle, M.D., Ireland, T., Bonning, G., Baeza, L., Alrefay, T.Y., Horner, J., Swindle, T.D., Hergenrother, C.W., Fries, M.D., Tomkins, A., Langendam, A., Rushmer, T., O'Neill, C., Janches, D., Hormaechea, J.L., Shaw, C., Young, J.S., Alexander, M., Mardon, A.D., Tate, J.R., 2020. A Global Fireball Observatory. Planetary and Space Science 191, 105036,10.1016/j.pss.2020.105036.

Dooren, J.M., 2026. NASA Welcomes Record-Setting Artemis II Moonfarers Back to Earth. NASA

Drob, D.P., Broutman, D., Hedlin, M.A., Winslow, N.W., Gibson, R.G., 2013. A method for specifying atmospheric gravity wavefields for long-range infrasound propagation calculations. Journal of Geophysical Research: Atmospheres 118, 3933-3943,10.1029/2012JD018077.

Drob, D.P., Garces, M., Hedlin, M., Brachet, N., 2010. The Temporal Morphology of Infrasound Propagation. Pure and Applied Geophysics 167, 437-453

Drob, D.P., Picone, J.M., Garces, M., 2003. Global morphology of infrasound propagation. Journal of Geophysical Research 108, 1-12,10.1029/2002JD003307.

Dumond, J.W.M., Cohen, R., Panofsky, W.K.H., Deeds, E., 1946. A Determination of the Wave Forms and Laws of Propagation and Dissipation of Ballistic Shock Waves. J. Acoust. Soc. America 18, 97-118

Edwards, W.N., 2009. Meteor Generated Infrasound: Theory and Observation, in: Le Pichon, A., Blanc, E., Hauchecorne, A. (Eds.), Infrasound Monitoring for Atmospheric Studies. Springer Netherlands, pp. 361-414,10.1007/978-1-4020-9508-5_12.

Efron, B., Tibshirani, R.J., 1993. An introduction to the bootstrap. Chapman and Hall, New York, NY

Ens, T.A., Brown, P.G., Edwards, W.N., Silber, E.A., 2012. Infrasound production by bolides: A global statistical study. Journal of Atmospheric and Solar-Terrestrial Physics 80, 208-229,10.1016/j.jastp.2012.01.018.

Evans, L.B., Bass, H.E., Sutherland, L.C., 1972. Atmospheric Absorption of Sound: Theoretical Predictions. The Journal of the Acoustical Society of America 51, 1565-1575,10.1121/1.1913000.

Evers, L.G., Assink, J.D., Smets, P.S., 2018. Infrasound from the 2009 and 2017 DPRK rocket launches. Geophysical Journal International 213, 1785-1791,10.1093/gji/ggy092.

Few, A.A., 1969. Power Spectrum of Thunder. Journal of Geophysical Research 74, 6926-6934,10.1029/JC074i028p06926.

Francis, S.R., Johnson, M.A., Queen, E., Williams, R.A., 2024. Entry, Descent, and Landing Analysis for the OSIRIS-REx Sample Return Capsule, 46th Annual AAS Guidance, Navigation and Control (GN&C) Conference, Breckenridge, CO

Friedman, M., 1937. The Use of Ranks to Avoid the Assumption of Normality Implicit in the Analysis of Variance. Journal of the American Statistical Association 32, 675-701,10.1080/01621459.1937.10503522.

Gi, N., Brown, P., 2017. Refinement of bolide characteristics from infrasound measurements. Planetary and Space Science 143, 169-181,10.1016/j.pss.2017.04.021.

Green, D.N., Vergoz, J., Gibson, R., Le Pichon, A., Ceranna, L., 2011. Infrasound radiated by the Gerdec and Chelopechene explosions: propagation along unexpected paths. Geophysical Journal International 185, 890-910,10.1111/j.1365-246X.2011.04975.x.

Hayes, W.D., Probstein, R.F., Frenkiel, F.N., Street, R.E., 1968. Hypersonic Flow Theory. Physics Today

Hetzer, C.H., 2024. The NCPAG2S command line client,10.5281/zenodo.13345069.

Hodges Jr, J.L., Lehmann, E.L., 2011. Estimates of location based on rank tests, in: Rojo, J. (Ed.), Selected works of EL Lehmann. Springer, pp. 287-300,10.1007/978-1-4614-1412-4_25.

Holm, S., 1979. A Simple Sequentially Rejective Multiple Test Procedure. Scandinavian Journal of Statistics 6, 65-70

Hupe, P., Ceranna, L., Le Pichon, A., Matoza, R.S., Mialle, P., 2022. International Monitoring System infrasound data products for atmospheric studies and civilian applications. Earth Syst. Sci. Data 14, 4201-4230,10.5194/essd-14-4201-2022.

Jones, D., Goyer, G., Plooster, M., 1968. Shock wave from a lightning discharge. Journal of Geophysical Research 73, 3121-3127

KC, R.J., Wilson, T.C., Fox, D., Spillman, K.B., Garcés, M.A., Elbing, B.R., 2025. Acoustic Observations of the OSIRIS-REx Sample Return Capsule Re-Entry from Wendover Airport. Seismological Research Letters,10.1785/0220250019.

Koschny, D., Drolshagen, E., Drolshagen, S., Kretschmer, J., Ott, T., Drolshagen, G., Poppe, B., 2017. Flux densities of meteoroids derived from optical double-station observations. Planetary and Space Science 143, 230-237,10.1016/j.pss.2016.12.007.

Krehl, P.O.K., 2009. History of Shock Waves, Explosions and Impact, 1 ed. Springer Berlin, Heidelberg, Berlin, Germany,0.1007/978-3-540-30421-0.

Lacanna, G., Ichihara, M., Iwakuni, M., Takeo, M., Iguchi, M., Ripepe, M., 2014. Influence of atmospheric structure and topography on infrasonic wave propagation. Journal of Geophysical Research: Solid Earth 119, 2988-3005,10.1002/2013JB010827.

Landau, L.D., 1945. On shock waves at large distances from the place of their origin. J. Phys. USSR 9, 496-500

Lauretta, D., Balram-Knutson, S., Beshore, E., Boynton, W., Drouet d'Aubigny, C., DellaGiustina, D., Enos, H., Golish, D., Hergenrother, C., Howell, E., 2017. OSIRIS-REx: sample return from asteroid (101955) Bennu. Space Science Reviews 212, 925-984

Lin, S.C., 1954. Cylindrical shock waves produced by instantaneous energy release. Journal of Applied Physics 25, 54

Maglieri, D.J., Plotkin, K.J., 1991. Sonic boom, in: Hubbard, H.H. (Ed.), In Aeroacoustics of flight vehicles: theory and practice. Volume 1: noise sources. NASA Langley Research Center, Hampton, Virginia, USA, pp. 519-561

Mitcheltree, R.A., Wilmoth, R.G., Cheatwood, F., Brauckmann, G., Greene, F.A., 1999. Aerodynamics of Stardust sample return capsule. Journal of Spacecraft and Rockets 36, 429-435,10.2514/2.3463.

Morse, P.M., Ingard, K.U., 1968. Theoretical acoustics. McGraw-Hill, New York, NY, USA

National Centers for Environmental Prediction, 2025. NCEP GFS 0.25 Degree Global Forecast Grids Historical Archive. National Centers for Environmental Prediction/National Weather Service/NOAA/U.S. Department of Commerce, Geoscience Data Exchange,10.5065/D65D8PWK.

Needham, C.E., 2018. Blast Waves, 2 ed. Springer Cham, New York, NY, USA,10.1007/978-3-319-65382-2.

Nippress, A., Green, D.N., 2017. Sensitivity of the International Monitoring System infrasound network to elevated sources: a western Eurasia case study. Geophysical Journal International 211, 920-935,10.1093/gji/ggx342.

Nishikawa, Y., Hamama, I., Elbehiri, H.S., Hasumi, Y., Sansom, E.K., Devillepoix, H., Yamamoto, M.-y., Silber, E.A., 2026. Optimizing Infrasound Observations for Sample Return Capsules re-entry: Insights from OSIRIS-REx and Hayabusa2. Publications of the Astronomical Society of Japan

Ostashev, V.E., Chunchuzov, I.P., Wilson, D.K., 2005. Sound propagation through and scattering by internal gravity waves in a stably stratified atmosphere. The Journal of the Acoustical Society of America 118, 3420-3429,10.1121/1.2126938.

Ott, T., Drolshagen, E., Koschny, D., Mialle, P., Pilger, C., Vaubaillon, J., Drolshagen, G., Poppe, B., 2019. Combination of infrasound signals and complementary data for the analysis of bright fireballs. Planetary and Space Science 179, 104715,10.1016/j.pss.2019.104715.

Pan, Y., Sotomayer, W., 1972. Sonic boom of hypersonic vehicles. AIAA Journal 10, 550-551

Picone, J.M., 2002. NRLMSISE-00 empirical model of the atmosphere: Statistical comparisons and scientific issues. Journal of Geophysical Research 107, 101029/

Pierce, A.D., 2019. Acoustics: an introduction to its physical principles and applications, 3rd Edition ed. Springer, Cham, Switzerland,10.1007/978-3-030-11214-1.

Pilger, C., Ceranna, L., Le Pichon, A., Brown, P., 2019. Large Meteoroids as Global Infrasound Reference Events, in: Le Pichon, A., Blanc, E., Hauchecorne, A. (Eds.), Infrasound Monitoring for Atmospheric Studies: Challenges in Middle Atmosphere Dynamics and Societal Benefits. Springer International Publishing, Cham, pp. 451-470,10.1007/978-3-319-75140-5_12.

Plooster, M.N., 1968. Shock Waves from Line Sources National Center for Atmospheric Research, Boulder, CO, p. 80

Plooster, M.N., 1970. Shock Waves from Line Sources. Numerical Solutions and Experimental Measurements. Physics of Fluids 13, 2665-2675,10.1063/1.1692848.

Randles, C., Da Silva, A., Buchard, V., Colarco, P., Darmenov, A., Govindaraju, R., Smirnov, A., Holben, B., Ferrare, R., Hair, J., 2017. The MERRA-2 aerosol reanalysis, 1980 onward. Part I: System description and data assimilation evaluation. Journal of climate 30, 6823-6850,10.1175/JCLI-D-16-0609.1.

ReVelle, D.O., 1974. Acoustics of meteors - effects of the atmospheric temperature and wind structure on the sounds produced by meteors. University of Michigan, Ann Arbor,10.7302/21181.

ReVelle, D.O., 1976. On meteor-generated infrasound. Journal of Geophysical Research 81, 1217-1230,10.1029/JA081i007p01217.

ReVelle, D.O., Edwards, W.N., 2006. Stardust—An artificial, low-velocity “meteor” fall and recovery: 15 January 2006. Meteoritics & Planetary Science 42, 271-299,10.1111/j.1945-5100.2007.tb00232.x.

Rudenko, O.V., Soluyan, S.I., 1977. Theoretical Foundations of Nonlinear Acoustics. Consultants Bureau, New York

Sachdev, P.L., 1972. Propagation of a blast wave in uniform or non-uniform media: a uniformly valid analytic solution. Journal of Fluid Mechanics 52, 369-378,10.1017/S0022112072001478.

Sachdev, P.L., 2004. Shock Waves and Explosions. CRC Press

Sakurai, A., 1965. Blast wave theory, in: Holt, M. (Ed.), Basic developments in fluid mechanics. Academic Press, pp. 309-375

Scamfer, L.T., Silber, E.A., Fries, M.D., Vida, D., Šegon, D., Jenniskens, P., Nishikawa, Y., Sawal, V., Rector, T., 2026. Multi-Sensor Trajectory Reconstruction of the 24 April 2025 Alaska Fireball and Implications for Planetary Defense. JGR Planets,10.1029/2025JE009440.

Sedov, L.I., 1946. Propagation of intense blast waves. Prikladnaya Matematika i Mekhanika (PMM) 10, 241-250

Silber, E., Bowman, D.C., 2025. Along-trajectory acoustic signal variations observed during the hypersonic reentry of the OSIRIS-REx Sample Return Capsule. Seismological Research Letters,10.1785/0220250014.

Silber, E.A., 2024. Perspectives and Challenges in Bolide Infrasound Processing and Interpretation: A Focused Review with Case Studies. Remote Sensing 16, 3628,10.3390/rs16193628.

Silber, E.A., Boslough, M., Hocking, W.K., Gritsevich, M., Whitaker, R.W., 2018. Physics of meteor generated shock waves in the Earth's atmosphere – A review. Advances in Space Research 62, 489-532,10.1016/j.asr.2018.05.010.

Silber, E.A., Bowman, D.C., 2023. Isolating the Source Region of Infrasound Travel Time Variability Using Acoustic Sensors on High-Altitude Balloons. Remote Sensing 15, 3661

Silber, E.A., Bowman, D.C., Carr, C.G., Eisenberg, D.P., Elbing, B.R., Fernando, B., Garces, M.A., Haaser, R., Krishnamoorthy, S., Langston, C.A., Nishikawa, Y., Webster, J., Anderson, J.F., Arrowsmith, S., Bazargan, S., Beardslee, L., Beck, B., Bishop, J.W., Blom, P., Bracht, G., Chichester, D.L., Christe, A., Clarke, J., Cummins, K., Cutts, J., Danielson, L., Donahue, C., Eack, K., Fleigle, M., Fox, D., Goel, A., Green, D., Hasumi, Y., Hayward, C., Hicks, D., Hix, J., Horton, S., Hough, E., Huber, D.P., Hunt, M.A., Inman, J., Ariful Islam, S.M., Izraelevitz, J., Jacob, J.D., Johnson, J., KC, R.J., Komjathy, A., Lam, E., LaPierre, J., Lewis, K., Lewis, R.D., Liu, P., Martire, L., McCleary, M., McGhee, E.A., Mitra, I.N.A., Ocampo Giraldo, L., Pearson, K., Plaisir, M., Popenhagen, S.K., Rassoul, H., Ronac Giannone, M., Samnani, M., Schmerr, N., Spillman, K., Srinivas, G., Takazawa, S.K., Tempert, A., Turley, R., Van Beek, C., Viens, L., Walsh, O.A., Weinstein, N., White, R., Williams, B., Wilson, T.C., Wyckoff, S., Yamamoto, M.-Y., Yap, Z., Yoshiyama, T., Zeiler, C., 2024. Geophysical Observations of the 24 September 2023 OSIRIS-REx Sample Return Capsule Re-Entry. The Planetary Science Journal 5,10.3847/PSJ/ad5b5e.

Silber, E.A., Bowman, D.C., Egan, S., Burkett, L., Fleigle, M., Kim, K., Newton, T., Schaible, L.P., Sonnenfeld, R., Wynn, N., Snively, J.B., 2026. Observational evidence for wind-driven low-pass filtering of infrasound at short range. Geophysical Research Letters 53,10.1029/2025GL120042.

Silber, E.A., Brown, P., 2019. Infrasound Monitoring as a Tool to Characterize Impacting Near-Earth Objects (NEOs), in: Le Pichon, A., Blanc, E., Hauchecorne, A. (Eds.), Infrasound Monitoring for Atmospheric Studies: Challenges in Middle Atmosphere Dynamics and Societal Benefits. Springer International Publishing, Cham, pp. 939-986,10.1007/978-3-319-75140-5_31.

Silber, E.A., Brown, P.G., 2014. Optical observations of meteors generating infrasound—I: Acoustic signal identification and phenomenology. Journal of Atmospheric and Solar-Terrestrial Physics 119, 116-128,10.1016/j.jastp.2014.07.005.

Silber, E.A., Brown, P.G., Krzeminski, Z., 2015. Optical observations of meteors generating infrasound: Weak shock theory and validation. Journal of Geophysical Research: Planets 120, 413-428,10.1002/2014JE004680.

Silber, E.A., Le Pichon, A., Brown, P.G., 2011. Infrasonic detection of a near-Earth object impact over Indonesia on 8 October 2009. Geophys. Res. Lett. 38, L12201,10.1029/2011gl047633.

Silber, E.A., Trigo-Rodriguez, J., Oseghae, I., Peña Asensio, E., Boslough, M.B., Whitaker, R., Pilger, C., Lubin, P., Sawal, V., Hetzer, C., Longenbaugh, R., Jenniskens, P., Bailey, B., Mas Sanz, E., Hupe, P., Cohen, A.N., Edwards, T.R., Egan, S., Silber, R.E., Czarnowski, S., Ronac Giannone, M., 2025. Multiparameter constraints on empirical infrasound period–yield relations for bolides and implications for planetary defense. The Astronomical Journal 170,10.3847/1538-3881/add47d.

Spearman, C., 1904. The proof and measurement of association between two things. The American Journal of Psychology 100, 72-101,10.2307/1422689.

Sutherland, L.C., Bass, H., E., 2004. Atmospheric absorption in the atmosphere up to 160 km. J. Acoustic. Soc. Am. 115, 1012-1032,10.1121/1.1631937.

Tahira, M., Donn, W.L., 1983. Anomalous infrasound from Space Shuttle II and Skylab I. The Journal of the Acoustical Society of America 73, 461-464,10.1121/1.388995.

Taylor, G., 1950. The formation of a blast wave by a very intense explosion. I. Theoretical discussion. Proceedings of the Royal Society of London. Series A, Mathematical and Physical Sciences 201, 159-174

Thomas, D., 2026. Failure Analysis of Heat Shield Tiles in Spacecraft Thermal Protection Systems. Journal of Failure Analysis and Prevention 26, 1-3,10.1007/s11668-025-02339-9.

Towne, D.H., 1967. Wave phenomena. Courier Dover Publications, New York, NY, USA

Tran, H., Johnson, C., Rasky, D., Hui, F., Hsu, M.-T., Chen, Y., 1996. Phenolic impregnated carbon ablators (PICA) for discovery class missions, 31st Thermophysics Conference, New Orleans, LA, p. 1911,10.2514/6.1996-1911.

Trigo-Rodríguez, J.M., Dergham, J., Gritsevich, M., Lyytinen, E., Silber, E.A., Williams, I.P., 2021. A Numerical Approach to Study Ablation of Large Bolides: Application to Chelyabinsk. Advances in Astronomy 2021, 8852772,10.1155/2021/8852772.

Tsikulin, M., 1970. Shock waves during the movement of large meteorites in the atmosphere. DTIC Document AD 715-537, Nat. Tech. Inform. Serv., Springfield, Va

Vergoz, J., Hupe, P., Listowski, C., Le Pichon, A., Garcés, M.A., Marchetti, E., Labazuy, P., Ceranna, L., Pilger, C., Gaebler, P., Näsholm, S.P., Brissaud, Q., Poli, P., Shapiro, N., De Negri, R., Mialle, P., 2022. IMS observations of infrasound and acoustic-gravity waves produced by the January 2022 volcanic eruption of Hunga, Tonga: A global analysis. Earth and Planetary Science Letters 591, 117639,10.1016/j.epsl.2022.117639.

Virtanen, P., Gommers, R., Oliphant, T.E., Haberland, M., Reddy, T., Cournapeau, D., Burovski, E., Peterson, P., Weckesser, W., Bright, J., 2020. SciPy 1.0: fundamental algorithms for scientific computing in Python. Nature methods 17, 261-272

Wilcoxon, F., 1945. Individual Comparisons by Ranking Methods. Biometrics Bulletin 1, 80-83,10.2307/3001968.

Wilson, T.C., Silber, E.A., Colston, T.A., Elbing, B.R., Edwards, T.R., 2025. Bolide infrasound signal morphology and yield estimates: A case study of two events detected by a dense acoustic sensor network. The Astronomical Journal 169,10.3847/1538-3881/adbb70.

Woffinden, D., Eckman, R., Robinson, S., 2023. Optimized trajectory correction burn placement for the NASA Artemis II Mission, 45th Annual AAS Guidance, Navigation and Control (GN&C) Conference

Wright, W.M., 1983. Propagation in air of N waves produced by sparks. The Journal of the Acoustical Society of America 73, 1948-1955,10.1121/1.389585.

Yamamoto, M.-y., Ishihara, Y., Hiramatsu, Y., Kitamura, K., Ueda, M., Shiba, Y., Furumoto, M., Fujita, K., 2011. Detection of acoustic/infrasonic/seismic waves generated by hypersonic re-entry of the HAYABUSA capsule and fragmented parts of the spacecraft. Publications of the Astronomical Society of Japan 63, 971-978,10.1093/pasj/63.5.971.